\documentclass[11pt,a4paper]{article}
\usepackage{jheppub}
\usepackage{mathrsfs}
\usepackage{amsfonts}
\usepackage{setspace}
\usepackage{cellspace}
\usepackage{amsmath,amssymb,bm}
\usepackage[colorlinks=true,linkcolor=blue]{hyperref}
\usepackage{xcolor}
\usepackage{slashed}
\usepackage{hhline,multirow,tabularx}  
\usepackage{dcolumn}    
\usepackage{url}        
\usepackage{braket}     
\renewcommand{\bra}[1]{\left<#1\left|}
\renewcommand{\ket}[1]{\right|#1\right>}
\preprint{CNF-UMD-2021}

\title{Higher-Order Kinematical Effects in Deeply Virtual Compton Scattering}

\author[a]{Yuxun~Guo}
\author[a,b]{, Xiangdong~Ji}
\author[b]{and Kyle~Shiells}
\affiliation[a]{University of Maryland,\\ College Park, MD 20742 USA}
\affiliation[b]{Center for Nuclear Femtography,\\ 1201 New York Ave., NW, Washington DC, 20005, USA}
\emailAdd{yuxunguo@umd.edu}
\emailAdd{xji@umd.edu}
\emailAdd{kshiells@sura.org}
\abstract{We study the deeply virtual Compton scattering cross-section in twist-two generalized parton distribution (GPD) approximation, and show that different choices of light-cone vectors and gauges for the final photon polarization will lead to different higher-order kinematical corrections to the cross-section formula. 
The choice of light-cone vectors 
affects kinematic corrections at the twist-three level, accounting for the differences between the cross-section formulas in the literature. On the other hand, kinematical corrections from higher-twist GPDs should eliminate the light-cone dependence at twist three. Those light-cone dependencies are studied systematically at JLab 12 GeV and future EIC kinematics. They serve as the intrinsic systematic uncertainties in extracting the Compton form factors through the cross-section formula. More importantly, they are also necessary for understanding cross-section measurements with higher-twist precision and to reconstruct higher-order Compton form factors.}
\date{\today}
\begin{document}
\maketitle

\section{Introduction}

Deeply Virtual Compton Scattering (DVCS)~\cite{Ji:1996ek,Ji:1996nm}, namely the process of leptoproduction of a photon off the nucleon with the virtual photon momentum square, or the virtuality, $Q^2=-q^2$ much larger than the momentum transfer square $t$, provides a clean probe of the Generalized Parton Distributions (GPDs)~\cite{Muller:1994ses, Ji:1996ek} of the nucleon, which reveals important information of the nucleon such as the mass, angular momentum and mechanical properties~\cite{Ji:1994av, Ji:1996ek, Polyakov:2002yz} and describes the three-dimensional structure of the nucleon~\cite{Burkardt:2000za, Ji:2003ak, Belitsky:2003nz}. They enter the DVCS cross-section in the form of Compton Form Factors (CFFs), which can be generally expressed in terms of the convolution of GPDs and complex-valued Wilson coefficients~\cite{Belitsky:2001ns,Diehl:2003ny,Belitsky:2005qn}. Therefore, the DVCS cross-section measurements that allow us to measure the CFFs and reconstruct the GPDs play an important role in studying the internal structure of the nucleon. There have been measurements from HERA (H1 \cite{H1:2001nez,H1:2005gdw,H1:2007vrx,H1:2009wnw}, ZEUS~\cite{ZEUS:2003pwh,ZEUS:2008hcd} and HERMES~\cite{HERMES:2012gbh,HERMES:2012idp}) and Jefferson Lab (JLab) (CLAS~\cite{CLAS:2007clm,CLAS:2008ahu} and Hall A~\cite{JeffersonLabHallA:2006prd,JeffersonLabHallA:2007jdm}) and more
programs are planned in the future such as the JLab 24 GeV, EIcC\cite{Anderle:2021wcy} and EIC\cite{AbdulKhalek:2021gbh}. Those cross-section measurements as well as lattice calculations are crucial in order to understand the nucleon three-dimensional structure.

The theoretical foundation of studying the DVCS process is the collinear factorization proven in Quantum Chromodynamics (QCD) to the leading power accuracy of $Q$~\cite{Ji:1997nk,Ji:1998xh,Collins:1998be}, where twist expansion is introduced. However, the twist expansion depends on the convention one chooses, similar to the case in which one can do a series expansion for a given function at different points and get different series. Those series converge to the same original function that does not depend on the expansion points chosen, but their finite sums do. Therefore, intrinsic uncertainties due to different conventions will enter the cross-section formulas at higher order, which can be significant. As a rough estimation, at JLab 12 GeV kinematics where $Q^2 \sim 5~ \text{GeV}^2$, the twist-four parameter related to the target mass $M$ will be around $M^2/Q^2\sim 20\%$ which is quite substantial for precision measurements of CFFs, while the twist-three effects are even more considerable.

The twist-three effects in the DVCS process have been studied in details~\cite{Anikin:2000em,Penttinen:2000dg,Belitsky:2000vx,Kivel:2000cn,Radyushkin:2000ap}. Besides, progresses have been made in understanding the kinematical twist-four effects, such as finite $t$ and target mass correction for both forward and off-forward processes~\cite{Blumlein:2006ia,Blumlein:2009hit,Braun:2011zr,Braun:2011dg,Braun:2012bg,Braun:2012hq,Braun:2014sta}. In this paper, we address another source of higher-twist effects related to the choice of light cone vectors. While the choice of light cone vectors are totally conventional, those different conventions will lead to different cross-section formulas. As we will show, the dependence on the choice of light cone vectors shows up at the twist three level, making it a non-negligible effect one must take into account in order to obtain the cross-section formula to twist-three accuracy. On the other hand, we also show that the light cone dependence gets canceled at twist three by considering kinematical corrections from twist-three GPDs.

In this paper, we focus on the case where only the leading twist CFFs enter the cross-section, whereas all higher-twist CFFs are taken to be zero. Therefore, the remaining higher-twist effects will be completely kinematical, and we call them kinematical higher-twist effects. A clear understanding of those kinematical higher-twist effects is the precondition for separating the higher-twist CFFs from the kinematical contributions and reconstructing the higher-twist GPDs, which will be left for future work.

The organization of the paper is as follows. In Sec. \ref{sec:general}, we briefly review the theoretical frameworks of the DVCS process. In Sec. \ref{sec:xsectionlf}, we show how to define the light cone vectors in general and solve the kinematics as well as the cross-section with given choice of light cone vectors. In Sec. \ref{sec:comparison}, we compare our cross-section with the previous formulas both numerically and analytically by twist expansion, and show their connections and differences. In Sec. \ref{sec:WWrelation}, we consider part of the kinematical corrections from twist-three GPDs and show that the light cone dependence at twist-three level get canceled. In the end, we conclude in Sec. \ref{sec:conclusion}.

\section{Notation and Twist Expansion in DVCS}
\label{sec:general}
In this section, we briefly review the general formalism for the DVCS cross-section, see for instance \cite{Ji:1996nm,Belitsky:2001ns,Belitsky:2005qn,Kriesten:2019jep} for more details. Consider the electroproduction of  a photon off a proton as,
\begin{equation}
    e(k,h)+N(P,S)\to e(k',h')+N(P',S')+\gamma(q',\Lambda')\ ,
\end{equation}
where the $k,k',P,P',q'$ are the momenta of the particles and $h,h',S,S',\Lambda'$ correspond to their helicities or polarization. The helicities $h$ and $h'$ take the value of $\pm 1/2$, $\Lambda'$ takes $\pm 1$ while the nucleon polarization vector $S$ satisfies $S^2=-1$ and $S\cdot P=0$ and similarly for $S'$. We also define the virtual photon momentum $q\equiv k-k'$ and its helicity $\Lambda$. 

\begin{figure}[ht]
\centering
\begin{minipage}[b]{0.45\textwidth}
\centering
\includegraphics[width=0.8\textwidth]{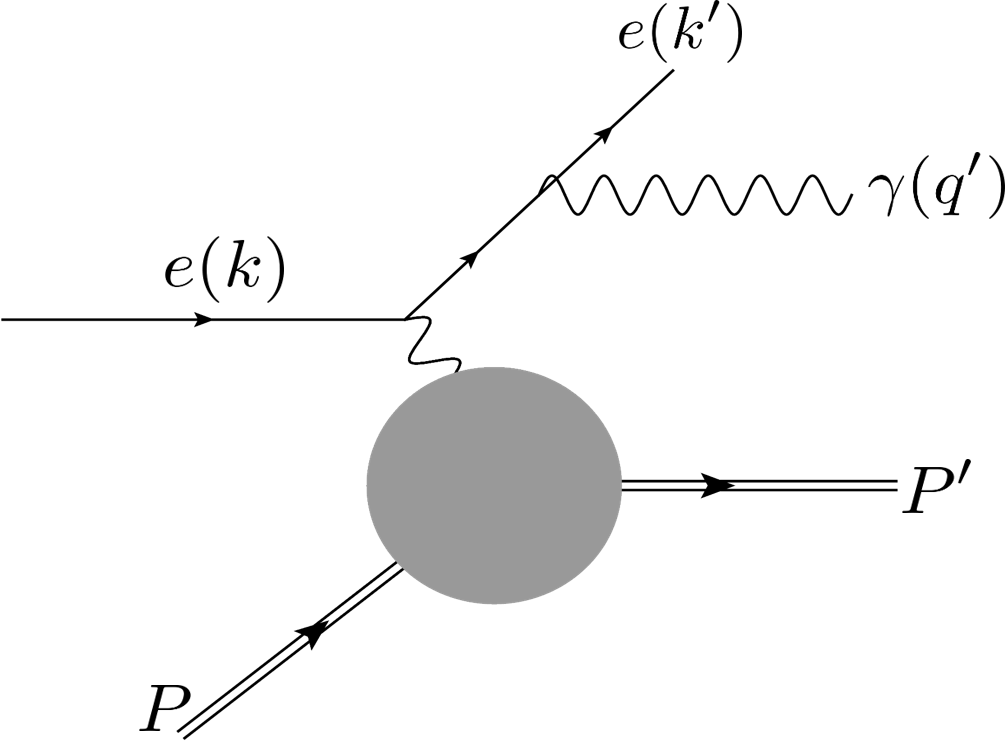}
\end{minipage}
\begin{minipage}[b]{0.45\textwidth}
\centering
\includegraphics[width=0.85\textwidth]{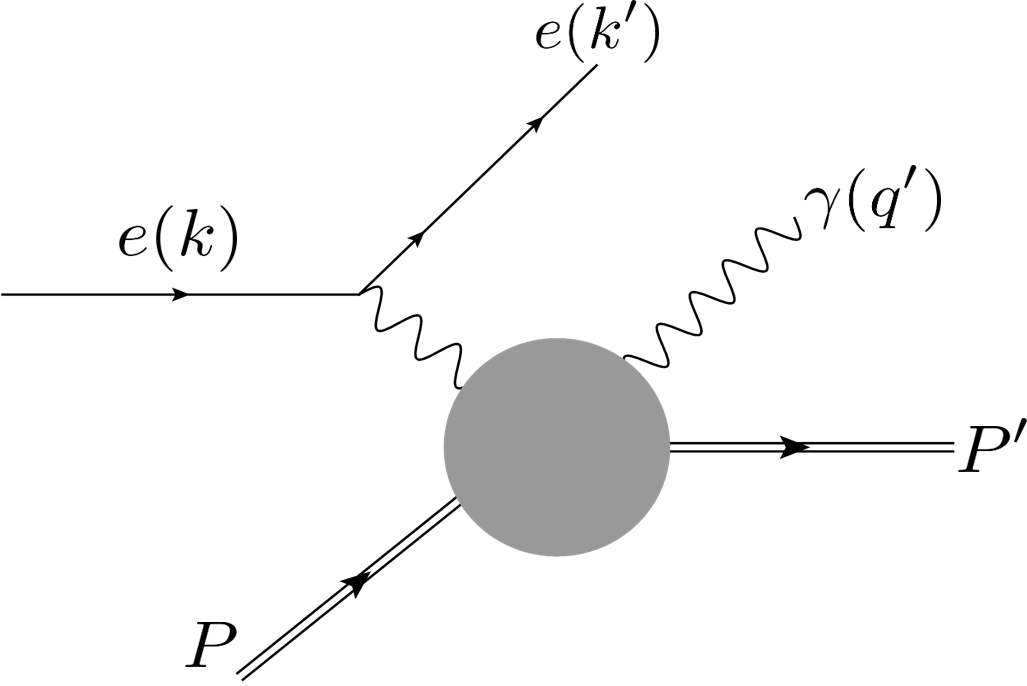}
\end{minipage}
\caption{\label{fig:amplitudeplot} The Bethe-Heitler (left) and the DVCS (right) processes that contribute to the real photon production.} 
\end{figure}
Two sub-processes, Bethe-Heitler (BH) and deeply virtual Compton scattering (DVCS), mainly contribute as shown in Fig. \ref{fig:amplitudeplot}, and the full amplitude will be the sum of them:
\begin{equation}
    \mathcal T=\mathcal T_{\rm{BH}}+\mathcal T_{\rm{DVCS}}\ .
\end{equation}
The electroproduction cross-section in the lab frame can be expressed in terms of the squared amplitude combined with some kinematical prefactors as \cite{Belitsky:2001ns,Kriesten:2019jep},
\begin{equation}
    \frac{\text{d}^5\sigma}{\text{d}x_B \text{d}Q^2\text{d}|t|\text{d}\phi \text{d}\phi_S}=\frac{\alpha_{\rm{EM}}^3 x_B y^2}{16\pi^2 Q^4\sqrt{1+\gamma^2}} \left|\mathcal T\right|^2 \ ,
\end{equation}
where we have the following definitions: fine structure constant $\alpha_{\rm{EM}}\equiv e^2/(4\pi)$ , photon virtuality $Q^2\equiv -q^2$, the Bjorken scaling variable $x_B\equiv Q^2/(2 P\cdot q)$, the electron energy loss variable $y \equiv P\cdot q/(P\cdot k)$ and momentum transfer square $t\equiv (P'-P)^2$. $\phi$ and $\phi_S$ describe the azimuthal angle between the leptonic plane and the reaction plane and the azimuthal angle between the leptonic plane and the target polarization vector in the case of transversely polarized target, respectively, see for instance the Fig. 47 in Ref. \cite{Belitsky:2005qn} where $\phi_S$ is denoted as $\Phi$. The notation 
\begin{equation}
   \gamma\equiv \frac{2M x_B}{Q} \ ,
\end{equation}
is introduced with $M$ the target mass. We also define the notation $\bar P\equiv (P+P')/2$, $\bar q\equiv (q+q')/2$ and $\Delta\equiv P'-P=q-q'$.

The squared amplitude consists of three parts,
\begin{equation}
\label{eq:sqrtamp}
    \left|\mathcal T \right|^2 =\left|\mathcal T_{\rm{BH}}\right|^2 +\left|\mathcal T_{\rm{DVCS}}\right|^2+T_{\rm{BH}}^*\mathcal T_{\rm{DVCS}}+\mathcal T_{\rm{DVCS}}^*\mathcal T_{\rm{BH}}\ .
\end{equation}
The interference contribution defined by the last term
\begin{equation}
    \mathcal I \equiv \mathcal T_{\rm{BH}}^*\mathcal T_{\rm{DVCS}}+\mathcal T_{\rm{DVCS}}^*\mathcal T_{\rm{BH}}=2 \text{Re}\left[\mathcal T_{\rm{BH}}^*\mathcal T_{\rm{DVCS}}\right]\ ,
\end{equation}
is linear in both the BH and DVCS amplitudes, and will be of more interest because of its rich angular structure, whereas the BH amplitude can be carried out relatively easily and will not be discussed in details. The notation
and convention for the laboratory frame can be found in Appendix \ref{app:labframe}, which follows from Ref. \cite{Kriesten:2019jep}.

\subsection{Twist-expansion of Compton tensor and light cone vectors}

The DVCS amplitude can be expressed in terms of the Compton tensor as,
\begin{equation}
\label{eq:DVCSamp}
    \mathcal T_{\rm{DVCS}}=\frac{e_l}{Q^2}\bar u(k',h') \gamma^\nu u(k,h) T^{\mu\nu} \varepsilon^*_{\mu}(q',\Lambda')\ ,
\end{equation}
with $e_l$ the lepton charge in the unit of electron charge that is positive (negative) for electron (positron), while the Compton tensor $T^{\mu\nu}$ is defined as,
\begin{equation}
   T^{\mu\nu}\equiv i \int \text{d}^4z e^{i(q+q')z/2} \bra{P',S'} \text{T}\left\{J^{\mu}\left(\frac{z}{2}\right)J^{\nu}\left(-\frac{z}{2}\right)\right\} \ket{P,S}\ ,
\end{equation}
with $J^\mu(z)$ the electromagnetic current operator and T the time order operator.
\begin{figure}[ht]
\centering
\begin{minipage}[b]{0.45\textwidth}
\centering
\includegraphics[width=\textwidth]{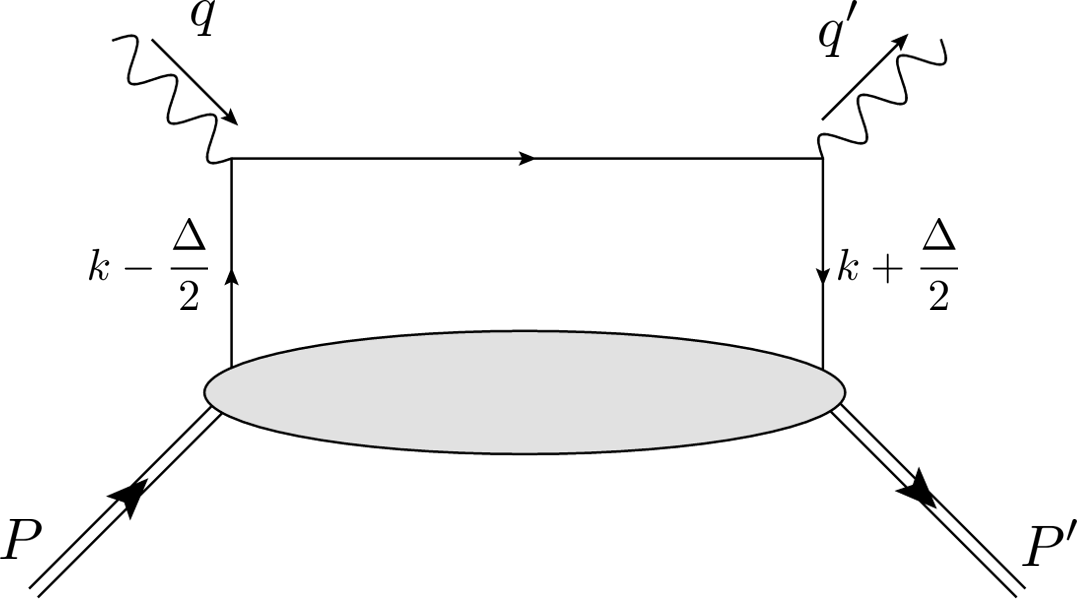}
\end{minipage}
\begin{minipage}[b]{0.47\textwidth}
\centering
\includegraphics[width=\textwidth]{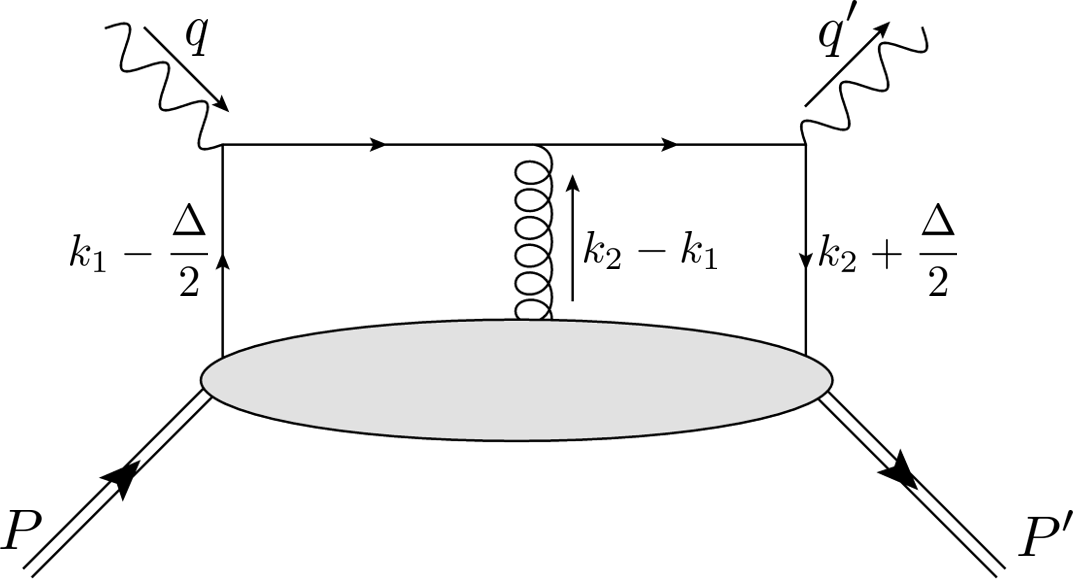}
\end{minipage}
\caption{\label{fig:handbagdiag} Examples of twist-two (left) and twist-three (right) handbag diagrams that contribute to the twist-three Compton tensor. Other diagrams that are  related to them by permutation are not shown here.} 
\end{figure}

The Compton tensor $T^{\mu\nu}$ is in general non-perturbative, but with QCD factorization in the Bjorken limit, it can be expressed in terms of the combination of perturbative coefficients and CFFs with Feynman diagrams as shown in Fig. \ref{fig:handbagdiag}, see also Ref.~\cite{Ji:1996nm,Belitsky:2005qn}. Then the Compton tensor $T^{\mu\nu}$ can be expressed in terms of twist expansion as,
\begin{align}
\begin{split}
   T^{\mu\nu}=T^{\mu\nu}_{(2)}+T^{\mu\nu}_{(3)}+\cdots\ ,
\end{split}
\end{align}
where higher twist contribution will not be considered, and the leading and next-to-leading Compton tensor can be written as
\cite{Belitsky:2005qn},
\begin{align}
\label{eq:comptontensor}
\begin{split}
   T^{\mu\nu}_{(2)}=&\int_{-1}^1\text{d}x \sum_{q}\left(\mathscr {T}^{\mu\nu}_{(2)} C^{q[-]}_{(0)}(x,\xi) n^\rho W^{\left[\gamma_\rho\right]}(x,\xi,t)  +\widetilde{\mathscr {T}}^{\mu\nu}_{(2)}C^{q[+]}_{(0)}(x,\xi) n^\rho W^{\left[\gamma_\rho\gamma^5\right]}(x,\xi,t)\right) \ ,\\
              T^{\mu\nu}_{(3)}=&\int_{-1}^1\text{d}x \sum_{q}\left(\mathscr {T}^{\mu\nu,\rho}_{(3)} C^{q[-]}_{(0)}(x,\xi)  W^{\left[\gamma_\rho^\perp\right]}(x,\xi,t)  +\widetilde{\mathscr {T}}^{\mu\nu,\rho}_{(3)}C^{q[+]}_{(0)}(x,\xi) W^{\left[\gamma_\rho^\perp\gamma^5\right]}(x,\xi,t)\right)\ ,
\end{split}
\end{align}
respectively, where $W^{\left[\Gamma\right]}$ are GPDs defined as
\begin{align}
\begin{split}
   W^{\left[\Gamma\right]}\equiv \int \frac{\text{d}\lambda}{2\pi} e^{i\lambda x} \bra{P',S'} \bar \psi\left(-\frac{\lambda n}{2}\right) \Gamma \psi\left(\frac{\lambda n}{2}\right)\ket{P,S}\ ,
\end{split}
\end{align}
with $\Gamma$ certain combination of Dirac matrices: $\Gamma=\{1,\gamma^\mu,\sigma^{\mu\nu},\gamma^\mu\gamma^5,\gamma^5\}$ and $\sigma^{\mu\nu} \equiv \frac{i}{2} \left[\gamma^\mu,\gamma^\nu\right]$. The tree-level Wilson coefficient functions $C^{q[\pm]}_{(0)}$ read~\cite{Ji:1996nm,Belitsky:2001ns},
\begin{align}
\begin{split}
C^{q[\pm]}_{(0)}=-Q_q^2\left(\frac{1}{x-\xi+i0}\mp \frac{1}{x+\xi-i0}\right)\ ,
\end{split}
\end{align}
where $Q_q$ is the charge of quarks in the unit of proton charge. The $(0)$ subscript stands for the tree-level coefficients, whereas in the next order of the strong coupling, much more complicated coefficients functions will be generated from loop correction. As for the GPDs, we use the definition in Ref. \cite{Ji:1996nm},
\begin{align}
   n_\mu W^{\left[\gamma^\mu\right]}&=\bar u(P',S')\left[
   \slashed n H(x,\xi,t)+\frac{i\sigma^{\mu\nu} n_\mu \Delta_\nu}{2M} E(x,\xi,t)\right] u(P,S) \ ,\\
   n_\mu W^{\left[\gamma^\mu\gamma^5\right]}&=\bar u(P',S')\left[
   \slashed n \gamma^5 \tilde H(x,\xi,t)+\frac{n^\mu \Delta_\mu \gamma^5 }{2M} \tilde E(x,\xi,t)\right] u(P,S)\ ,
\end{align}
and the corresponding CFFs are defined from the convolution of GPDs and Wilson coefficients as \cite{Belitsky:2001ns,Kriesten:2019jep},
\begin{equation}
    \begin{aligned}
   \mathcal H (\xi,t) \equiv \int_{-1}^1 \text{d}x C^{q[-]}_{(0)}(x,\xi) H (x,\xi,t)\ ,&\quad \mathcal E (\xi,t) \equiv \int_{-1}^1 \text{d}x C^{q[-]}_{(0)}(x,\xi) E (x,\xi,t)\ ,\\
  \tilde{ \mathcal H} (\xi,t) \equiv \int_{-1}^1 \text{d}x C^{q[+]}_{(0)}(x,\xi) \tilde H (x,\xi,t)\ ,&\quad \tilde{\mathcal E} (\xi,t) \equiv \int_{-1}^1 \text{d}x C^{q[+]}_{(0)}(x,\xi) \tilde E (x,\xi,t)\ ,
\end{aligned}
\end{equation}
where summing over all the quarks is implicit here. Note that we focus on the cross-section formula related to the leading twist CFFs in this paper, so the second line of Eq. (\ref{eq:comptontensor}) is set to zero hereafter unless mentioned otherwise.

In the Bjorken limit, both $q^2$ and $P\cdot q$ go to infinity while their ratio is kept finite. Therefore, both initial and final photon momenta reach the light cone and define the light cone vector $n$ in the expressions above (mass dimension $-1$), and the skewness parameter $\xi$ is defined as $\xi\equiv -\Delta \cdot n/(2\bar P \cdot n)$. With the light cone vector $n$ and its conjugate light cone vector $p$ (mass dimension +1) that satisfies $p^2=0$ and $n\cdot p=1$, we could define the transverse projection operator as,
\begin{align}
\label{eq:gperpdef}
\begin{split}
   g^{\mu\nu}_\perp\equiv g^{\mu\nu}-p^\mu n^\nu -n^\mu p^\nu\ .
\end{split}
\end{align}
Then the $\gamma^\perp_\rho$ which indicates terms that are non-zero only in the transverse direction can be expressed with the transverse projection operator as $\gamma^\perp_\rho \equiv g^\perp_{\rho\sigma} \gamma^\sigma$. We also define the following  antisymmetric symbol in the transverse space as,
\begin{align}
\label{eq:epsilonperpdef}
\begin{split}
   \epsilon^{\mu\nu}_\perp\equiv \epsilon^{\mu\nu\rho\sigma} p_\rho n_\sigma \ .
\end{split}
\end{align}
Then the Compton tensor coefficients $\mathscr {T}^{\mu\nu}$s in the Compton tensor can be expressed in terms of the light cone vectors as~\cite{Belitsky:2005qn},
\begin{align}
\label{eq:LFcompton}
\begin{split}
\mathscr {T}^{\mu\nu}_{(2)} &=-\frac{1}{2}\left[g^{\mu\nu}_\perp- \frac{1}{p \cdot \bar q}\left(p^\mu {q'}^\nu_\perp+q^\mu_\perp p^\nu\right) \right]\ ,\\
\widetilde{\mathscr {T}}^{\mu\nu}_{(2)} &=\frac{i}{2}\left[\epsilon^{\mu\nu}_\perp- \frac{1}{p \cdot \bar q}\left(-p^\mu \epsilon^{\nu\rho}_\perp {q'}_\rho^\perp+\epsilon^{\mu\rho}_\perp q_\rho^\perp p^\nu\right) \right]\ ,\\
\mathscr {T}^{\mu\nu,\rho}_{(3)} &=\frac{1}{2 p \cdot \bar q}\left[{q'}^\mu g_\perp^{\nu\rho}+g_\perp^{\mu\rho}\left(q^\nu+4\xi p^\nu\right)\right]\ ,\\
\widetilde{\mathscr {T}}^{\mu\nu,\rho}_{(3)} &=\frac{i}{2 p \cdot \bar q}\left[\epsilon^{\mu\nu\rho\sigma}\bar q_\sigma+\xi \left(p^\mu \epsilon_\perp^{\rho\nu}+p^\nu \epsilon_\perp^{\rho\mu}\right)\right]\ ,
\end{split}
\end{align}
with twist-three accuracy, where the leading terms are first derived in Ref. \cite{Ji:1996nm}. It is worth noting that $\mathscr {T}^{\mu\nu}$'s are ambiguous unless the light cone vectors are clearly defined, and we will leave the detailed discussion on this for the next section.

We notice that there are other set-ups without explicit light cone vectors found in the literature. One of the commonly used is that of Belitsky-Kirchner-M\"{u}ller (BKM)~\cite{Belitsky:2001ns}, while a similar set-up also is used in the Belitsky-M\"{u}ller-Ji (BMJ) results~\cite{Belitsky:2012ch}. Instead of the light cone vectors $n$ and $p$, the effective light cone vectors are expressed in terms of the covariant physical momenta as, 
\begin{align}
\begin{split}
    p_{\rm{BKM}}^\mu = \bar P^\mu \quad, n_{\rm{BKM}}^\mu= \frac{\bar q^\mu}{\bar P\cdot \bar q}\ ,
\end{split}
\end{align}
which are not light-like: $p_{\rm{BKM}}^2 \not =0, n_{\rm{BKM}}^2 \not =0$. Those vectors are not ideal for rigorous twist expansions, and the GPDs defined with them will mix with higher-order GPDs. But still, one can
calculate the Compton tensor coefficients with those effective light cone vectors, which gives~\cite{Belitsky:2005qn},
\begin{align}
\label{eq:BKMcompton}
\begin{split}
    \mathscr{T}^{\mu\nu}_{\rm{BKM},(2)} &=-\frac{1}{2}\mathcal P^{\mu\rho}\left[ g_{\rho\sigma} + \frac{\bar{q}^2}{(\bar P \cdot \bar q)^2}\bar P_\rho \bar P_\sigma \right]\mathcal P^{\sigma\nu}\ ,\\
\widetilde{\mathscr{T}}^{\mu\nu}_{\rm{BKM},(2)} &=\frac{i}{2\bar P\cdot \bar q}\mathcal{E}^{\mu\nu\rho\sigma}\bar P_\rho \bar q_\sigma \ ,\\
\mathscr{T}^{\mu\nu,\rho}_{\rm{BKM},(3)} &=-\frac{\bar {q}^2}{2 (\bar P \cdot \bar q)^2}\left[\mathcal P^{\mu\rho} \bar P_\sigma \mathcal P^{\sigma\nu} +\mathcal P^{\mu\sigma}\bar P_\sigma \mathcal P^{\rho\nu}\right]\ ,\\
\widetilde{\mathscr {T}}^{\mu\nu,\rho}_{\rm{BKM},(3)} &=\frac{i}{2\bar P\cdot \bar q}\mathcal{E}^{\mu\nu\rho\sigma}\bar q_\sigma \ ,
\end{split}
\end{align}
where the projectors are defined as,
\begin{align}
\begin{split}
   \mathcal P^{\mu\nu}\equiv g^{\mu\nu}-\frac{q^\mu q'^\nu}{q\cdot q'}\ ,\quad \mathcal{E}^{\mu\nu\rho\sigma}\equiv\left(g^{\mu\alpha}-\frac{\bar P^\mu q'^\alpha}{\bar P\cdot q'}\right)\left(g^{\nu\beta}-\frac{\bar P^\nu q^\beta}{\bar P\cdot q}\right) \epsilon_{\alpha\beta}^{~~~~\rho\sigma}\ ,
\end{split}
\end{align}
and we rewrite the original expression in terms of our notations. The BKM Compton tensor coefficients are equivalent to the Compton tensor coefficients expressed with light cone vectors $p^\mu$ in the direction of $\bar P$ and $n^\mu$ in the direction of $\bar q$ up to twist-four corrections. Without loss of generality, we will still use our light cone vectors $p$ and $n$ unless otherwise specified. A careful comparison between those two different set-ups is left to the next section.

\subsection{Gauge invariance, helicity amplitude and higher order effects}

\label{subsec:gaugeinvariance}
Up to now, we have not used the explicit photon polarization vectors. It is simple to sum over the final photon polarization, for which we replace,
\begin{align}
     \sum_{\Lambda'=\pm1}  \varepsilon^\mu(q',\Lambda')\varepsilon^{*\nu}(q',\Lambda') \to -g^{\mu\nu}\ ,
\end{align}
and then the cross-sections will be independent of the specific form of polarization vectors of the photon. However, the substitution of photon polarization vectors needs more care, since the Compton tensor $T^{\mu\nu}$ only has twist-three accuracy, and it does not satisfy the current conservation condition exactly as $q'_\mu T^{\mu\nu}= T^{\mu\nu} q_\nu= 0$. Instead, it only conserves up to a higher twist term,
\begin{align}
\label{eq:compgauge}
    q'_\mu T^{\mu\nu}\sim T^{\mu\nu} q_\nu\sim \mathcal O\left(\Delta_\perp^2\right)\ .
\end{align}
Therefore, the exact gauge invariance cannot be maintained, and the gauge dependence of the polarization sum is non-trivial.

To be more specific, consider an on-shell photon $A^\mu(q')$, for which one additional gauge fixing condition is needed besides the physical requirement $q'^\mu A_\mu (q')=0$, since there are only two degrees of freedom.  If we choose in general $V^\mu A_\mu (q')=0$ with $V^\mu$ any four-vector, the photon polarization sum can be written as
\begin{align}
    \label{eq:completeness2}
     &\sum_{\Lambda'=\pm1}  \varepsilon^\mu(q',\Lambda')\varepsilon^{*\nu}(q',\Lambda') =-g^{\mu\nu}+\frac{V^\mu q'^\nu+V^\nu q'^\mu}{V\cdot q'}-\frac{V^2 q'^\mu q'^\nu}{(V\cdot q')^2}\ .
\end{align}
Besides the $-g^{\mu\nu}$ term, we also have terms that depend explicitly on the gauge choice $V^\mu$. Those terms are proportional to $q'^\mu$ or $q'^\nu$, so they do not contribute if the Compton tensor conserves to all orders. However, since the Compton tensor only conserves up to some higher order twist-four terms, ignoring those extra gauge-dependent terms in the polarization sum will introduce extra twist-four contributions to the cross-section and is not preferred. It is worth noting that the necessity of using Eq. (\ref{eq:completeness2}) results from the unconserved Compton tensor. Thus, one might regard
the usage of Eq. (\ref{eq:completeness2}) as a gauge improving process, since it eliminate the terms in the Compton tensor that are proportional to $q'^\mu$, which are unphysical. The effect of the gauge dependence of the cross-section will be discussed more carefully later.

Another way of taking care of the gauge dependence is to use the helicity amplitude method. Since the Compton tensor is always connected to two photons, one could rewrite Eq. (\ref{eq:DVCSamp}) in terms of the photon polarization vectors as~\cite{Belitsky:2008bz,Belitsky:2010jw,Kriesten:2019jep}
\begin{equation}
    \mathcal T_{\rm{DVCS}}=\frac{1}{Q^2}\bar u(k',h') \gamma^\rho  u(k,h) \sum_{\Lambda} \varepsilon^{*}_\rho(q,\Lambda) \varepsilon_\nu (q,\Lambda)  T^{\mu\nu} \varepsilon^{*}_{\mu}(q',\Lambda')\ ,
\end{equation} 
where $\Lambda$ sums over the three possible polarizations of the virtual photon. If we define the helicity amplitude as,
\begin{equation}
  T_{\Lambda\Lambda'}\equiv \varepsilon_\nu (q,\Lambda)  T^{\mu\nu} \varepsilon^{*}_{\mu}(q',\Lambda')\ ,
\end{equation} 
the DVCS amplitude can then be expressed in helicity amplitude as,
\begin{equation}
    \mathcal T_{\rm{DVCS}}=\frac{1}{Q^2}\sum_{\Lambda}\bar u(k',h') \gamma^\rho  u(k,h)  \varepsilon^{*}_\rho(q,\Lambda) T_{\Lambda\Lambda'} \ .
\end{equation}
The DVCS amplitude expressed this way will be independent of frame, since helicity is an invariant scalar. 
{\it However, the definition of helicity depends on the choice of light cone vectors, as the transverse directions are defined by the light cone vectors, and thus each helicity amplitude will depend on the convention for the light cone vectors.} Meanwhile, the DVCS amplitude still depends on the gauge choice implicitly, because the photon polarization vectors $\varepsilon^{*}_{\mu}(q',\Lambda')$ do. The helicity amplitude method does not resolve the gauge dependence issue, it only makes it implicit. Still, we will need to fix the gauge in order to calculate the helicity amplitude, and the helicity amplitude will be different under different gauge choices, though it will be a higher-twist effect.

In the literature, the radiation gauge in the lab frame is commonly chosen, where we set $V^\mu \propto (1,0,0,0)$ in the lab frame, for which we could simply set $V^\mu=P^\mu$, since Eq. (\ref{eq:completeness2}) is invariant under a rescaling of $V^\mu$. In order to calculate the helicity amplitude, we set up our coordinates in the lab frame by setting the virtual photon momentum in the $-z$ direction and the lepton beam momentum in the $x-z$ plane. The details of all the physical momenta are given in Appendix \ref{app:labframe} which follows Ref.~\cite{Belitsky:2001ns,Kriesten:2019jep}. Then the photon polarization vectors can be written as
\begin{align}
\begin{split}
  \varepsilon(q,\Lambda=\pm1)&= \frac{e^{-i \Lambda \phi}}{\sqrt{2}}(0,1,i\Lambda,0)\ ,\\
  \varepsilon(q,\Lambda=0)&= \frac{Q}{\sqrt{2 x_B M}}(-\sqrt{1+\gamma^2},0,0,1)\ ,\\
  \varepsilon^* (q',\Lambda'=\pm1) &= \frac{1}{\sqrt{2}}\left(0,-\cos\theta\cos\phi+i\Lambda' \sin\phi,-i\Lambda' \cos\phi- \cos\theta\sin\phi,\sin\theta\right)\ ,
\end{split}
\end{align} 
and the completeness relation for the virtual photon reads
\begin{align}
\label{eq:completeness1}
     &\sum_{\Lambda=\pm1}  \varepsilon^\mu(q,\Lambda)\varepsilon^{*\nu}(q,\Lambda) -\varepsilon^\mu(q,\Lambda=0)\varepsilon^{*\nu}(q,\Lambda=0)=-g^{\mu\nu}+\frac{q^\mu q^\nu}{q^2}\ .
\end{align}
The polarization sum of the virtual photon polarization can be replaced with $-g^{\mu\nu}$, since the leptonic part always satisfies the relation $\bar u(k')\gamma^\nu u(k) q_\nu=0$, and the extra $q^\mu q^\nu$ term will not contribute. 

We calculate the helicity amplitudes using the Compton tensor coefficients in Eq. (\ref{eq:BKMcompton}) for the BKM set-up as an example, which reads,
\begin{align}
\label{eq:BKMhelicity}
\begin{split}
    \mathscr {T}^{\Lambda\Lambda'=++}_{\rm{BKM},(2)}=\mathscr {T}^{\Lambda\Lambda'=--}_{\rm{BKM},(2)}&=\frac{1}{2}-\frac{x_B^2(t(1- x_B)+M^2x_B^2)}{2(2-x_B)^2Q^2}+\mathcal{O}(Q^{-3})\ ,\\
    \mathscr {T}^{\Lambda\Lambda'=+-}_{\rm{BKM},(2)}=\mathscr {T}^{\Lambda\Lambda'=-+}_{\rm{BKM},(2)}&=-\frac{x_B^2(t(1- x_B)+M^2x_B^2)}{2(2-x_B)^2Q^2}+\mathcal{O}(Q^{-3})\ ,\\
    \mathscr {T}^{\Lambda\Lambda'=0+}_{\rm{BKM},(2)}=\mathscr {T}^{\Lambda\Lambda'=0-}_{\rm{BKM},(2)}&=\frac{\sqrt{-x_B^2(t(1- x_B)+M^2x_B^2)}}{\sqrt{2}(x_B-2)Q}+\mathcal{O}(Q^{-3})\ ,
\end{split}
\end{align}
where we defined $ \mathscr {T}^{\Lambda\Lambda'}_{(2)}\equiv \varepsilon_\nu (q,\Lambda)  \mathscr {T}^{\mu\nu}_{(2)} \varepsilon^{*}_{\mu}(q',\Lambda')$ and they are consistent with the results in Ref. \cite{Belitsky:2008bz}. We leave the helicity amplitudes with light cone vectors and the comparison with the BKM ones to the next section, after we define our light cone vectors there.

\section{Cross-Section Formalism With General Light Cone Vectors}
\label{sec:xsectionlf}
In the last section, we introduce the light cone vectors $n$ and $p$ without defining them, since their definitions are conventional. In this section, we define those light cone vectors in a general form such that light cone vectors under different conventions can be studied systematically. Then we can use them to work out the cross-section formula explicitly.

Note that the BH process does not involve the Compton tensor, so it does not depend on the light cone vectors defined. With the BH amplitude which can be written as, 
\begin{align}
\begin{split}
    \mathcal T_{\rm{BH}}=-\frac{1}{t}\varepsilon^*_{\mu}(q',\Lambda') l_{\rm{BH}}^{\mu\nu} \bar u(P',S')\left[(F_1+F_2)\gamma_\nu-\frac{\bar P_\nu}{M}F_2\right]u(P,S)\ ,
\end{split}
\end{align}
where the BH leptonic amplitude is defined as,
\begin{align}
    l_{\rm{BH}}^{\mu\nu}\equiv \bar u(k',h') \left[\frac{\gamma^{\mu}(\slashed k'+\slashed q')\gamma^\nu}{(k'+q')^2}+\frac{\gamma^{\nu}(\slashed k-\slashed q')\gamma^\mu}{(k-q')^2}\right] u(k,h) \ ,
\end{align}
one can write down the BH cross-section without light cone vectors, see for instance Ref. \cite{Ji:1996nm,Belitsky:2001ns,Kriesten:2019jep} for more details. With that in mind, we will focus on the pure DVCS and interference cross-sections and treat and the BH contribution as the background in this paper.

\subsection{Choice of light cone directions}

Since each four vector has four degrees of freedom, we will need 8 constraints to determine the two light cone vectors $n$ and $p$. To start with, we have 3 constraints $n^2=0$, $p^2=0$ and $p\cdot n=1$ from the definitions of the light cone vectors. In addition, since the light cone vectors should always lie in the space spanned by the four physical momenta $P,P',q,q'$ who only span a three-dimensional subspace due to momentum conservation, we have two extra constraints that
\begin{align}
n_\mu  \epsilon^{\mu\nu\rho\sigma}P_\nu P'_\rho q_\sigma=p_\mu  \epsilon^{\mu\nu\rho\sigma}P_\nu P'_\rho q_\sigma=0\ .
\end{align}
Therefore, there remain $3=2\times 4-5$ degrees of freedom unfixed in defining the light cone vectors which are totally conventional. First, we could fix the scale of the light cone vectors by asking that $\bar P \cdot n =1$, which can always be done by boosting in the $n$ direction. Then we are left with two extra degrees of freedom that describe the direction of the two light cone vectors $n$ and $p$. Two parameters, $\alpha$ and $\beta$, are introduced to define the directions of two light cone vectors, with which we define 
\begin{align}
q_L\equiv \alpha q^\mu+\left(1-\alpha\right) q'^\mu\ , \qquad P_L \equiv \beta P^\mu+\left(1-\beta\right) P'^\mu\ .
\end{align}
Then our light cone vectors $p$ and $n$ are defined such that $q_L$ and $P_L$ are longitudinal:
\begin{align}
q_L^\mu g_{\mu\nu}^\perp = P_L^\mu g_{\mu\nu}^\perp=0\ .
\end{align}
The above two conditions completely fix the light cone vectors, though implicitly, and thus all the physical momenta can be expressed in terms of light cone vectors and their transverse components using the general decomposition,
\begin{align}
V^\mu = p^\mu (V\cdot n )+ n^\mu (V\cdot p)+V^\mu_\perp\ .
\end{align}
The following expansions can be worked out for all the physical momenta,
\begin{align}
\label{eq:momentumlf}
\begin{split}
    P^{\mu}&= (1+\xi) p^\mu + \frac{M^2+(\beta-1)^2|\Delta_\perp|^2}{2(1+\xi)} n^\mu+(\beta-1)\Delta^\mu_\perp\ ,\\
P'^{\mu}&= (1-\xi) p^\mu +\frac{M^2+\beta^2|\Delta_\perp|^2}{2(1-\xi)} n^\mu+\beta\Delta^\mu_\perp\ ,\\
\bar P^{\mu}&= p^\mu + \frac{1}{2}\left(\frac{M^2+(\beta-1)^2|\Delta_\perp|^2}{2(1+\xi)}+\frac{M^2+\beta^2|\Delta_\perp|^2}{2(1-\xi)}\right)n^\mu+(\beta-1/2)\Delta^\mu_\perp\ ,\\
\Delta^{\mu}&= -2\xi p^\mu -\frac{|\Delta_\perp|^2+t}{4\xi}n^\mu+\Delta^\mu_\perp\ ,\\
q^{\mu}&= -2\eta p^\mu - \frac{-Q^2+(\alpha-1)^2|\Delta_\perp|^2}{4\eta} n^\mu+(1-\alpha)\Delta^\mu_\perp\ ,\\
q'^{\mu}&= -2(\eta-\xi) p^\mu - \left(\frac{-Q^2+(\alpha-1)^2|\Delta_\perp|^2}{4\eta}-\frac{|\Delta_\perp|^2+t}{4\xi}\right) n^\mu-\alpha\Delta^\mu_\perp\ .
\end{split}
\end{align}

With all the physical constraints on the momenta such as $P^2=P'^2=M^2, q^2=-Q^2, q'^2=0$ and $\Delta^2=t$ and the momentum conservation $\Delta=P'-P=q-q'$, the $|\Delta_\perp|^2$ and the parameter $\eta \equiv -q \cdot n /(2\bar P \cdot n)$ can be solved, for which we have
\begin{align}
    |\Delta_\perp|^2&=\frac{-4\xi^2 M^2+(\xi^2-1) t}{\left[1+(2\beta-1)\xi\right]^2} \ ,\\
\begin{split}
    \eta&=\xi\frac{\sqrt{(Q^2+t)^2+4\alpha(Q^2+(1-\alpha)t)|\Delta_\perp|^2}-Q^2+t+2(1-\alpha)|\Delta_\perp|^2}{2(t+|\Delta_\perp|^2)} \\
    &=\xi\left(1-\frac{\alpha^2 |\Delta_\perp|^2}{Q^2}\right) +\mathcal {O}(Q^{-4})\ .
\end{split}
\end{align}
The last undetermined constant in the above expansions is the skewness parameter $\xi$, which can be solved in terms of the Bjorken scaling parameter $x_B$ with $2 P\cdot q= Q^2/x_B$, and the solution reads
\begin{align}
\begin{split}
   \xi=\frac{x_B}{2-x_B}-\frac{2 x_B\left\{2\alpha M^2 x_B^2  +(x_B-1)\left[1-2\alpha+(\beta-1)x_B\right]t\right\}}{(x_B-2)^2(1+(\beta-1)x_B)Q^2}+\mathcal{O}(Q^{-4})\ .
\end{split}
\end{align}
An exact solution of $\xi$ is in general inaccessible, so we expand $\xi$ with twist-four accuracy. However, in the case where $\alpha=0$ we can solve them exactly to get,
\begin{align}
\begin{split}
    \eta(\alpha=0)=\xi\ ,\qquad  \xi(\alpha=0)=\frac{x_B(1+t/Q^2)}{2-x_B(1-t/Q^2)}\ .
\end{split}
\end{align}
which corresponds to choosing the light cone vector $n$ in the direction $q'$. The above definition of $\xi$ with $\alpha=0$ agrees with the one in Ref. \cite{Braun:2011dg}. This simplification can be easily understood considering that $q'^2 =0$, and thus the choice $n\propto q'$ builds a simple connection between the physical momenta and light cone vectors.

\begin{figure}[ht]
\centering
\includegraphics[width=0.4\textwidth]{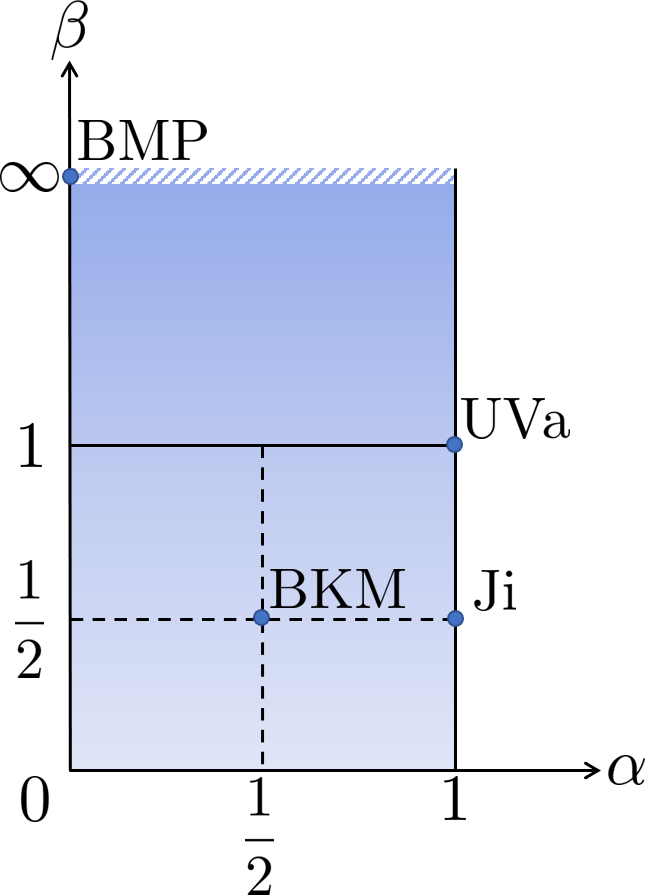}
\caption{\label{fig:parameter} A plot illustrating that different conventions in the literatures corresponds to different choice of $\alpha$ and $\beta$.}
\end{figure}

Other different conventions have been used in the literatures. Besides the Braun-Manashov-Pirnay (BMP) convention~\cite{Braun:2011dg}, corresponding to $\alpha=0,\beta\to\infty$,  in Ref.~\cite{Ji:1996nm} (Ji), $\alpha=1,\beta=1/2$ , in Ref.~\cite{Belitsky:2001ns} (BKM), $\alpha=\beta=1/2$ and in Ref.\cite{Kriesten:2019jep} (UVa), $\alpha=\beta=1$ is chosen respectively. Those different conventions are plotted together in Fig. \ref{fig:parameter}.

The above vector expansions on the light cone also allow us to explicitly express the light cone vectors in terms of the physical momenta as
\begin{align}
\begin{split}
    p^\mu = \frac{-D P_L +B q_L}{BC-DA}\ ,\\
    n^\mu = \frac{C P_L - A q_L}{BC-DA}\ ,
\end{split}
\end{align}
where we define the constants as
\begin{align}
\begin{split}
   A&=(1-\xi)+2\xi\beta\ ,\\
   B&=\frac{M^2+\beta^2 |\Delta_\perp|^2}{2(1-\xi)}+\beta \frac{t+|\Delta_\perp|^2}{4\xi}\ ,\\
   C&=-2 (\eta-\xi)-2\xi \alpha \ ,\\
   D&=-\frac{-Q^2+(\alpha-1)^2|\Delta_\perp|^2}{4\eta}+(1-\alpha)\frac{|\Delta_\perp|^2+t}{4\xi}\ .
\end{split}
\end{align}
Therefore, all the variables in the light cone coordinates can be expressed in terms of physical momenta and lab frame kinematical variables with twist-four accuracy. 

While each set of value of $\{\alpha,\beta\}$ represents a different choice of light cone vectors, the value of $\{\alpha,\beta\}$ should not be chosen arbitrarily. The twist expansion assumes that the photon momenta $q$ and $q'$ are hard and approach the light cone $n$ direction in the Bjorken limit, so $\alpha$ should satisfy $|\alpha|^2 |\Delta_\perp|^2 \ll Q^2$ and $|1-\alpha|^2 |\Delta_\perp|^2 \ll Q^2$. Since $|\Delta_\perp|^2/Q^2$ is already twist-four suppressed, we can simply take 
\begin{align}
\begin{split}
   0\le \alpha\le 1\ ,
\end{split}
\end{align}
which has very simple physical meaning that when $\alpha=0$ the light cone vector $n$ is in the direction of $q'$ and when $\alpha=1$, it basically aligns with the $q$.

As for the choice of $\beta$, while it is not as constrained, the twist expansion has a singularity at $1+(2\beta-1)\xi=0$. At the singularity, we have $n \cdot P_L=0$ and thus $P_L$ has to be proportional to $n$. Then in the Bjorken limit, $q,q'$ and $P_L$ will all approach the $n$ direction and the $p$ vector is not well-defined, whereas all physical quantities are regular when $\beta \to \infty$ and it is safe to let $\beta\to\infty$. With that in mind, we require that $|\Delta_\perp|^2\ll Q^2$, which gives  $|1+(2\beta-1)\xi| \gg \left(-4\xi^2 M^2+(\xi^2-1) t\right)/Q^2$. Again, since $t/Q^2$ and $M^2/Q^2$ are already twist-four suppressed, we can simply have $|1+(2\beta-1)\xi| \gtrsim \mathcal{O}(1)$. In practice, we choose
\begin{align}
\begin{split}
  0\le \beta \le 2 \ .
\end{split}
\end{align}
For $0<\xi<1$, it satisfies $|1+(2\beta-1)\xi| \ge 1-\xi$. Though large $\beta$ will not cause any singularity, we drop that region from our choice of $\beta$ since it differs too much from the other choices in the literatures with $\beta$ from $0$ to $1$. Apparently, the larger range we choose for $\alpha$ and $\beta$, the more conventions will be covered, but the more uncertainties will appear in the final cross-section consequently. The above ranges of $\alpha$ and $\beta$ define the confidence bands we used in this paper.

\subsection{Light cone dependence of helicity amplitudes}

With all the light cone vectors defined, we can then perform the calculation of the helicity amplitudes in the
radiation gauge. For comparison, we first calculate the helicity amplitudes using the Compton tensor coefficients in Eq. (\ref{eq:LFcompton}) expressed with light cone vectors, and get
\begin{align}
\begin{split}
    \mathscr {T}^{\Lambda\Lambda'=++}_{(2)}=\mathscr {T}^{\Lambda\Lambda'=--}_{(2)}&=\frac{1}{2}-\frac{x_B^2(t(1- x_B)+M^2x_B^2)(\beta-1)^2}{2[1+(\beta-1)x_B]^2Q^2}+\mathcal{O}(Q^{-3})\ ,\\
    \mathscr {T}^{\Lambda\Lambda'=+-}_{(2)}=\mathscr {T}^{\Lambda\Lambda'=-+}_{(2)}&=-\frac{x_B^2(t(1- x_B)+M^2x_B^2)(\beta-1)^2}{2[1+(\beta-1)x_B]^2Q^2}+\mathcal{O}(Q^{-3})\ ,\\
    \mathscr {T}^{\Lambda\Lambda'=0+}_{(2)}=\mathscr {T}^{\Lambda\Lambda'=0-}_{(2)}&=\frac{(\beta-1)\sqrt{-x_B^2(t(1- x_B)+M^2x_B^2)}}{\sqrt{2}[1+(\beta-1)x_B]Q}+\mathcal{O}(Q^{-3})\ ,
\end{split}
\end{align}
which indeed reduces to the BKM helicity amplitudes in Eq. (\ref{eq:BKMhelicity}) when one takes $\beta=1/2$, consistent with the choice made in the BKM set-up. The helicity amplitudes turn out to be independent of $\alpha$ at twist four, whereas it depends on $\beta$ starting at twist three. The dependence of the helicity amplitudes on the choice of light cone vectors is a mixture, originated from two sources. The obvious one is the Compton tensor coefficients in Eq. (\ref{eq:LFcompton}) that depends on the light cone vectors. Besides that, the polarization vectors also depend on the choice of light cone vectors, because the transverse and longitudinal directions are conventional, as one could see in the definition of $g_\perp^{\mu\nu}$ and $\epsilon^{\mu\nu}_\perp$ in Eqs. (\ref{eq:gperpdef}) and (\ref{eq:epsilonperpdef}).

More specifically, in the Bjorken limit all physical momenta reach the light cone and instead of a three-dimensional subspace, they only span a two-dimensional space expanded by the light cone vectors $n$ and $p$. The remaining two-dimensional space orthogonal to the physical momenta defines the transverse space, and the helicity can be defined unambiguously in the two-dimensional transverse space. However, the above picture ceases to work when higher twist effects are taken into account. The four physical momenta will then span a three-dimensional subspace, and only one transverse direction, the direction of $\epsilon^{\mu \nu\rho \sigma} P_\nu P'_\rho q_\sigma$,  is well-defined. Then the helicity corresponding to a circular polarization that involves two transverse directions inevitably mixes the well-defined transverse direction with another transverse direction that lies within the $P,P',q,q'$ plane, which are totally conventional.

In the end, when we sum over all the polarizations, the dependence of polarization vectors on the light cone vectors will vanish, since they reduce to the tensor in Eq. (\ref{eq:completeness2}), while the dependence of the Compton tensor on the light cone vectors will persist in the final cross-section formula. On the other hand, the gauge dependence is always there, explicit or not, which cannot be avoided.

With those thoughts in mind, we will not use the helicity amplitude method for the cross-section formula in this paper. Instead, we simply apply the gauge fixing condition directly when summing over the final photon polarization, which has the same effect as the helicity amplitude method in terms of the final cross-section but does not involve specific expressions of the polarization vectors.

\subsection{Pure DVCS cross-section and polarization asymmetry}

\label{subsec:puredvcs}

We start with the pure DVCS cross-section, which is relatively simpler. The DVCS squared amplitude $\left| {\mathcal T}_{\rm{DVCS}}\right|^2$  in Eq. (\ref{eq:sqrtamp}) can be written into the leptonic and hadronic parts as,
\begin{align}
    \left| {\mathcal T}_{\rm{DVCS}}\right|^2= \frac{1}{Q^4}L_{\rm{DVCS}}^{\rho\sigma} H^{\rm{DVCS}}_{\rho\sigma} \ ,
\end{align}
where we have
\begin{align}
       L_{\rm{DVCS}}^{\rho\sigma}\equiv&\sum_{h'}\bar u(k,h)\gamma^\rho u(k',h')  \bar u(k',h') \gamma^\sigma u(k,h) \ ,\\
   \label{eq:hdvcs}
   H_{\rm{DVCS}}^{\rho\sigma}\equiv & \sum_{S',\Lambda'} T^{* \mu\rho} T^{ \nu\sigma} \varepsilon_\mu(q',\Lambda')\varepsilon^*_{\nu}(q',\Lambda') \ .
\end{align}
The leptonic tensor $L_{\rm{DVCS}}^{\rho\sigma}$ can always be expressed in terms of the polarized part and the unpolarized part as,
\begin{align}
\label{eq:ldvcspol}
\begin{split}
   L_{\rm{DVCS}}^{\rho\sigma}= L_{\rm{DVCS,U}}^{\rho\sigma}+ i2h L_{\rm{DVCS,L}}^{\rho\sigma}\ ,
\end{split}
\end{align}
where each term can be solved as \cite{Ji:1996nm},
\begin{align}
   L_{\rm{DVCS,U}}^{\rho\sigma}&= 2\left(k^\rho k'^\sigma+k'^\rho k^\sigma-g^{\rho\sigma}k\cdot k' \right)\ ,\\
   L_{\rm{DVCS,L}}^{\rho\sigma}&= 2 \epsilon^{\rho\sigma\alpha\beta} k_\alpha k'_\beta \ .
\end{align}
As for the hadronic part, it can be worked out with Eq. (\ref{eq:hdvcs}) and the Compton tensor given in Eq. (\ref{eq:comptontensor}). It turns out that the hadronic tensor contains four terms that have a different dependence on the target polarization, which can be written as,
\begin{align}
\label{eq:hdvcspol1}
\begin{split}
   H_{\rm{DVCS}}^{\rho\sigma}=&H_{\rm{DVCS,U}}^{\rho\sigma}+\frac{1}{M} H_{\rm{DVCS,T}}^{\rho\sigma}\epsilon^{n \bar P\Delta S}+ M H_{\rm{DVCS,n}}^{\rho\sigma} (n\cdot S)+\frac{1}{M}H_{\rm{DVCS,\Delta}}^{\rho\sigma}( \Delta \cdot S)\ ,
\end{split}
\end{align}
where we introduce the notation $\epsilon^{n \bar P\Delta S}\equiv \epsilon^{\mu\nu\alpha\beta}n_\mu \bar P_\nu \Delta_\alpha S_\beta$ when contracting with the $\epsilon$ tensor, which will be used hereafter. The explicit expressions of those $H^{\rho\sigma}_{\rm{DVCS}}$s of different polarization configurations are not of interest for now.

In order to further study the polarization dependence of the cross-section, it is helpful to write down a basis for the target polarization $S$ and express the cross-section in terms of the base vectors, as the polarized cross-section is linear in $S$. We define the longitudinal polarization vector to be a spatial vector along the direction of $n^\mu$ such that $S_{\rm{L}}^{\mu}\sim n^\mu$, where $\sim$ indicates projecting out the temporal component and normalization. As for the transverse polarization, note that the transverse components of all the physical momenta in Eq. (\ref{eq:momentumlf}) are in the same direction, the direction of $\Delta_{\perp}^\mu$. Thus, we define the two transverse polarization vectors as $S^\mu_{\rm{T,in}}\sim\Delta_\perp^{\mu}$ and $S^\mu_{\rm{T,out}}\sim \epsilon^{\mu\nu}_\perp \Delta^\perp_{\nu}$, where $\Delta^{\mu}_\perp$ is defined as $\Delta^{\mu}_\perp\equiv g^{\mu\nu}_\perp \Delta_{\nu}$ and satisfies $\Delta^{2}_\perp=-|\Delta_\perp|^2$. The $\rm{in}$ and $\rm{out}$ subscripts indicate that the two vectors are parallel and perpendicular to the hadronic plane, respectively. Then we 
define our polarization vectors as,
\begin{equation}
\label{eq:polvecdef}
    \begin{aligned}
        S_{\rm{L}}^{\mu}&=\frac{M}{1+\xi}\left(-n^\mu +\frac{1+\xi}{M^2}P^\mu\right)\ ,\\
        S_{\rm{T,in}}^{\mu}&=\frac{1}{NM (1+\xi)} \epsilon^{\mu\nu Pn }\epsilon^{nPP'}_{~~~~~~\nu}\ ,\\
        S_{\rm{T,out}}^{\mu}&=\frac{1}{NM}\epsilon^{n PP'\mu}\propto\epsilon^{\mu\nu}_\perp \Delta^\perp_\nu\ ,
    \end{aligned}
\end{equation}
where we introduce the notation 
\begin{equation}
    \begin{aligned}
       N=\frac{\sqrt{-4M^2\xi^2-t(1-\xi^2)}}{M}\ ,
    \end{aligned}
\end{equation}
and an overall plus or minus sign is conventional for all the polarization vectors. Note that while we have the relation $S_{\rm{T,out}}^{\mu} \propto\epsilon^{\mu\nu}_\perp \Delta^\perp_\nu$, it does NOT apply to $S_{\rm{T,in}}$ as $\Delta_\perp^\mu$ might have a temporal component depending on the choice of light cone vectors. Consequently, we use the definition in Eq. (\ref{eq:polvecdef}) such that $S_{\rm{T,in}}\cdot P =0$ is manifest. The three polarization vectors form an orthonormal basis for the polarization vector. With a specific choice of parameters $\alpha=\beta=1$, they reduce to
\begin{equation}
    \begin{aligned}
        S_{\rm{L}}^{\mu}&=(0,0,0,1)\ ,\\
        S_{\rm{T,in}}^{\mu}&=(0,\cos \phi,\sin\phi,0)\ ,\\
        S_{\rm{T,out}}^{\mu}&=(0,-\sin \phi,\cos\phi,0)\ ,
    \end{aligned}
\end{equation}
in the lab frame, whereas higher twist corrections to those expressions are needed if different $\alpha$ and $\beta$ are chosen.
Then, if we write the polarization vector as~\cite{Kriesten:2019jep}
\begin{equation}
    \begin{aligned}
        S^\mu=2\Lambda_L(0,0,0,1) +2\Lambda_T \left(0,\cos\phi_S,\sin\phi_S,0\right)\ ,
    \end{aligned}
\end{equation}
where $\Lambda_L,\Lambda_T$ stands for the spin projection in the longitudinal/transverse direction respectively and $\phi_S$ is the azimuthal angle between the transverse polarization vector and the leptonic plane, it can be decomposed as,
\begin{equation}
\label{eq:poldecomp}
    \begin{aligned}
        S^\mu=(2\Lambda_L)  S^\mu_{\rm{L}}+ 2\Lambda_T \cos\left(\phi_S-\phi\right) S^\mu_{\rm{T,in}}+ 2\Lambda_T \sin\left(\phi_S-\phi\right) S^\mu_{\rm{T,Out}}\ ,
    \end{aligned}
\end{equation}
and thus the hadronic tensor for the pure DVCS cross-section can be written as,
\begin{equation}
\label{eq:hdvcspol}
    \begin{aligned}
         H_{\rm{DVCS}}^{\rho\sigma}= &H_{\rm{DVCS,U}}^{\rho\sigma} + (2\Lambda_L)   H_{\rm{DVCS,L}}^{\rho\sigma}\\
         &+2\Lambda_T \left[H_{\rm{DVCS,T,in}}^{\rho\sigma}\cos\left(\phi_S-\phi\right) + H_{\rm{DVCS,T,out}}^{\rho\sigma}\sin\left(\phi_S-\phi\right)\right]\ ,
    \end{aligned}
\end{equation}
where we defined
\begin{align}
 H_{\rm{DVCS,U}}^{\rho\sigma}&\equiv \frac{1}{2}\Big[ H_{\rm{DVCS}}^{\rho\sigma}(S)+H_{\rm{DVCS}}^{\rho\sigma}(-S)\Big]\ ,\\
 H_{\rm{DVCS,L}}^{\rho\sigma}&\equiv \frac{1}{2}\Big[ H_{\rm{DVCS}}^{\rho\sigma}(S_{\rm{L}})-H_{\rm{DVCS}}^{\rho\sigma}(-S_{\rm{L}})\Big]\ ,\\
  H_{\rm{DVCS,T,in}}^{\rho\sigma}&\equiv \frac{1}{2}\Big[ H_{\rm{DVCS}}^{\rho\sigma}(S_{\rm{T,in}})-H_{\rm{DVCS}}^{\rho\sigma}(-S_{\rm{T,in}})\Big]\ ,\\
  H_{\rm{DVCS,T,out}}^{\rho\sigma}&\equiv \frac{1}{2}\Big[ H_{\rm{DVCS}}^{\rho\sigma}(S_{\rm{T,out}})-H_{\rm{DVCS}}^{\rho\sigma}(-S_{\rm{T,out}})\Big]\ .
\end{align}
Plugging the polarization vectors $S$ into the definition of  the hadronic tensor above, we have for $H_{\rm{DVCS, U}}^{\rho\sigma}$,
\begin{align}
\label{eq:dvcsu}
\begin{split}
  H_{\rm{DVCS, U}}^{\rho\sigma}=&4\Bigg[(1-\xi^2)\left({\mathscr H}^{\rho\sigma}\mathcal{H}^*\mathcal{H}+\widetilde{{\mathscr H}}^{\rho\sigma}\widetilde{\mathcal{H}}^*\widetilde{\mathcal{H}}\right) -\frac{t}{4M^2}\left( {\mathscr H}^{\rho\sigma}\mathcal{E}^*\mathcal{E}+\xi^2 \widetilde{{\mathscr H}}^{\rho\sigma}\widetilde{\mathcal{E}}^*\widetilde{\mathcal{E}}\right)\\
   &\quad-\xi^2\left({\mathscr H}^{\rho\sigma}\mathcal{E}^*\mathcal{E}+{\mathscr H}^{\rho\sigma} (\mathcal{E}^*\mathcal{H} +\mathcal{H}^*\mathcal{E})+\widetilde{\mathscr H}^{\rho\sigma} (\widetilde{\mathcal{E}}^*\widetilde{\mathcal{H}} +\widetilde{\mathcal{H}}^*\widetilde{\mathcal{E}})\right)\Bigg]\ ,
\end{split}
\end{align}
and for $H_{\rm{DVCS, L}}^{\rho\sigma}$
\begin{align}
\begin{split}
  H_{\rm{DVCS,L}}^{\rho\sigma}=&-8i  {\mathscr H}_{-}^{\rho\sigma}\text{Re}\Bigg\{\left(1-\xi^2\right)\widetilde{\mathcal{H}}^*\mathcal{H}-\xi^2\left(\widetilde{\mathcal{H}}^*\mathcal{E}+\widetilde{\mathcal{E}}^*\mathcal{H}\right)-\left(\frac{\xi^2}{1+\xi}+\frac{t}{4M^2}\right)\xi\widetilde{\mathcal{E}}^*\mathcal{E}\Bigg\}\\
  &+8 {\mathscr H}_{+}^{\rho\sigma} \text{Im}\Bigg\{\left(1-\xi^2\right)\widetilde{\mathcal{H}}^*\mathcal{H}-\xi^2 \left(\widetilde{\mathcal{H}}^*\mathcal{E}+\widetilde{\mathcal{E}}^*\mathcal{H}\right)-\left(\frac{\xi^2}{1+\xi}+\frac{t}{4M^2}\right)\xi\widetilde{\mathcal{E}}^*\mathcal{E}\Bigg\}\ ,
\end{split}
\end{align}
and for $H_{\rm{DVCS,T,in}}^{\rho\sigma}$
\begin{align}
\begin{split}
  H_{\rm{DVCS,T,in}}^{\rho\sigma}=&-4Ni {\mathscr H}_{-}^{\rho\sigma}\text{Re}\Bigg\{\widetilde{\mathcal{H}}^*\mathcal{E}-\xi\widetilde{\mathcal{E}}^*\mathcal{H}-\frac{\xi^2}{1+\xi}\widetilde{\mathcal{E}}^*\mathcal{E} \Bigg\}\\
  &+4N{\mathscr H}_{+}^{\rho\sigma} \text{Im}\Bigg\{\widetilde{\mathcal{H}}^*\mathcal{E}-\xi\widetilde{\mathcal{E}}^*\mathcal{H}-\frac{\xi^2}{1+\xi}\widetilde{\mathcal{E}}^*\mathcal{E} \Bigg\}\ ,
\end{split}
\end{align}
and for $H_{\rm{DVCS,T,out}}^{\rho\sigma}$
\begin{align}
  H_{\rm{DVCS, T,out}}^{\rho\sigma}=&4 N \text{Im}\Bigg[{\mathscr H}^{\rho\sigma}\mathcal{H}^*\mathcal{E}-\xi \widetilde{{\mathscr H}}^{\rho\sigma}\widetilde{\mathcal{H}}^*\widetilde{\mathcal{E}}\Bigg]\ , 
\end{align}
where four tensor structures emerge in the hadronic tensor and we define them as
\begin{equation}
\label{eq:hampdef}
    \begin{aligned}
   {\mathscr H}^{\rho\sigma}&\equiv {\mathscr T}^{\mu\rho}_{(2)} {\mathscr T}^{\nu\sigma}_{(2)}\mathcal{G}_{\mu\nu} \  ,& {\mathscr H}_{+}^{\rho\sigma}&\equiv \frac{-i}{2}\left( {\mathscr T}^{\mu\rho}_{(2)} \widetilde{{\mathscr T}}^{\nu\sigma}_{(2)}+{\mathscr T}^{\nu\sigma}_{(2)} \widetilde{{\mathscr T}}^{\mu\rho}_{(2)}\right)\mathcal{G}_{\mu\nu} \ ,\\
 \widetilde{{\mathscr H}}^{\rho\sigma}&\equiv  -\widetilde{{\mathscr T}}^{\mu\rho}_{(2)} \widetilde{{\mathscr T}}^{\nu\sigma}_{(2)}\mathcal{G}_{\mu\nu} \ ,&{\mathscr H}_{-}^{\rho\sigma}&\equiv \frac{-i}{2}\left( {\mathscr T}^{\mu\rho}_{(2)} \widetilde{{\mathscr T}}^{\nu\sigma}_{(2)}-{\mathscr T}^{\nu\sigma}_{(2)} \widetilde{{\mathscr T}}^{\mu\rho}_{(2)}\right)\mathcal{G}_{\mu\nu}\ .
\end{aligned}
\end{equation}
We also define the projection tensor
\begin{align}
  \mathcal{G}^{\mu\nu} \equiv \sum_{\Lambda'=\pm1}  \varepsilon^\mu(q',\Lambda')\varepsilon^{*\nu}(q',\Lambda') =-g^{\mu\nu}+\frac{P^\mu q'^\nu+P^\nu q'^\mu}{P\cdot q'}-\frac{P^2 q'^\mu q'^\nu}{(P\cdot q')^2}\ ,
\end{align}
resulting from the sum of final photon polarizations. As discussed in the last section, $\mathcal{G}_{\mu\nu}$ reduces to $-g_{\mu\nu}$ only when the Compton tensor is exactly conserved. 

The last step is to contract our hadronic part in Eq. (\ref{eq:hdvcspol}) with the leptonic part in Eq. (\ref{eq:ldvcspol}), and from this we have
\begin{equation}
\label{eq:amp2dvcs}
    \begin{aligned}
         \left|\mathcal T_{\rm{DVCS}}\right|^2=&\frac{1}{Q^4} \Bigg\{F_{UU}+(2\Lambda_L) F_{UL}+(2\Lambda_T)\big(\cos\left(\phi_S-\phi\right) F_{U T,\rm{in}}+\sin\left(\phi_S-\phi\right)F_{U T,\rm{out}}\big)\\
         +&(2h)\Big[F_{LU}+(2\Lambda_L) F_{LL}+(2\Lambda_T) \big(\cos\left(\phi_S-\phi\right) F_{L T,\rm{in}}+\sin\left(\phi_S-\phi\right)F_{LT,\rm{out}}\big)\Big] \Bigg\}\ ,
    \end{aligned}
\end{equation}
where we define
\begin{equation}
\label{eq:dvcsstruct}
    \begin{aligned}
        F_{UU} &\equiv  L^{\rm{DVCS,U}}_{\rho\sigma}H_{\rm{DVCS,U}}^{\rho\sigma}\ ,&\qquad  F_{LU} &\equiv  i L^{\rm{DVCS,L}}_{\rho\sigma}H_{\rm{DVCS,U}}^{\rho\sigma} \ ,\\
        F_{UL} &\equiv  L^{\rm{DVCS,U}}_{\rho\sigma} H_{\rm{DVCS,L}}^{\rho\sigma}\ ,&\qquad F_{LL} &\equiv  i L^{\rm{DVCS,L}}_{\rho\sigma} H_{\rm{DVCS,L}}^{\rho\sigma}\ ,\\
         F_{UT,\rm{in}} &\equiv   L^{\rm{DVCS,U}}_{\rho\sigma}H_{\rm{DVCS,T,in}}^{\rho\sigma}\ ,&F_{LT,\rm{in}} &\equiv  i L^{\rm{DVCS,L}}_{\rho\sigma}H_{\rm{DVCS,T,in}}^{\rho\sigma}\ ,\\
         F_{UT,\rm{out}} &\equiv  L^{\rm{DVCS,U}}_{\rho\sigma} H_{\rm{DVCS,T,out}}^{\rho\sigma}\ ,&\qquad  F_{LT,\rm{out}} &\equiv  iL^{\rm{DVCS,L}}_{\rho\sigma} H_{\rm{DVCS,T,out}}^{\rho\sigma}\ ,
    \end{aligned}
\end{equation}
analogous to those in Ref. \cite{Kriesten:2019jep}. The two structure functions $F_{LU}$ and $F_{LT,\rm{out}}$ vanish with leading twist CFFs, and we list them here for completeness. Since those hadronic tensor $H_{\rm{DVCS}}^{\rho\sigma}$s only contain the tensor structures defined in Eq. (\ref{eq:hampdef}), it is only those tensor structures contracted with the leptonic tensor that will enter the cross-section.
Thus, we define the following four scalar coefficients for the pure DVCS cross-section
\begin{equation}
\label{eq:dvcsscalaramp}
    \begin{aligned}
   h^{\rm{U}}\equiv L^{\rm{DVCS,U}}_{\rho\sigma} {\mathscr H}^{\rho\sigma} \  ,&\qquad h^{+,\rm{U}}\equiv L^{\rm{DVCS,U}}_{\rho\sigma} {\mathscr H}_{+}^{\rho\sigma}\ ,\\
 \tilde {h}^{\rm{U}}\equiv L^{\rm{DVCS,U}}_{\rho\sigma} \widetilde{{\mathscr H}}^{\rho\sigma}\ ,&\qquad h^{-,\rm{L}}\equiv L^{\rm{DVCS,L}}_{\rho\sigma}{\mathscr H}_{-}^{\rho\sigma}\ .
\end{aligned}
\end{equation}
Since the four tensor structures are either symmetric/antisymmetric in their $\rho,\sigma$ indices, they only couple to the unpolarized/polarized leptonic tensor respectively, whereas coefficients such as $h^{\rm{L}}$ or $ \tilde {h}^{\rm{L}}$ vanish. Therefore, only four independent coefficients can be defined.
Then we can express the structure functions in terms of them, which are given in Appendix. \ref{app:dvcsstructurefunc}.

In the leading twist approximation, it can be easily checked that our $F_{UU}$ and $F_{LL}$ agrees with the unpolarized and double spin process result in \cite{Ji:1996nm}. Besides, our $F_{UU}$, $F_{LL}$, $F_{LT,\rm{in}}$ and $F_{UT,\rm{out}}$ agree with the  $\mathcal C_{\rm{unp}}^{\rm{DVCS}}$,$\mathcal C_{\rm{LP}}^{\rm{DVCS}}$, $\mathcal C_{\rm{TP,+}}^{\rm{DVCS}}$, $\mathcal C_{\rm{TP,-}}^{\rm{DVCS}}$ in Ref. \cite{Belitsky:2001ns,Belitsky:2010jw}. In addition, they correspond to the $F_{UU,T},F_{LL},F_{LT,T}^{\cos(\phi-\phi_S)},F_{UT,T}^{\sin(\phi-\phi_S)}$ terms in Ref. \cite{Kriesten:2019jep} respectively. Therefore, our polarized cross-section reduces to the results in Ref.~\cite{Ji:1996nm,Belitsky:2001ns,Belitsky:2010jw,Kriesten:2019jep} at leading twist. 

\subsection{Interference cross-section and polarization asymmetry}
\label{subsec:intxsection}

The interference cross-section can be carried out similarly. Again, we write the interference squared amplitude as a product of their leptonic and hadronic parts as
\begin{equation}
     \mathcal I  =-\frac{e_l}{Q^2 t} L_{\rm{INT}}^{\mu\rho\sigma} H^{\rm{INT}}_{\mu\rho\sigma} +\text{c.c.}\ .
\end{equation}
where we have,
\begin{align}
\label{eq:lint}
L_{\rm{INT}}^{\mu\rho\sigma}\equiv & \sum_{h'} \bar u(k,h)\gamma^\rho u(k',h') l_{\rm{BH}}^{\mu\sigma}\ ,\\
       \label{eq:hint}
   H_{\rm{INT}}^{\mu\rho\sigma }\equiv&\sum_{S',\Lambda'}  T^{*\nu\rho}\varepsilon_\nu(q',\Lambda')\varepsilon^*_{\mu}(q',\Lambda') \bar u(P',S')\left[(F_1+F_2)\gamma^\sigma-\frac{\bar P^\sigma}{M}F_2\right]u(P,S)  \ .
\end{align}
Then we split the leptonic term into a polarized and unpolarized part as, 
\begin{equation}
    L^{\rm{INT}}_{\mu\rho\sigma}=L^{\rm{INT,U}}_{\mu\rho\sigma}+i 2 h L^{\rm{INT,L}}_{\mu\rho\sigma}\ ,
\end{equation}
whose explicit expressions are given in App. \ref{app:structurefunc}. The hadronic tensor can be written as,
\begin{align}
\begin{split}
      H_{\rm{INT}}^{\mu\rho\sigma }= &\Bigg\{4\bar P^\sigma \left(F_1 \mathcal {H}^*-\frac{t}{4M^2} F_2 \mathcal {E}^*\right)+\left[\left(t n^\sigma+2\xi \Delta^\sigma \right) +2i M\epsilon^{\sigma n \Delta S} \right](F_1+F_2)(\mathcal {H}^*+\mathcal {E}^*) 
      \\& -\frac{2i}{M}\epsilon^{\sigma \bar P \Delta S}(F_1+F_2)\mathcal {E}^* +\frac{2i}{M}\bar P^\sigma \epsilon^{n \bar P \Delta S} F_2(\mathcal {H}^*+\mathcal {E}^*)\Bigg\}\mathcal G^{\mu}_{~\nu} \mathscr{T}^{\nu\rho}_{(2)}\\
      &+\Bigg\{\left[2i \epsilon^{\sigma n \bar P \Delta}+2M \left(n^\sigma+\frac{2\xi}{t}\Delta^\sigma\right)(\Delta \cdot S)\right] (F_1+F_2)\widetilde{\mathcal {H}}^*\\
      &\qquad+4  M   \left(S^\sigma-\frac{(\Delta\cdot S)\Delta^\sigma}{t}\right) \xi (F_1+F_2) \left(\widetilde{\mathcal {H}}^*+\frac{t}{4M^2}\widetilde{\mathcal {E}}^*\right) \\
      &\qquad-4M \bar P^\sigma (n\cdot S) \left(F_1 +\frac{t}{4M^2}F_2\right)\widetilde{\mathcal {H}}^*\\
      &\qquad+\frac{2}{M} \bar P^\sigma (\Delta \cdot S)\left[\xi F_1 \widetilde{\mathcal {E}}^*-(1+\xi)F_2\widetilde{\mathcal {H}}^*\right]\Bigg\}\mathcal G^{\mu}_{~\nu} \widetilde{\mathscr{T}}^{\nu\rho}_{(2)}\ .
\end{split}
\end{align}
Terms proportional to $\Delta^\sigma$ can be dropped since $\Delta^\sigma L^{\rm{INT}}_{\mu\rho\sigma}=0$, though they are kept in the above expression such that $\Delta^\sigma H^{\rm{INT}}_{\mu\rho\sigma}=0$ is manifest. Then following the same polarization decomposition in Eq. (\ref{eq:poldecomp}), we can write the hadronic tensor in terms of four different terms as
\begin{equation}
\label{eq:hintpol}
    \begin{aligned}
         H_{\rm{INT}}^{\mu\rho\sigma}= &H_{\rm{INT,U}}^{\mu\rho\sigma} + (2\Lambda_L)   H_{\rm{INT,L}}^{\mu\rho\sigma}+2\Lambda_T  \left[H_{\rm{INT,T,in}}^{\mu\rho\sigma}\cos\left(\phi_S-\phi\right) + H_{\rm{INT,T,out}}^{\mu\rho\sigma}\sin\left(\phi_S-\phi\right)\right]\ ,
    \end{aligned}
\end{equation}
where we define similarly
\begin{align}
 H_{\rm{INT,U}}^{\mu\rho\sigma}&\equiv \frac{1}{2}\Big[ H_{\rm{INT}}^{\mu\rho\sigma}(S)+H_{\rm{INT}}^{\mu\rho\sigma}(-S)\Big]\ ,\\
 H_{\rm{INT,L}}^{\mu\rho\sigma}&\equiv \frac{1}{2}\Big[ H_{\rm{INT}}^{\mu\rho\sigma}(S_{\rm{L}})-H_{\rm{INT}}^{\mu\rho\sigma}(-S_{\rm{L}})\Big]\ ,\\
  H_{\rm{INT,T,in}}^{\rho\sigma}&\equiv \frac{1}{2}\Big[ H_{\rm{INT}}^{\mu\rho\sigma}(S_{\rm{T,in}})-H_{\rm{INT}}^{\mu\rho\sigma}(-S_{\rm{T,in}})\Big]\ ,\\
  H_{\rm{INT,T,out}}^{\rho\sigma}&\equiv\frac{1}{2}\Big[ H_{\rm{INT}}^{\mu\rho\sigma}(S_{\rm{T,out}})-H_{\rm{INT}}^{\mu\rho\sigma}(-S_{\rm{T,out}})\Big]\ .
\end{align}
The four hadronic tensor can be solved by inserting our $S$, and we have for $ H_{\rm{INT,U}}^{\mu\rho\sigma }$,
\begin{align}
\begin{split}
      H_{\rm{INT,U}}^{\mu\rho\sigma }= &\Big[4\bar P^\sigma \left(F_1 \mathcal {H}^*-\frac{t}{4M^2} F_2 \mathcal {E}^*\right)+t n^\sigma(F_1+F_2)(\mathcal {H}^*+\mathcal {E}^*) 
      \Big]\mathcal G^{\mu}_{~\nu} \mathscr{T}^{\nu\rho}_{(2)}\\
      &+2i \epsilon^{\sigma n \bar P \Delta} (F_1+F_2)\widetilde{\mathcal {H}}^*\mathcal G^{\mu}_{~\nu} \widetilde{\mathscr{T}}^{\nu\rho}_{(2)}\ ,
\end{split}
\end{align}
and for $H_{\rm{INT,L}}^{\mu\rho\sigma}$,
\begin{align}
    \begin{split}
      H_{\rm{INT,L}}^{\mu\rho\sigma}=&\Bigg\{-4\bar P^\sigma\left[F_1\left(\widetilde{ \mathcal {H}}^*-\frac{\xi^2}{1+\xi}\widetilde{ \mathcal {E}}^* \right)-F_2\frac{t}{4M^2}\xi\widetilde{ \mathcal {E}}^*\right]\\
       &\quad-t n^\sigma(F_1+F_2)\left(\widetilde{ \mathcal {H}}^*+\frac{\xi}{1+\xi} \widetilde{\mathcal {E}}^* \right)\Bigg\}\mathcal G^{\mu}_{~\nu} \widetilde{\mathscr{T}}^{\nu\rho}_{(2)}\\
       &-2i\epsilon^{\sigma n \bar P\Delta }(F_1+F_2)\left( \mathcal {H}^*+\frac{\xi}{1+\xi} \mathcal {E}^* \right)   \mathcal G^{\mu}_{~\nu} \mathscr{T}^{\nu\rho}_{(2)} \ ,
\end{split}
\end{align}
and for $H_{\rm{INT,T,in}}^{\mu\rho\sigma}$,
\begin{align}
    \begin{split}
      H_{\rm{INT,T,in}}^{\mu\rho\sigma}= &
      \frac{1}{N}\Bigg\{-8\bar P^\sigma\Bigg[\xi F_1\left(\xi \widetilde{ \mathcal {H}}^*+\left(\frac{\xi^2}{1+\xi}+\frac{t}{4M^2}\right)\widetilde{ \mathcal {E}}^* \right)+F_2\frac{t}{4M^2}\left((\xi^2-1)\widetilde{\mathcal {H}}^* +\xi^2\widetilde{ \mathcal {E}}^*\right)\Bigg]\\
       &\qquad-2 t n^\sigma(F_1+F_2)\left[\widetilde{ \mathcal {H}}^*+\left(\frac{t}{4M^2}-\frac{\xi}{1+\xi}\right)\xi\widetilde{\mathcal {E}}^* \right]\Bigg\}\mathcal G^{\mu}_{~\nu} \widetilde{\mathscr{T}}^{\nu\rho}_{(2)}\\
       &+\frac{4i}{N}\epsilon^{\sigma n \bar P\Delta }(F_1+F_2)\left[\xi \mathcal {H}^*+\left(\frac{\xi^2}{1+\xi}+\frac{t}{4M^2}\right) \mathcal {E}^* \right]   \mathcal G^{\mu}_{~\nu} \mathscr{T}^{\nu\rho}_{(2)}\ ,
\end{split}
\end{align}
and for $H_{\rm{INT,T,out}}^{\mu\rho\sigma}$,
\begin{align}
    \begin{split}
      H_{\rm{INT,T,out}}^{\mu\rho\sigma}= &\frac{1}{N}\Bigg\{-8 i \bar P^\sigma \left[F_1\left(\xi^2\mathcal H^*+\left(\xi^2+\frac{t}{4M^2}\right)\mathcal E^* \right)+\frac{t}{4M^2}F_2\left((\xi^2-1)\mathcal H^*+\xi^2\mathcal E^*\right)\right]\\
      &\qquad-2i t n^\sigma (F_1+F_2)\left(\mathcal H^*+\frac{t}{4M^2}\mathcal E^*\right)\Bigg\}   \mathcal G^{\mu}_{~\nu} \mathscr{T}^{\nu\rho}_{(2)}\\
      &-\frac{4}{N} \epsilon^{\sigma n\bar P \Delta }\xi(F_1+F_2)\left(\widetilde{\mathcal H}^*+\frac{t}{4M^2}\widetilde{\mathcal E}^*\right)\mathcal G^{\mu}_{~\nu} \widetilde{\mathscr{T}}^{\nu\rho}_{(2)}\ ,
\end{split}
\end{align}
repsectively. Again, the last step is to contract the leptonic and hadronic part to get the squared amplitude. We have the same eight polarization configurations as what we get for the DVCS in Eq. (\ref{eq:amp2dvcs}), which are defined as,
\begin{align}
\begin{split}
    \mathcal I=\frac{-e_l}{Q^2 t}\Bigg\{&F^{I}_{UU}+2\Lambda_L F^{I}_{UL}+2\Lambda_T \left(F^{I}_{UT,\rm{in}}\cos(\phi_S-\phi)+F^{I}_{UT,\rm{out}}\sin(\phi_S-\phi)\right)\\
    &+2h \Big[F^{I}_{LU}+2\Lambda_L F^{I}_{LL}+2\Lambda_T \left(F^{I}_{LT,\rm{in}}\cos(\phi_S-\phi)+F^{I}_{LT,\rm{out}}\sin(\phi_S-\phi)\right)\Big]\Bigg\}\ ,
\end{split} 
\end{align}
where those eight polarized cross-sections can be expressed as,
\begin{equation}
\label{eq:intxsecdef}
    \begin{aligned}
        F^I_{UU} &\equiv  L^{\rm{INT,U}}_{\mu\rho\sigma}H_{\rm{INT,U}}^{\mu\rho\sigma}+\text{c.c.}\ ,& F^I_{LU} &\equiv  i L^{\rm{INT,L}}_{\mu\rho\sigma}H_{\rm{INT,U}}^{\mu\rho\sigma}+ \text{c.c.}\\
        F^I_{UL} &\equiv  L^{\rm{INT,U}}_{\mu\rho\sigma} H_{\rm{INT,L}}^{\mu\rho\sigma}+\text{c.c.}\ ,&\qquad F^I_{LL} &\equiv  i L^{\rm{INT,L}}_{\mu\rho\sigma} H_{\rm{INT,L}}^{\mu\rho\sigma}+\text{c.c.}\ ,\\
        F^I_{UT,\rm{in}} &\equiv   L^{\rm{INT,U}}_{\mu\rho\sigma}H_{\rm{INT,T,in}}^{\mu\rho\sigma}+\text{c.c.}\ ,&\qquad F^I_{LT,\rm{in}} &\equiv  i L^{\rm{INT,L}}_{\mu\rho\sigma}H_{\rm{INT,T,in}}^{\mu\rho\sigma}+\text{c.c.}\ ,\\
       F^I_{UT,\rm{out}} &\equiv  L^{\rm{INT,U}}_{\mu\rho\sigma} H_{\rm{INT,T,out}}^{\mu\rho\sigma}+\text{c.c.}\ ,& \qquad F^I_{LT,\rm{out}} &\equiv  i L^{\rm{INT,L}}_{\mu\rho\sigma} H_{\rm{INT,T,out}}^{\mu\rho\sigma}+\text{c.c.}\ .
    \end{aligned}
\end{equation}
analogous to those in Ref. \cite{Kriesten:2019jep}. In order to express those cross-sections further, we define 6 coefficients $A^I,B^I,C^I,\tilde A^I,\tilde B^I$ and $\tilde C^I$ which are defined as
\begin{equation}
\label{eq:intstructfunc}
\begin{aligned}
    A^I& \equiv8 \bar P^\sigma \mathcal{G}^{\mu}_{~\nu} \mathscr{T}^{\nu\rho}_{(2)}   L^{\rm{INT}}_{\mu\rho\sigma} \ ,&\qquad \tilde{A}^I& \equiv8 i \bar P^\sigma \mathcal{G}^{\mu}_{~\nu} \widetilde{\mathscr{T}}^{\nu\rho}_{(2)}   L^{\rm{INT}}_{\mu\rho\sigma} \ ,\\
    B^I &\equiv 2 t n^\sigma \mathcal{G}^{\mu}_{~\nu}  \mathscr{T}^{\nu\rho}_{(2)}   L^{\rm{INT}}_{\mu\rho\sigma}\ ,& \qquad \tilde{B}^I &\equiv 2i t n^\sigma \mathcal{G}^{\mu}_{~\nu}  \widetilde{\mathscr{T}}^{\nu\rho}_{(2)}   L^{\rm{INT}}_{\mu\rho\sigma}\ ,\\
    C^I& \equiv4i \epsilon^{\sigma n \bar P \Delta} \mathcal{G}^{\mu}_{~\nu} \widetilde{\mathscr{T}}^{\nu\rho}_{(2)}  L^{\rm{INT}}_{\mu\rho\sigma} \ ,&\qquad \tilde{C}^I& \equiv 4  \epsilon^{\sigma n \bar P \Delta} \mathcal{G}^{\mu}_{~\nu} \mathscr{T}^{\nu\rho}_{(2)}  L^{\rm{INT}}_{\mu\rho\sigma}\ ,
\end{aligned}
\end{equation}
where each of them can be written as the sum of their unpolarized and polarized parts,
\begin{align}
\begin{split}
    \mathcal A^{I}\equiv \mathcal A^{I,\rm{U}} +i 2 h \mathcal A^{I,\rm{L}}\ ,
\end{split}
\end{align}
with $\mathcal A= \{A,B,C,\tilde A,\tilde B,\tilde C\}$ such that each $\mathcal A^{I,\rm{U/L}}$ is given by the same definition in Eq. (\ref{eq:intstructfunc}) but with $L^{\rm{INT}}$ replaced by $ L^{\rm{INT,U}}/ L^{\rm{INT,L}}$ respectively. The $A^{I,\rm{U}}$, $B^{I,\rm{U}}$ and $C^{I,\rm{U}}$ are the same as the coefficients defined in \cite{Kriesten:2019jep}. Then the eight polarized cross-sections can be written with these coefficients, which are given in App. \ref{app:structurefunc}. In the unpolarized case, our expression of $F^I_{UU}$ agrees with the $F^I_{UU}$ in \cite{Kriesten:2019jep} and the $C^{I}_{\rm{unp}}$ in \cite{Belitsky:2001ns,Belitsky:2010jw}. More detailed comparisons are left to the next section.

\section{Comparison of Different Cross-Section Formulas}
\label{sec:comparison}
In the last section, we expressed the square amplitude in terms of combinations of hadronic tensors and leptonic ones. Complete analytical expressions of those terms are cumbersome, and in this section we mainly compare them through twist expansion and numerical evaluation with the help of \textsc{FeynCalc}~\cite{Mertig:1990an,Shtabovenko:2016sxi,Shtabovenko:2020gxv}. As we discussed before, the BH contribution will be treated as the background and omitted for the comparison.

\subsection{Unpolarized pure DVCS cross-section comparison}

To start with, we compare our pure DVCS cross-section with the BKM10 results \cite{Belitsky:2010jw}, whereas the BKM01 \cite{Belitsky:2001ns} results lack all the higher-twist kinematics and do not agree with our formula nor the BKM10 one, so we will not have it for comparison.
The four-fold cross-section 
\begin{equation}
    \frac{\text{d}^4 \sigma^{UU}_{\rm{DVCS}}}{\text{d}x_B\text{d}Q^2\text{d}|t| \text{d}\phi}=\int_0^{2\pi} \text{d}\phi_S\frac{\text{d}^4 \sigma^{UU}_{\rm{DVCS}}}{\text{d}x_B\text{d}Q^2\text{d}|t| \text{d}\phi\text{d}\phi_S}\ ,
\end{equation}
is of interest, since a transverse polarized target will not be involved in this paper.

In the subsection \ref{subsec:puredvcs}, we worked out the pure DVCS cross-section formulas and argue that they agree with the BKM10 results to the leading order. Here we compare them beyond the leading order. The twist expansion of those coefficients $h^{\rm{U}}$, $\tilde{h}^{\rm{U}}$, $h^{+,\rm{U}}$ and $h^{-,\rm{L}}$ which are needed for the pure DVCS cross-section are given in Appendix. \ref{app:dvcsstructurefunc}, and it can be checked that in the unpolarized case, our leading term, corresponding to $h^{\rm{U}}_{(2)}=\tilde{h}^{\rm{U}}_{(2)}$, agrees with the $c_{0,\rm{unp}}^{\rm{DVCS}}$ term in BKM10~\cite{Belitsky:2010jw} and the subleading term, corresponding to $h^{\rm{U}}_{(3)}=\tilde{h}^{\rm{U}}_{(3)}$, agrees with the $c_{1,\rm{unp}}^{\rm{DVCS}}$ term in BKM10 if we put in $\beta=1/2$ which corresponds to the convention chosen there. Therefore, our result agrees with the BKM10 result up to twist-four terms. However, we do not have twist-four agreement due to the approximation made in BKM10, for which they use the light cone vectors $n_{\rm{BKM}}$ and $p_{\rm{BKM}}$ that are not light-like and drop the twist-four term in $\mathscr T^{\Lambda\Lambda'=++}_{(2)}$ in the helicity amplitude in Eq. (\ref{eq:BKMhelicity}). As a result, we do not have exactly the same hadronic matrix element as their $\mathcal C_{\rm{unp}}^{\rm{DVCS}}$, since their twist-four terms are incomplete, and we will focus on the numerical comparison at higher order. 

One comment on the hadronic structures such as $\mathcal C_{\rm{unp}}^{\rm{DVCS}}$ in the BKM10 results is that those structures by definition must be functions of the parton momentum fraction $x$ (which will be convoluted with Wilson coefficients and eventually vanish in CFFs), skewness parameter $\xi$, momentum transfer $t$ and proton mass $M$ only, whereas the $Q^2$ dependence only comes in through the evolution of GPD. Therefore, the hadronic part must be expressible in terms of $\xi,t$ and $M$. However, in the BKM10 results those structures are written with $x_B$ and it can not be completely converted into $\xi$ which make their expression unjustified, as $x_B$ are defined from the virtual photon momentum $q$ through $q\cdot P=Q^2/(2 x_B)$ which the hadronic part should not depend on. This results from the effectively light cone vectors $n^\mu_{\rm{BKM}}\propto{\bar q^\mu}$ used in their paper.

\begin{figure}[ht]
\centering
\begin{minipage}[b]{\textwidth}
\includegraphics[width=0.5\textwidth]{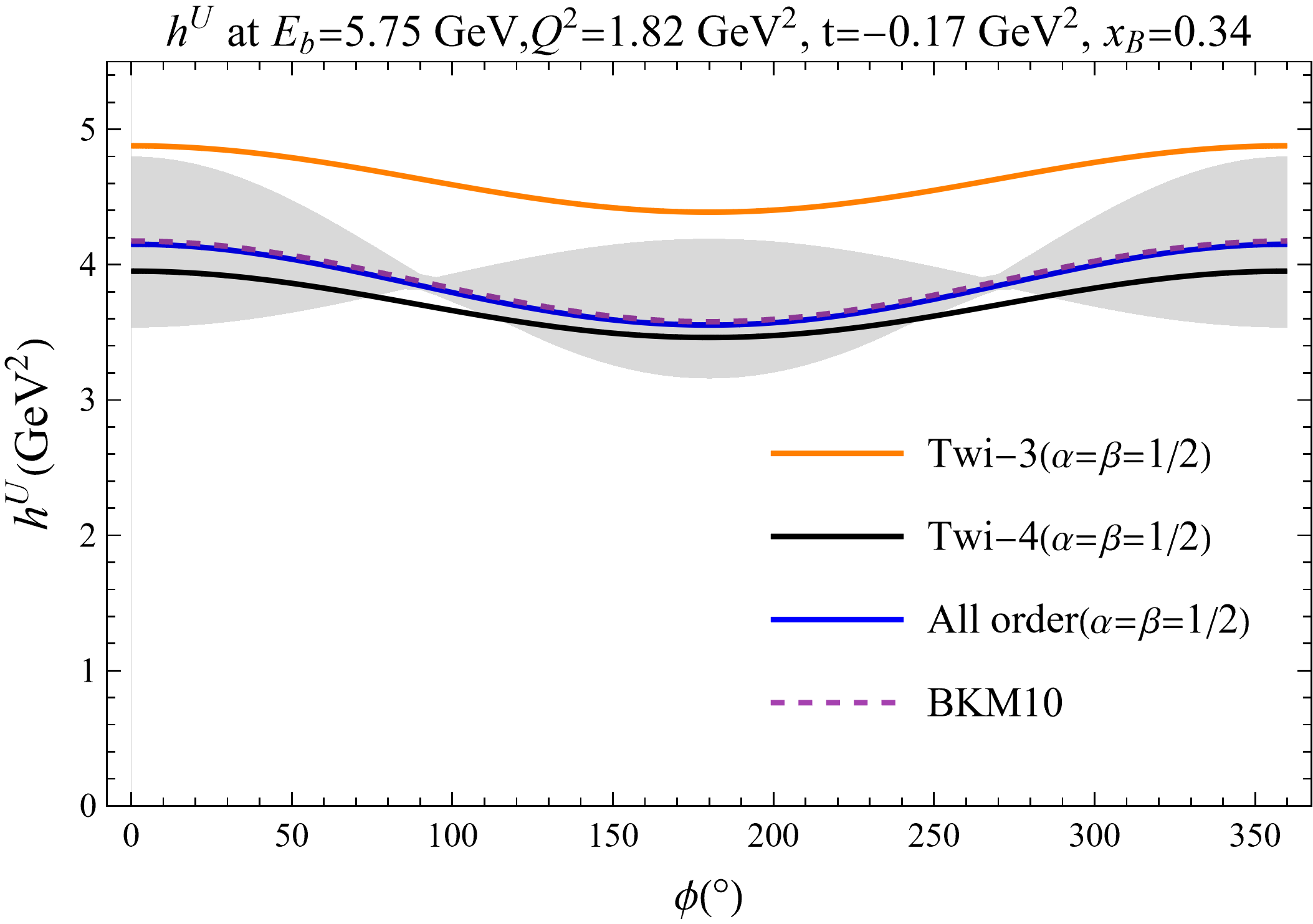}
\includegraphics[width=0.5\textwidth]{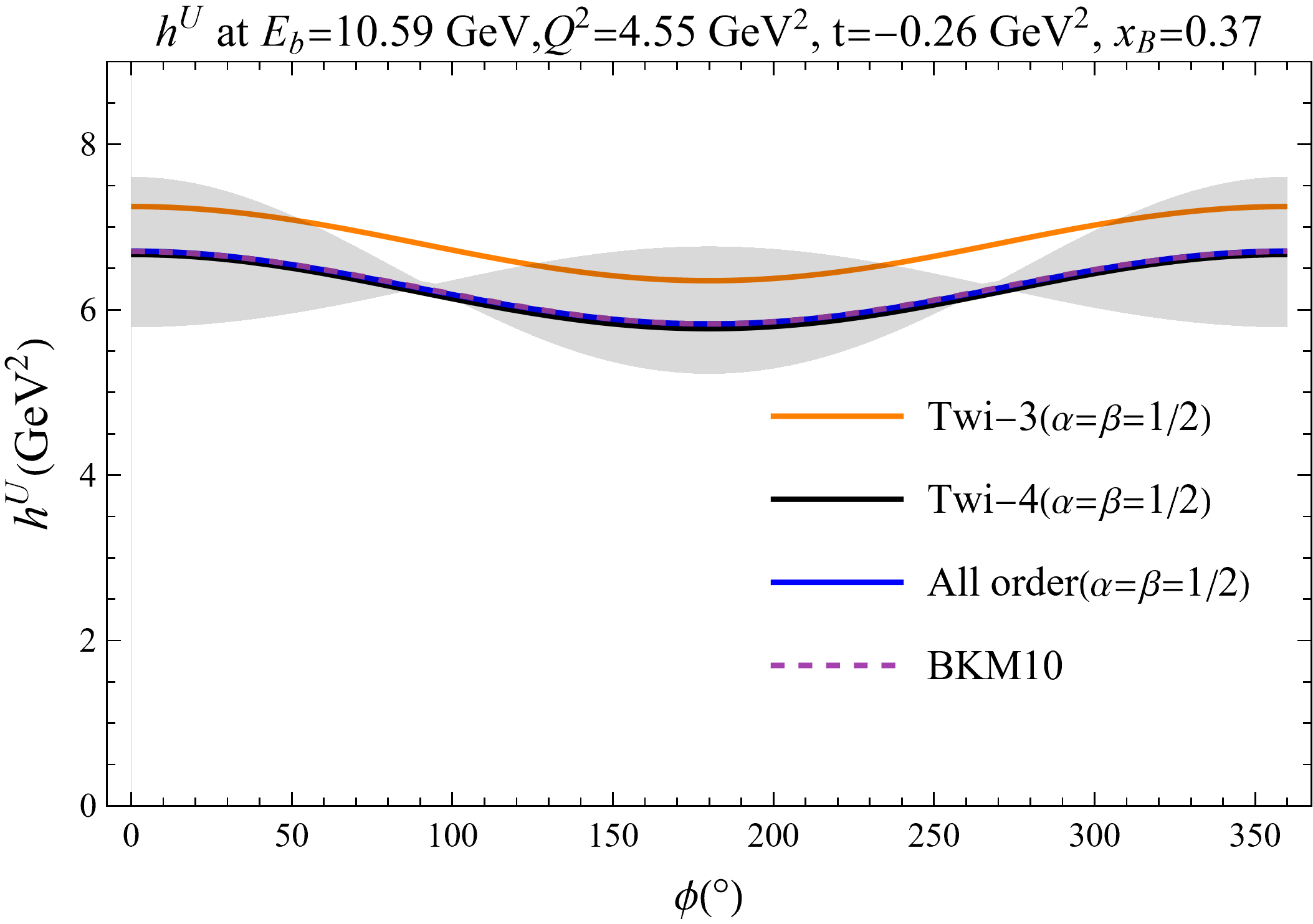}
\end{minipage}
\begin{minipage}[b]{\textwidth}
\includegraphics[width=0.5\textwidth]{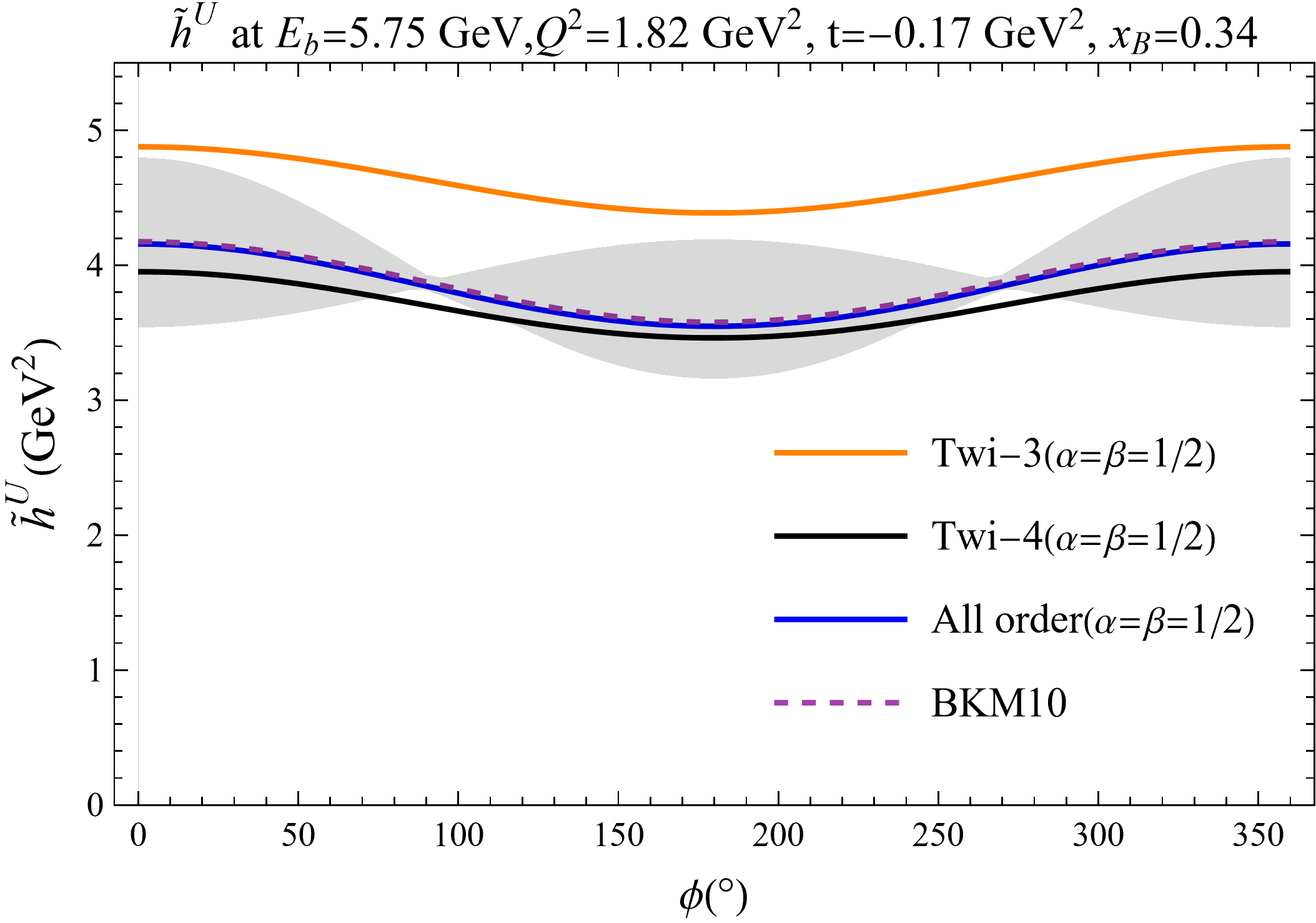}
\includegraphics[width=0.5\textwidth]{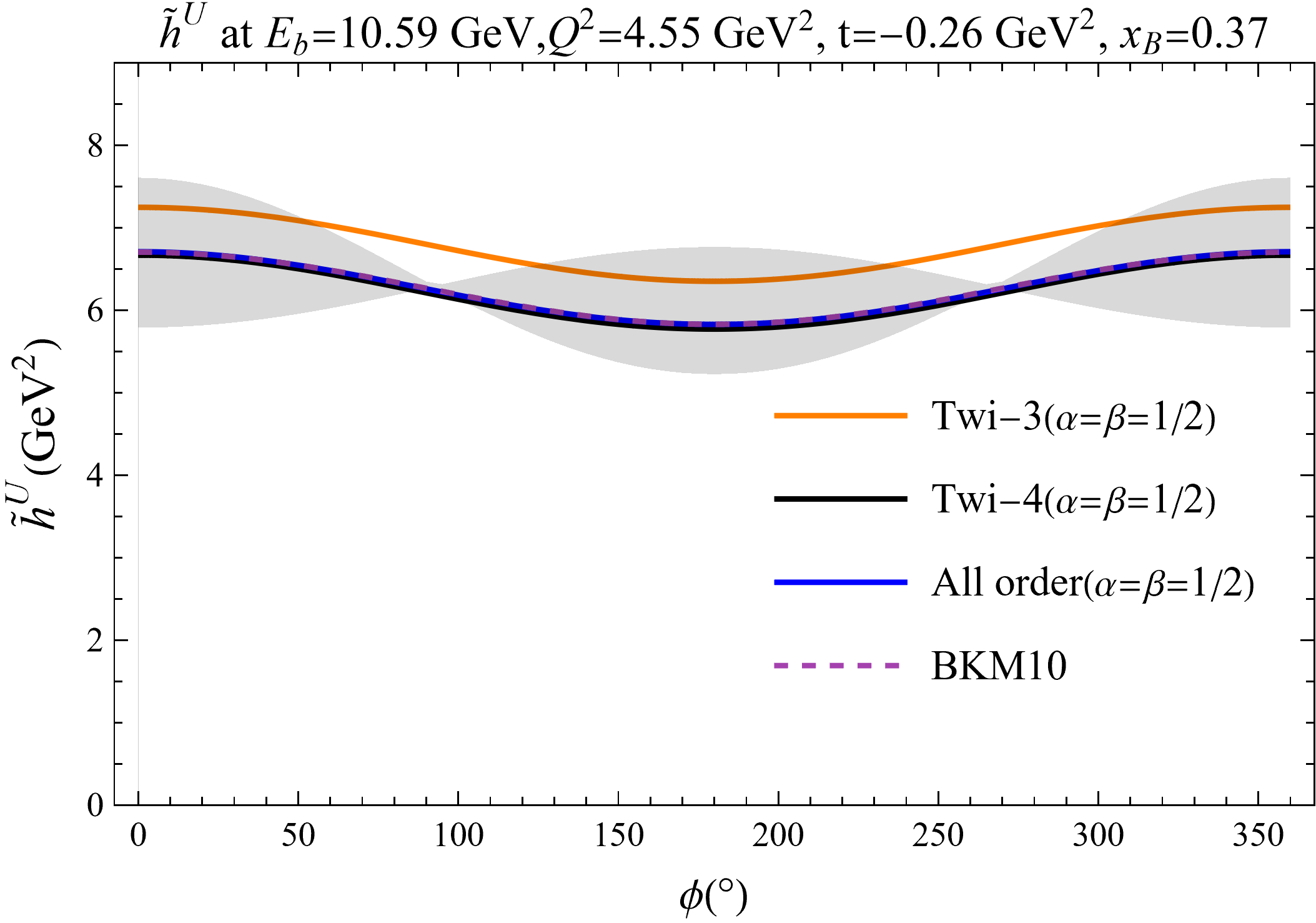}
\end{minipage}
\caption{\label{fig:dvcsstructurefuncexp} Comparison of the coefficients $h^{\rm{U}}$ and $\tilde{h}^{\rm{U}}$ from light cone calculation with the BKM10 result. The orange line stands for the results with only twist-three kinematical accuracy, the black line has twist-four kinematical accuracy, while the blue line contains kinematics of all order. The dashed line is the BKM10 result. The gray band consists of all possible values of those coefficients with $0\le \alpha\le 1$ and $0\le\beta\le 2$.}
\end{figure}

In Fig. \ref{fig:dvcsstructurefuncexp}, we compare those coefficients $h^{\rm{U}}$ and $\tilde{h}^{\rm{U}}$ with the BKM10 results \cite{Belitsky:2010jw}, which are almost the same with twist-four agreement and numerically very close to each other as we can see from the plot. To do so, we need to convert the BKM10 results into our form, which are given in Eq. (\ref{eq:hbkm}). It turns out that our twist expansion converges reasonably fast at twist four for both the JLab 6 GeV and 12 GeV points, where it converges faster for the 12 GeV which has a larger $Q^2=4.55 \text{ GeV}^2$. Besides, the all-order results and the twist-four results agree well with the BKM10 result \cite{Belitsky:2010jw}, when $\alpha=\beta=1/2$, corresponding to the choice of light cone vectors made there.

\begin{figure}[t]
\centering
\begin{minipage}[b]{0.48\textwidth}
\includegraphics[width=\textwidth]{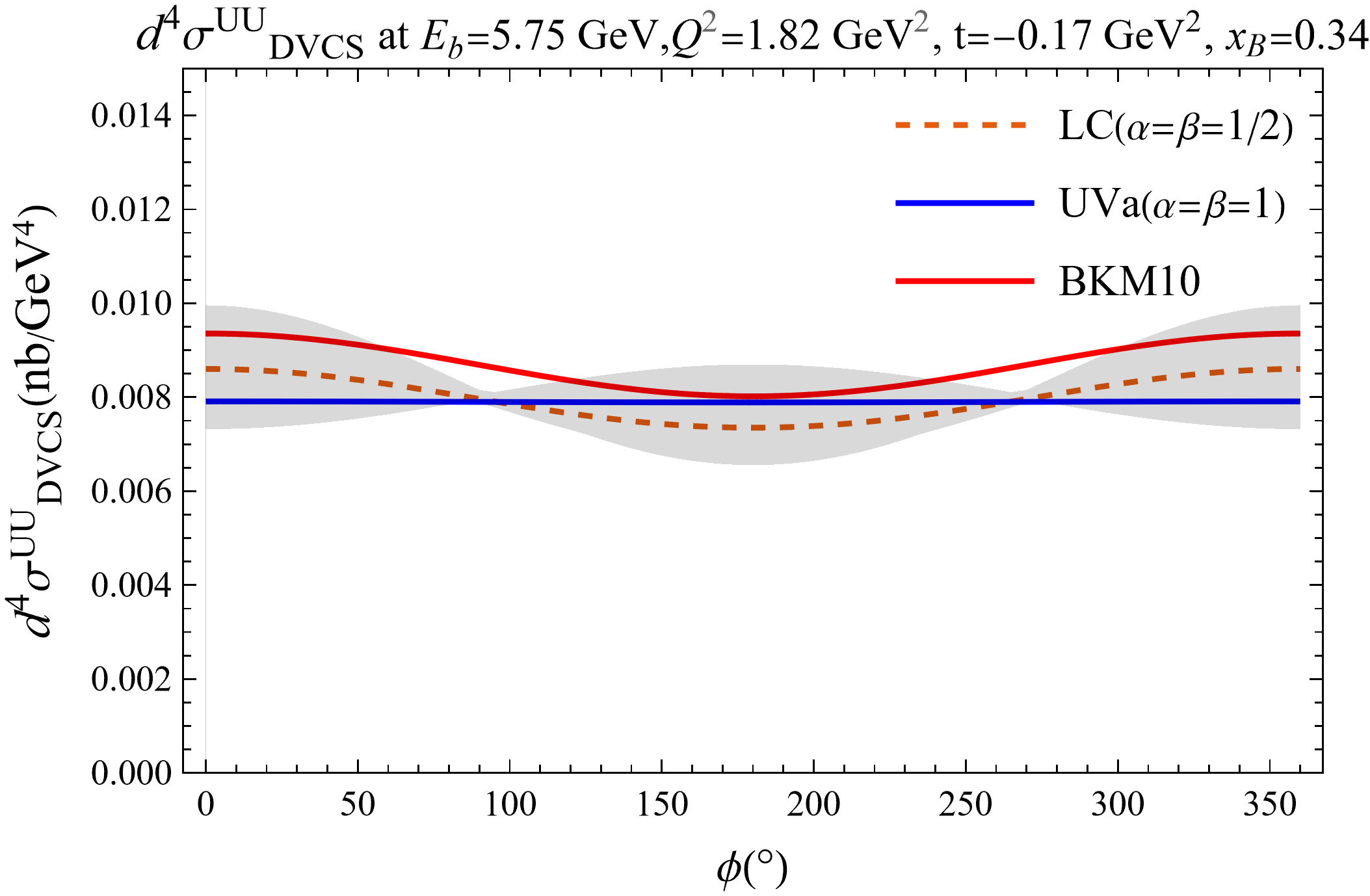}
\end{minipage}
\begin{minipage}[b]{0.5\textwidth}
\includegraphics[width=\textwidth]{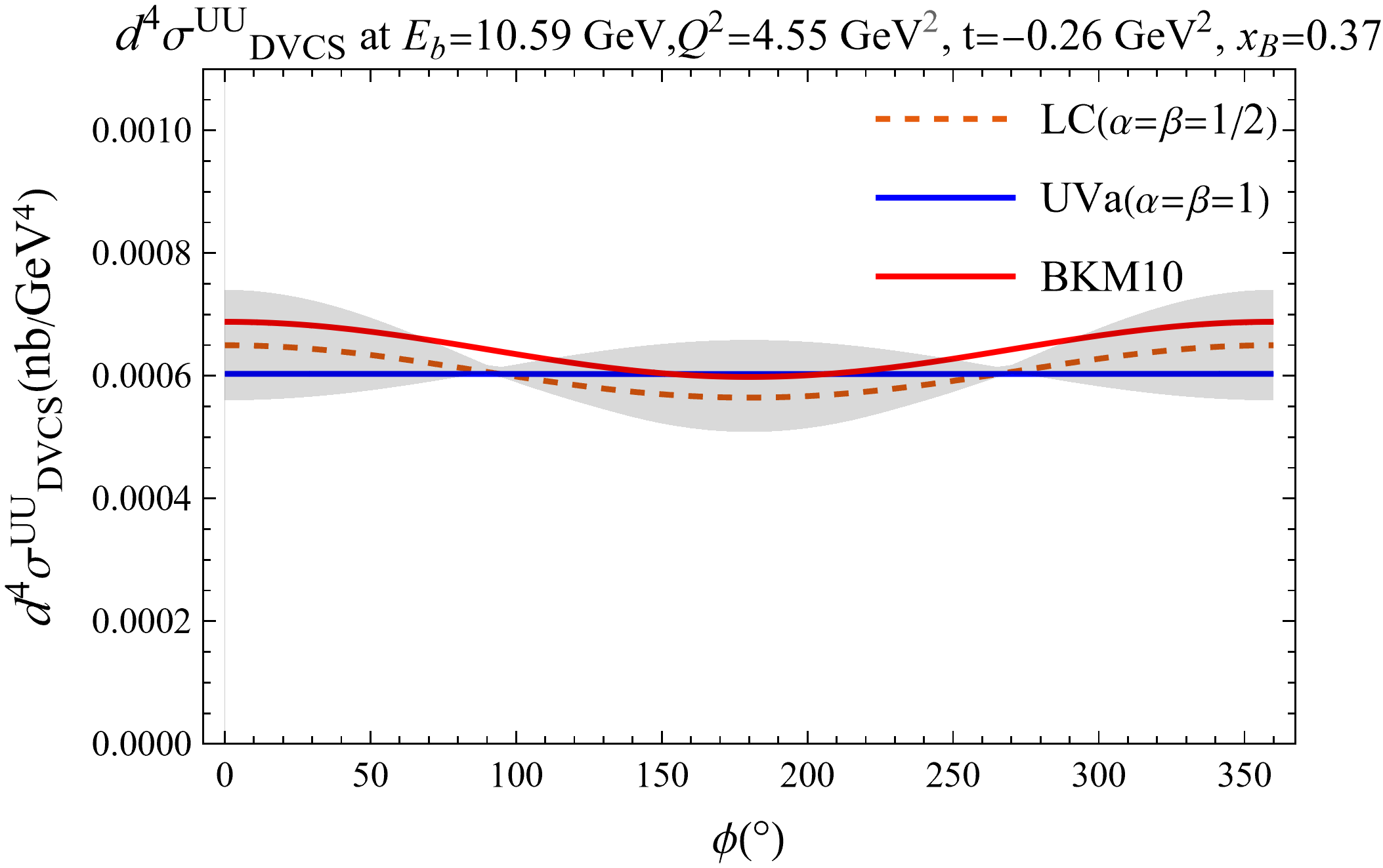}
\end{minipage}
\caption{\label{fig:DVCSxsection} Comparison of the unpolarized four-fold pure DVCS cross-section $\text{d}^4 \sigma^{UU}_{\rm{DVCS}}$. The dashed line corresponds to our light cone calculation with $\alpha=\beta=1/2$, the blue line corresponds to our reproduced UVa result with $\alpha=\beta=1$, and the red line is the BKM10 result. Besides, the gray band consists of  $0\le \alpha\le 1$ and $0\le\beta\le 2$. The three results shown here are reasonably close, while their difference are higher twist effects. }
\end{figure}

With those coefficients, we can move on to compare the cross-section. In order to perform the comparison of the cross-section, we need input for the CFFs. Here we used the UVa CFFs listed in~\cite{Kriesten:2020wcx} which are shown in Tab. \ref{tab:CFF}. With those CFFs, we can calculate and compare all the different DVCS cross-section formulas, which are plotted in Fig. \ref{fig:DVCSxsection}.  The light cone result with $\alpha=\beta=1/2$ has the same twist-three behavior as the BKM10, {while the light cone result at $\alpha=\beta=1$, corresponding to the UVa \cite{Kriesten:2019jep} formula, will have different twist-three behavior due to its different $\beta$, since the cross-section depends on $\beta$ at twist three as indicated by Eq. (\ref{eq:hutwist}}). One obvious difference is that the UVa results with $\beta=1$ actually have no angular dependence, while the BKM10 results with $\beta=1/2$ DO have. Both features are correctly reproduced using our general set-up. With that said, those results are reasonably close to each other numerically. In Fig. \ref{fig:DVCSxsection}, we also show a gray band consists of  all cross-section predictions with $0\le \alpha\le 1$ and $0\le\beta\le 2$. this band serves as an estimation of the intrinsic theoretical uncertainties for the pure DVCS cross-section.

\begin{table*}
\centering
\scalebox{0.85}{
\begin{tabular}{|c|c|c|c|c|c|c|c|c|c|c|c|}
\hline
 CFF & $x_{Bj}$ &  $|t|(\text{GeV}^{2})$ &  $Q^{2} (\text{GeV}^{2}) $ &Re$\mathcal H$  &  Re$\mathcal E$  &   Re$\widetilde{\mathcal H}$   &  Re$\widetilde{\mathcal E}$   & Im$\mathcal H$ &  Im$\mathcal E$  &  Im$\widetilde{\mathcal H}$  &   Im$\widetilde{\mathcal E}$   \\
 \hline
    VA & 0.34 & 0.17 & 1.82 & -0.897 & -0.541  & 2.444 & 2.207  & 2.421  & 0.903  & 1.131  & 5.383  \\
 \hline 
 VA & 0.37 & 0.26 & 4.55 &
   -0.884  & -0.424  & 3.118  & 2.900  & 1.851  & 0.649  & 0.911  & 3.915  \\
 \hline
\end{tabular}}
\caption{CFFs used for numerical comparison, see \cite{Kriesten:2020wcx}.} 
\label{tab:CFF}
\end{table*}

\subsection{Unpolarized interference cross-section comparison}
\label{subsect:interferencecomp}
Besides the pure DVCS cross-section, we also compare the four-fold interference cross-section $\text{d}^4 \sigma^{UU}_{\rm{INT}}$ with similar definition as the pure DVCS one with the BKM10 \cite{Belitsky:2010jw} result. Recall that in Eq. (\ref{eq:intstructfunc}), we define 6 coefficients that are independent of the CFFs, we first compare those coefficients.

\begin{figure}[ht]
\centering
\begin{minipage}[b]{\textwidth}
\includegraphics[width=0.5\textwidth]{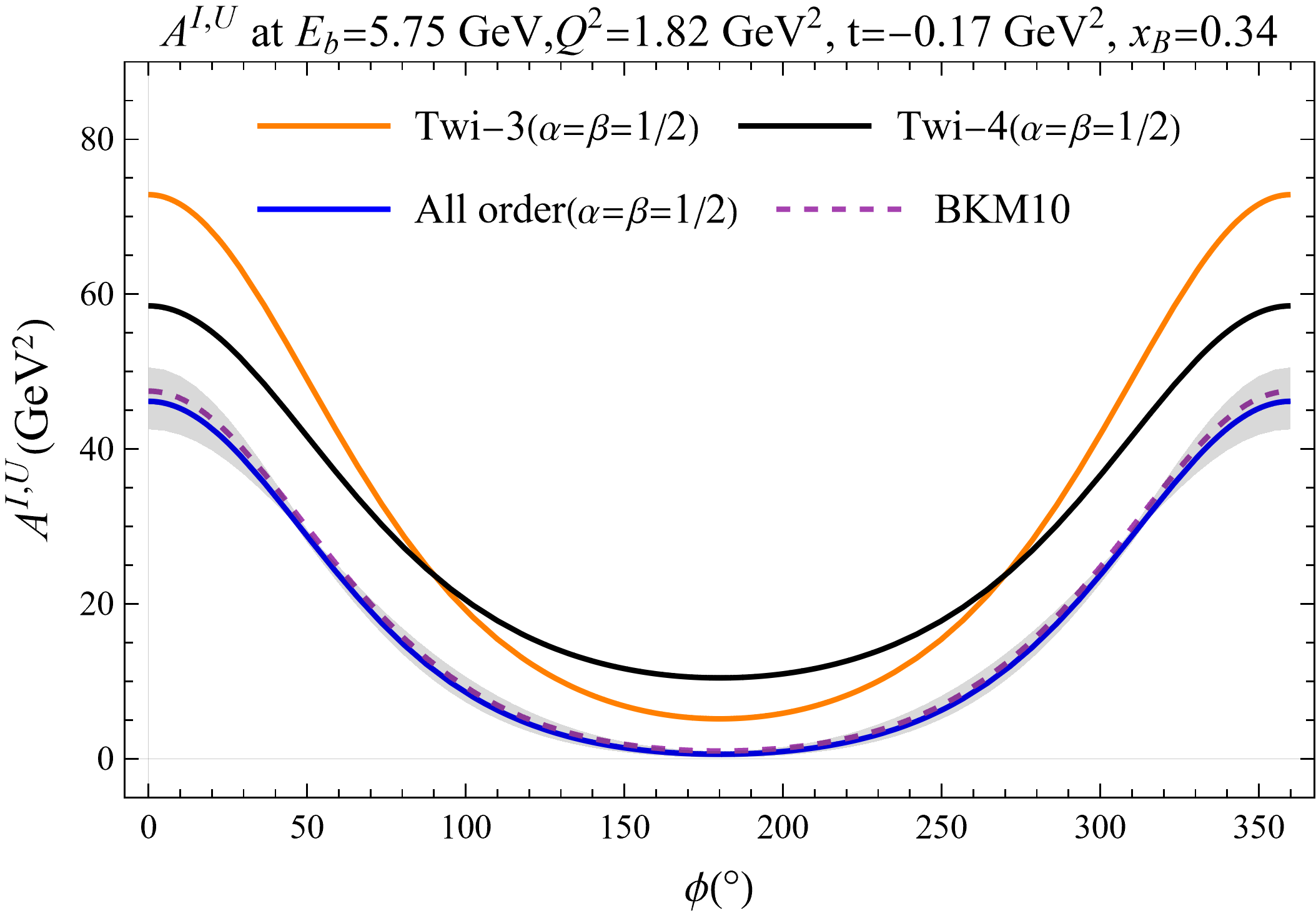}
\includegraphics[width=0.5\textwidth]{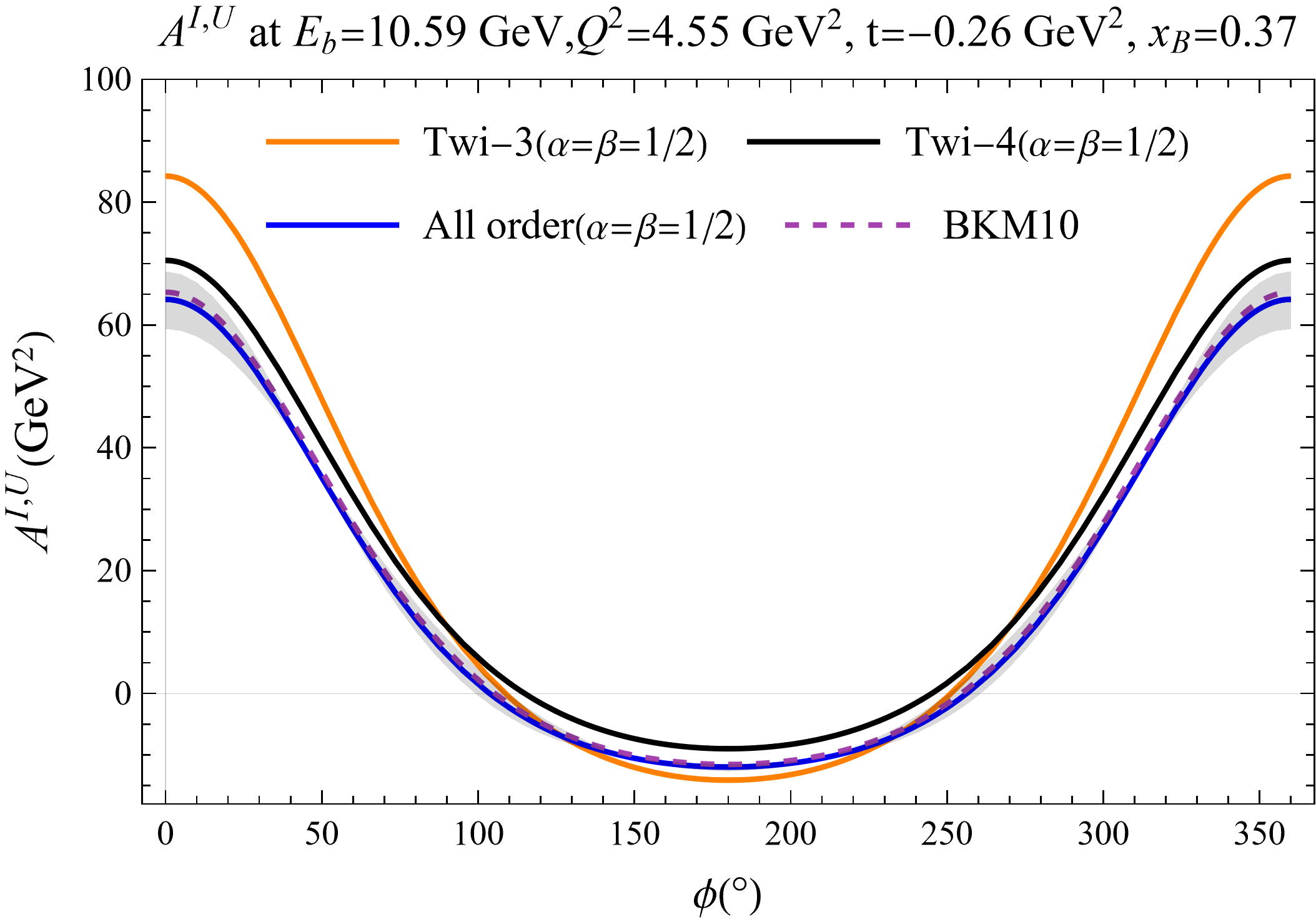}
\end{minipage}
\begin{minipage}[b]{\textwidth}
\includegraphics[width=0.5\textwidth]{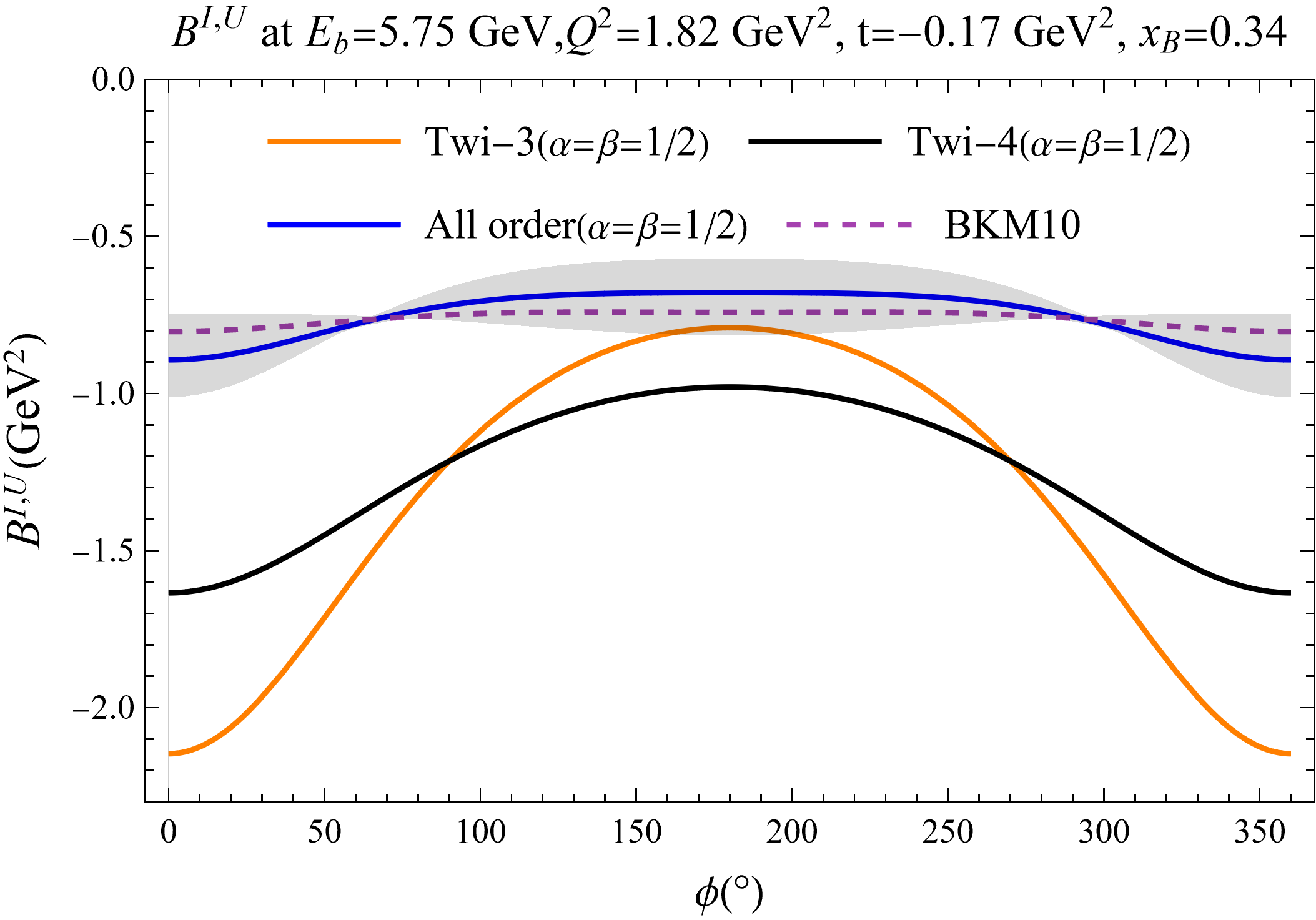}
\includegraphics[width=0.5\textwidth]{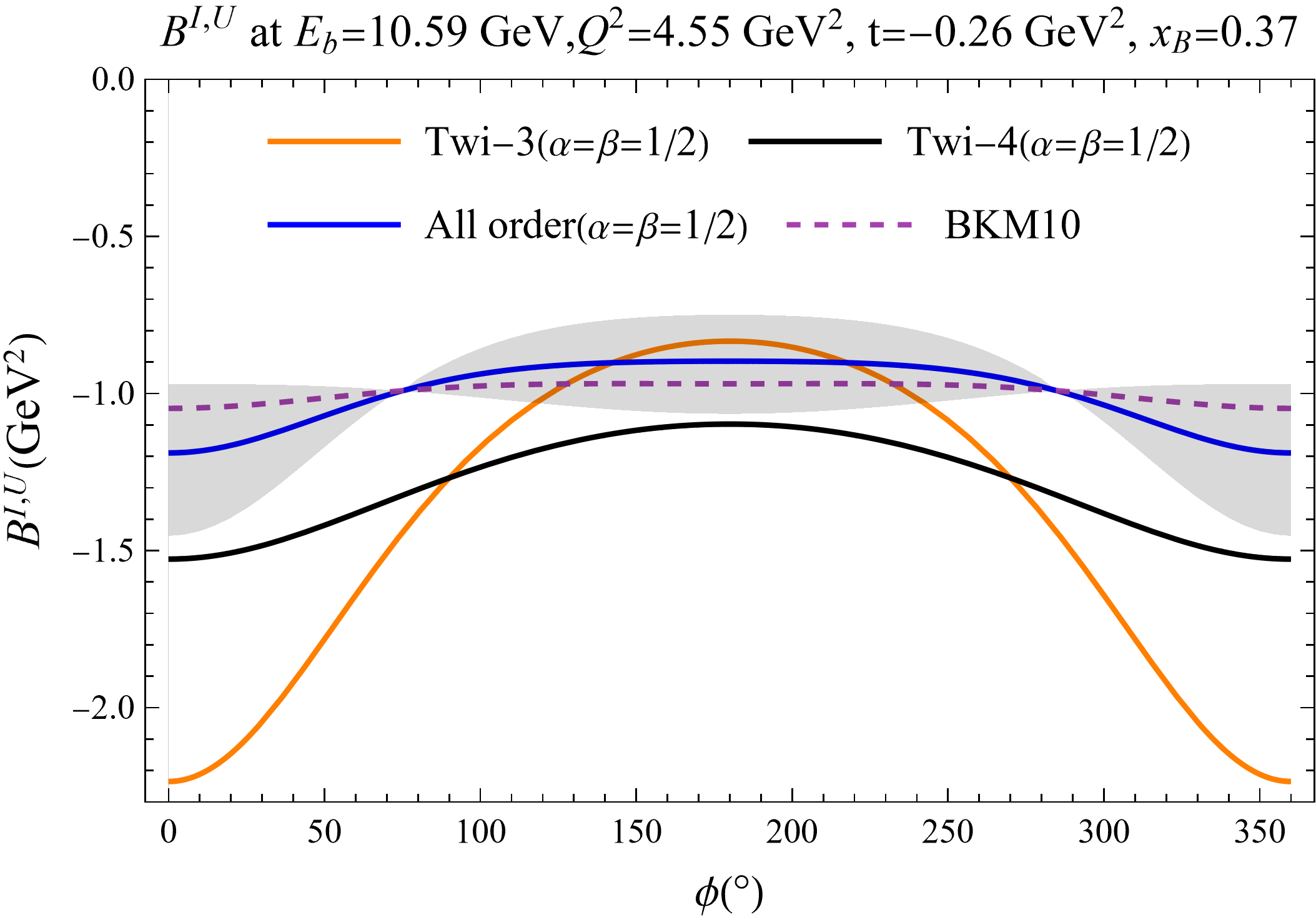}
\end{minipage}
\begin{minipage}[b]{\textwidth}
\includegraphics[width=0.5\textwidth]{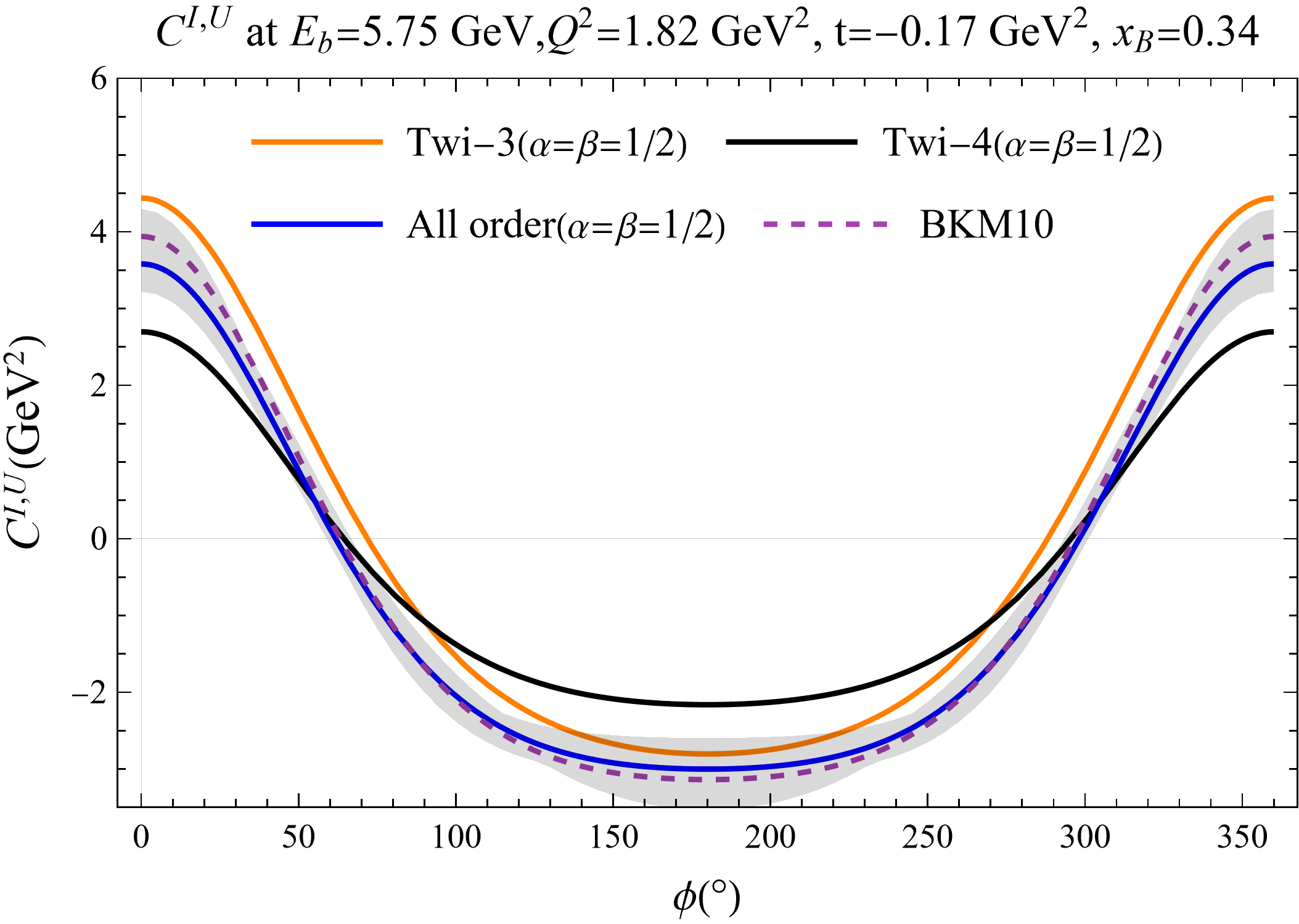}
\includegraphics[width=0.5\textwidth]{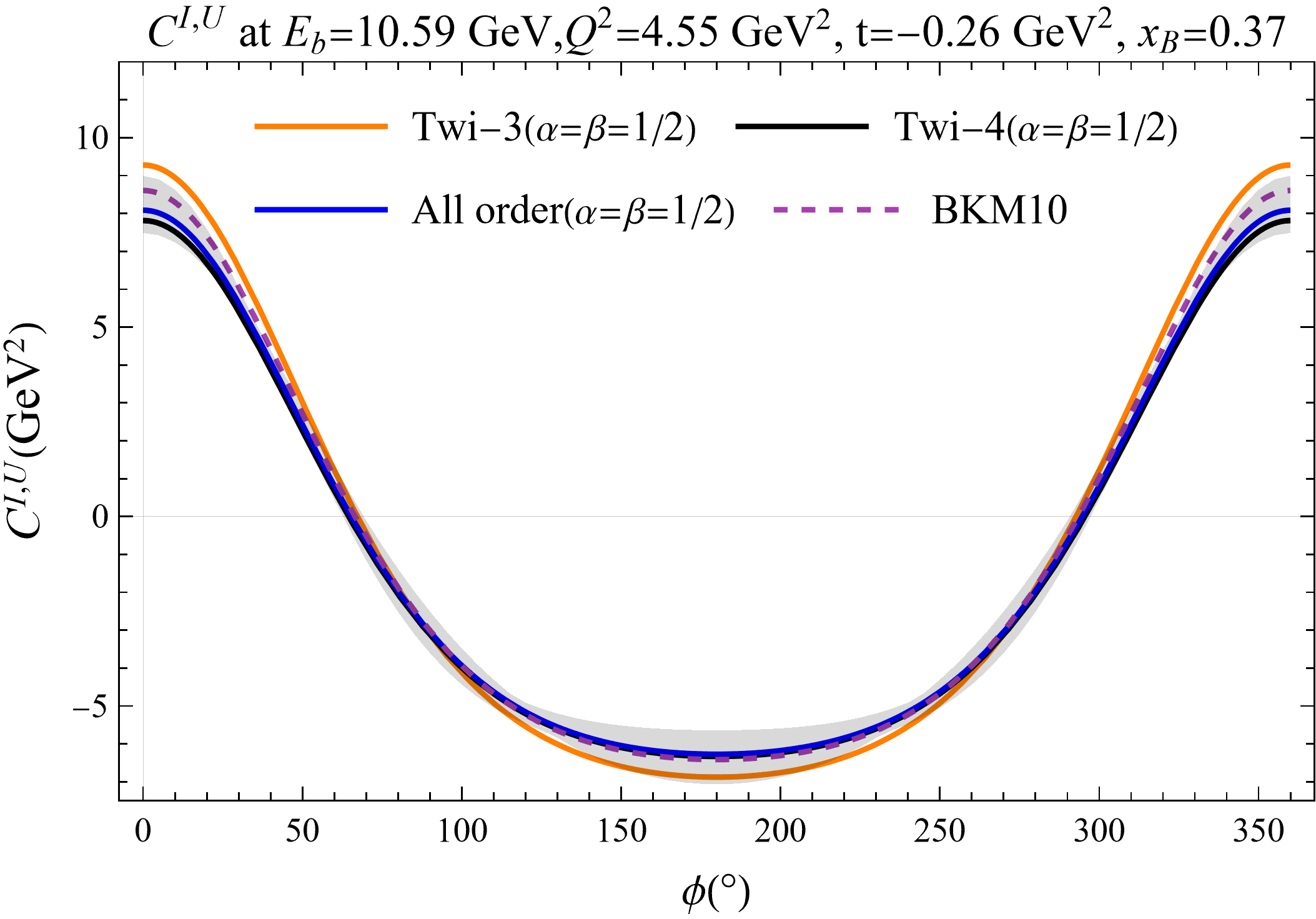}
\end{minipage}
\caption{\label{fig:structurefuncexp} Comparison of the coefficients $A^{I,\rm{U}},B^{I,\rm{U}},C^{I,\rm{U}}$ from light cone calculation with the BKM10 result. The orange line stands for the results with only twist-three kinematical accuracy, the black line has twist-four kinematical accuracy, while the blue line contains kinematics of all order. The dashed line is the BKM10 result. The gray band consists of all possible values of those coefficients with $0\le \alpha\le 1$ and $0\le\beta\le 2$.}
\end{figure}

For unpolarized target, the coefficients contain only the unpolarized part $A^{I,\rm{U}}$, $B^{I,\rm{U}}$ and $C^{I,\rm{U}}$. The coefficients of BKM10~\cite{Belitsky:2010jw} can be expressed into those coefficients as in Eqs. (\ref{eq:aiubkm}) - (\ref{eq:ciubkm}). Then similarly to the case of the pure DVCS cross-section, we perform a twist expansion comparison for those coefficients $A^{I,\rm{U}}$, $B^{I,\rm{U}}$ and $C^{I,\rm{U}}$ (see Eqs. (\ref{eq:aiutwist}) - (\ref{eq:ciutwist}) for the expansion). It can be easily checked that the expansions agree with the BKM10 results when taking $\beta =1/2$ at twist-three level, but not at twist-four. In Fig. \ref{fig:structurefuncexp}, we compare both our complete and expanded coefficients $A^{I,\rm{U}},B^{I,\rm{U}},C^{I,\rm{U}}$ with BKM10 ones. As we can see from the plot, the all-order results agree well with the BKM10 result, while the twist-four expansion works fine. It is also evident that for the JLab 12 GeV point with $Q^2=4.55 \text{GeV}^2$, the twist-expansion converges much faster than the JLab 6 GeV with $Q^2=1.82 \text{GeV}^2$, due to the larger $Q^2$.

Though the all-order light cone result at $\alpha=\beta=1/2$ agrees well with the BKM10 result numerically, an exact twist-four agreement is not expected due to the approximation made in BKM10. Given the effective light cone vectors $n_{\rm{BKM}}$ and $p_{\rm{BKM}}$ used in BKM, our light cone vectors $n$ and $p$ will not reduce to those off-light-cone vectors by setting $\alpha=\beta=1/2$.

\begin{figure}[ht]
\centering
\begin{minipage}[b]{0.48\textwidth}
\includegraphics[width=\textwidth]{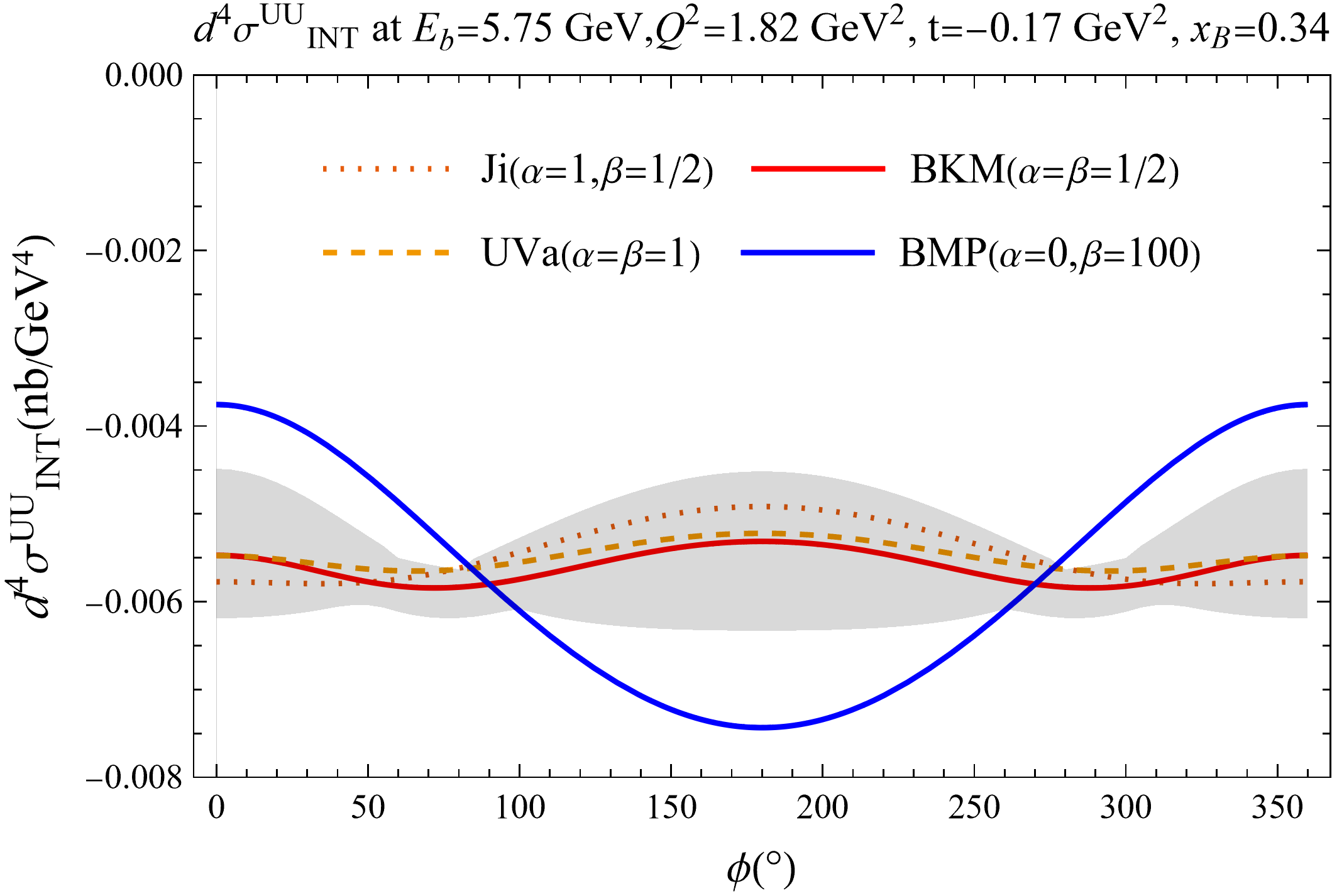}
\end{minipage}
\begin{minipage}[b]{0.5\textwidth}
\includegraphics[width=\textwidth]{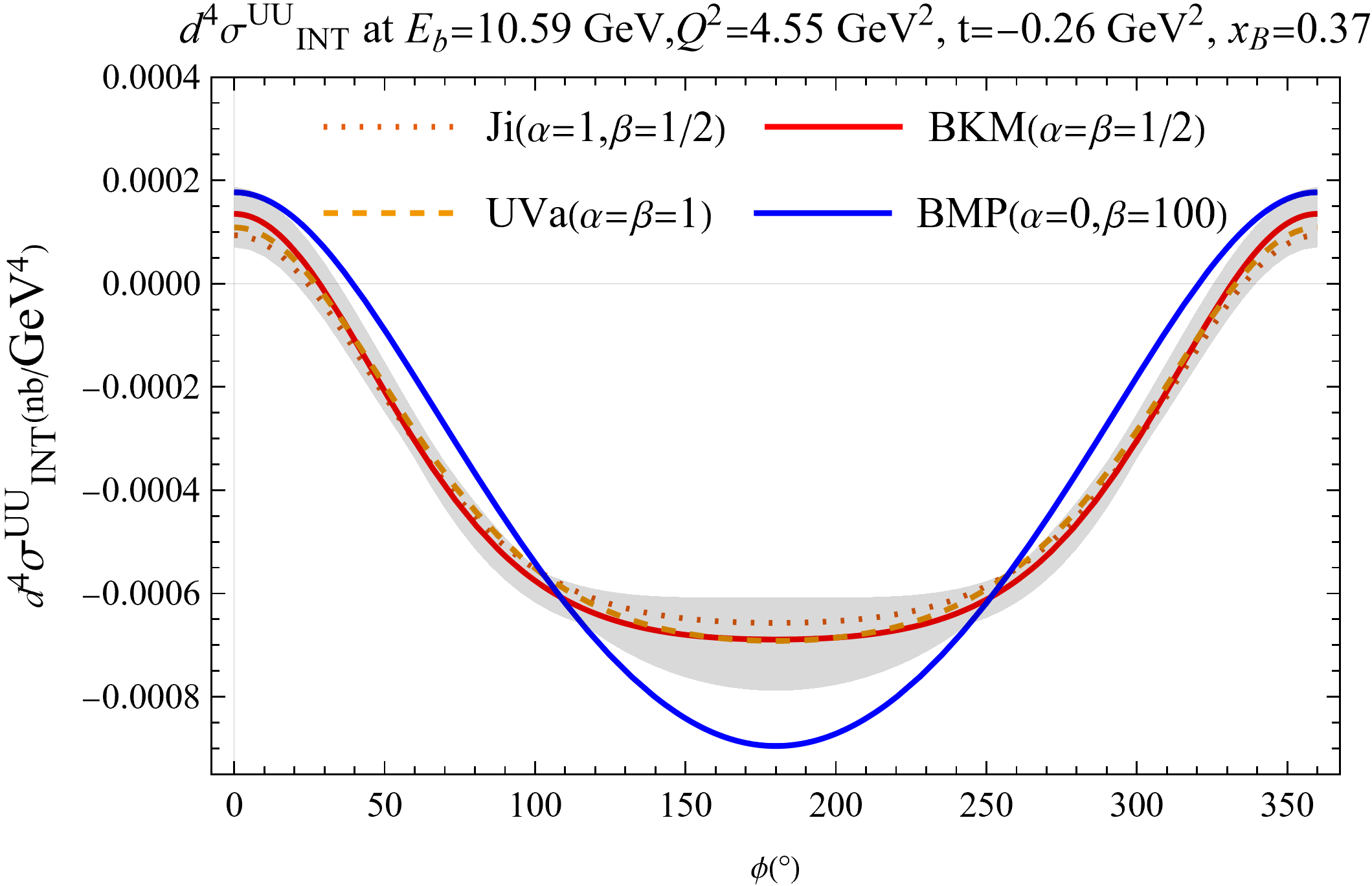}
\end{minipage}
\caption{\label{fig:paraplot} Comparison of the four-fold interference cross-section $\text{d}^4 \sigma^{UU}_{\rm{INT}}$ of different choice of $\alpha$ and $\beta$. Besides the different lines showing different choice of parameters, the gray band consists of all cross-section results with $0\le\alpha\le 1$ and $0\le\beta\le 2$. The UVa results shown here are obtained by taking $\alpha=\beta=1$ in our set-up and differ from the actual result in Ref \cite{Kriesten:2019jep} with an overall factor of $\cos \phi$.}
\end{figure}
With those coefficients compared, we could move on to the cross-section. In Fig. \ref{fig:paraplot}, we show the different cross-section prediction with different conventions ($\alpha$ and $\beta$) corresponding to the difference choice illustrated in Fig. \ref{fig:parameter}. While the BMP convention with $\beta\to \infty$ differs most from the other conventions, the others are quite close to each other. Again, we vary $\alpha$ from $0$ to $1$ and $\beta$ from $0$ to $2$, and then our predicted cross-sections will form a band as shown in Fig .\ref{fig:paraplot}. Since the choice of $\alpha$ and $\beta$ are conventional, those bands serve as an estimation of the theoretical intrinsic uncertainties of the cross-section prediction.

\subsection{Cross-section with longitudinally polarized target and unpolarized beam}
In the last two subsections, we compared the cross-section in the unpolarized case, and in this subsection we show that our formulas apply to the polarized cross-section which also agrees with the BKM10 results \cite{Belitsky:2010jw}. For simplicity, we consider the cross-section with longitudinally polarized target and unpolarized beam.

\begin{figure}[ht]
\centering
\begin{minipage}[b]{\textwidth}
\includegraphics[width=0.5\textwidth]{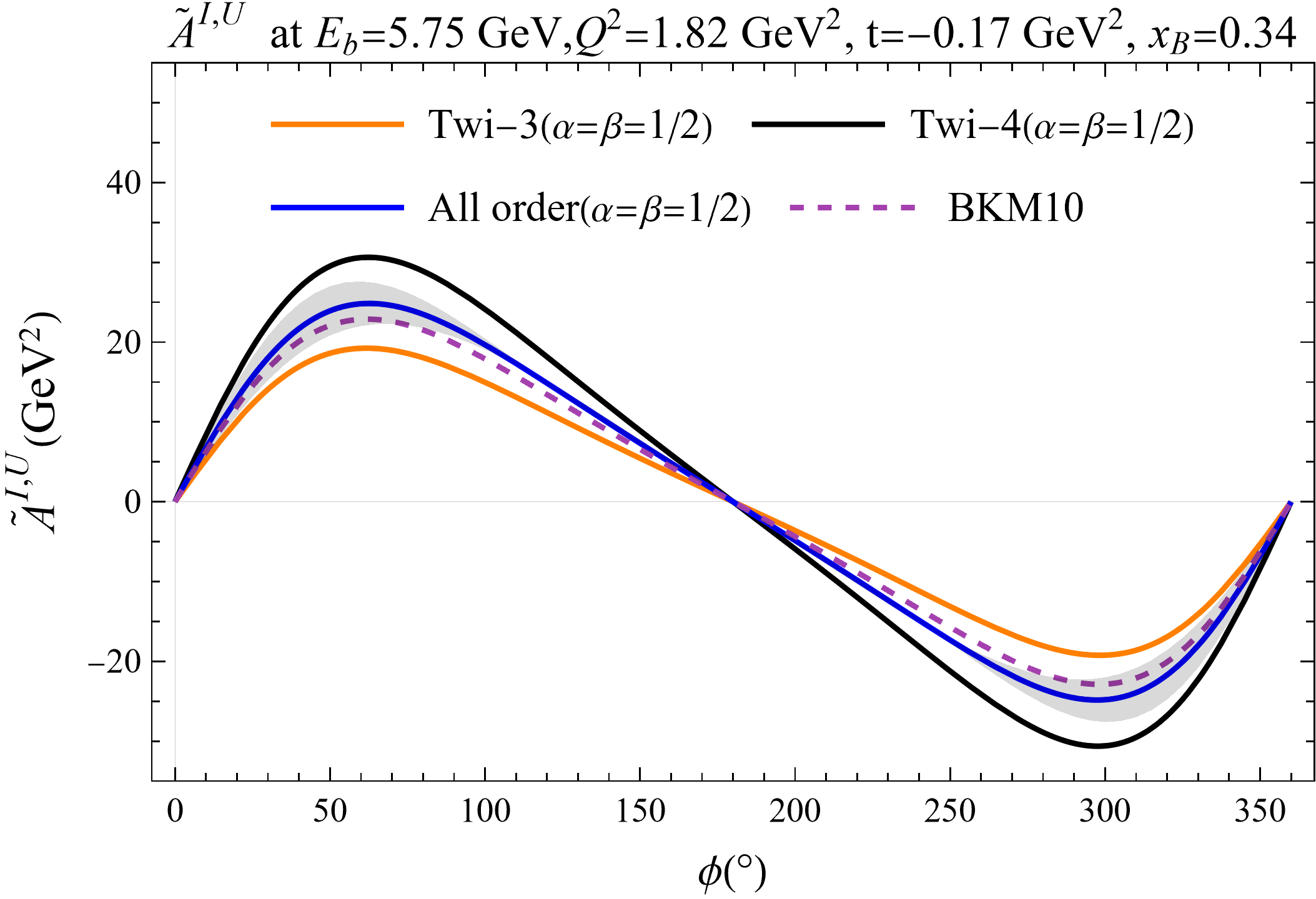}
\includegraphics[width=0.5\textwidth]{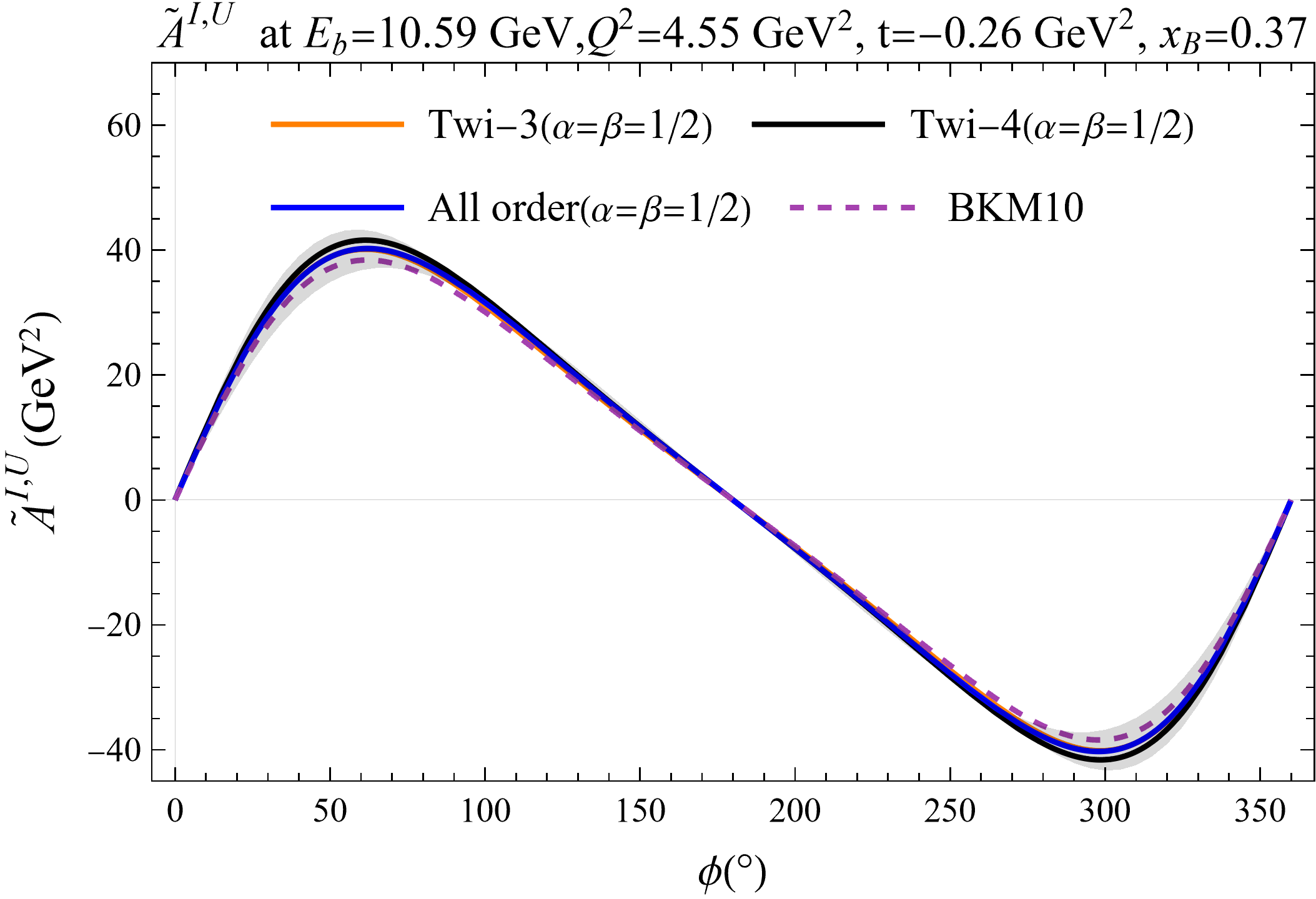}
\end{minipage}
\begin{minipage}[b]{\textwidth}
\includegraphics[width=0.5\textwidth]{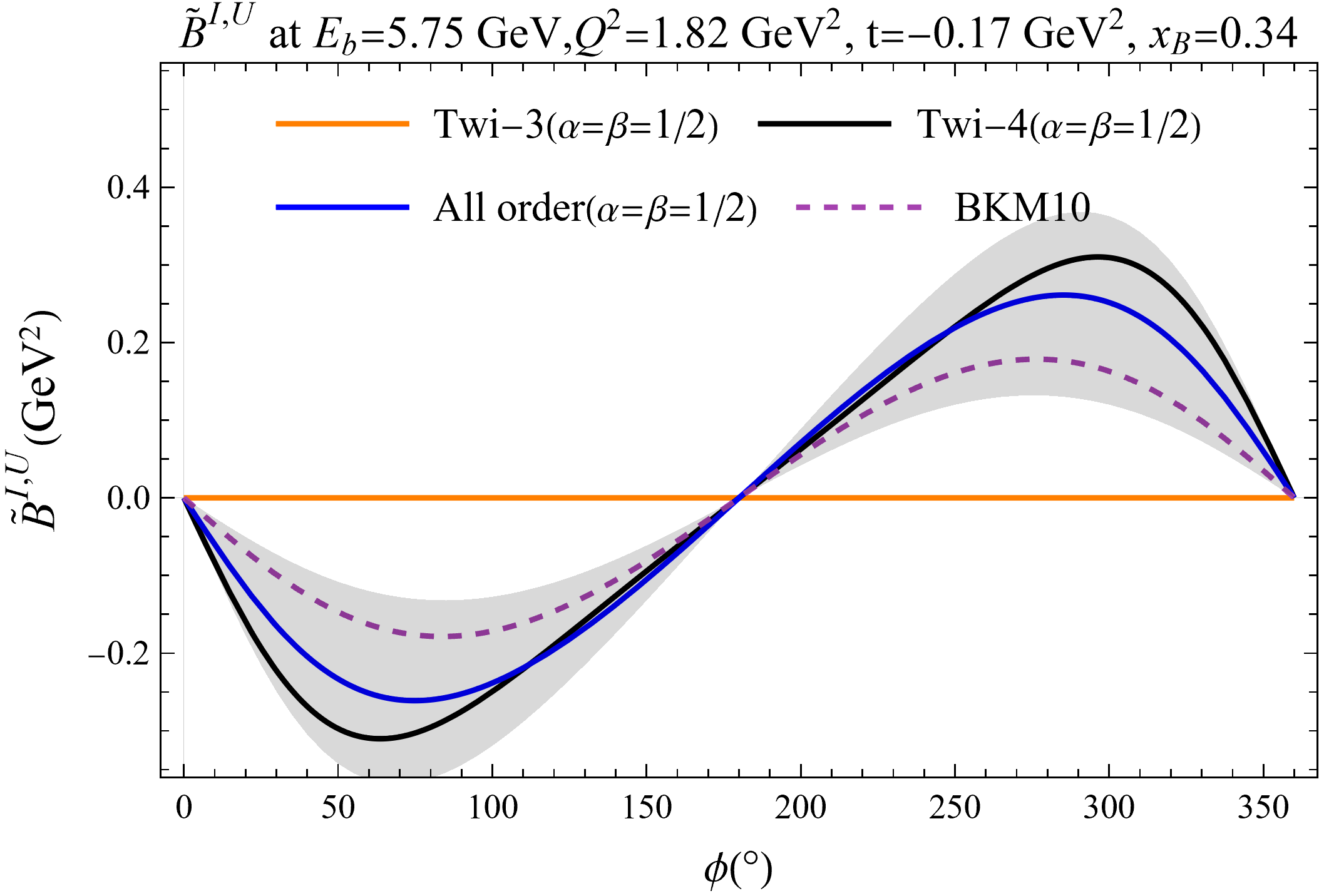}
\includegraphics[width=0.5\textwidth]{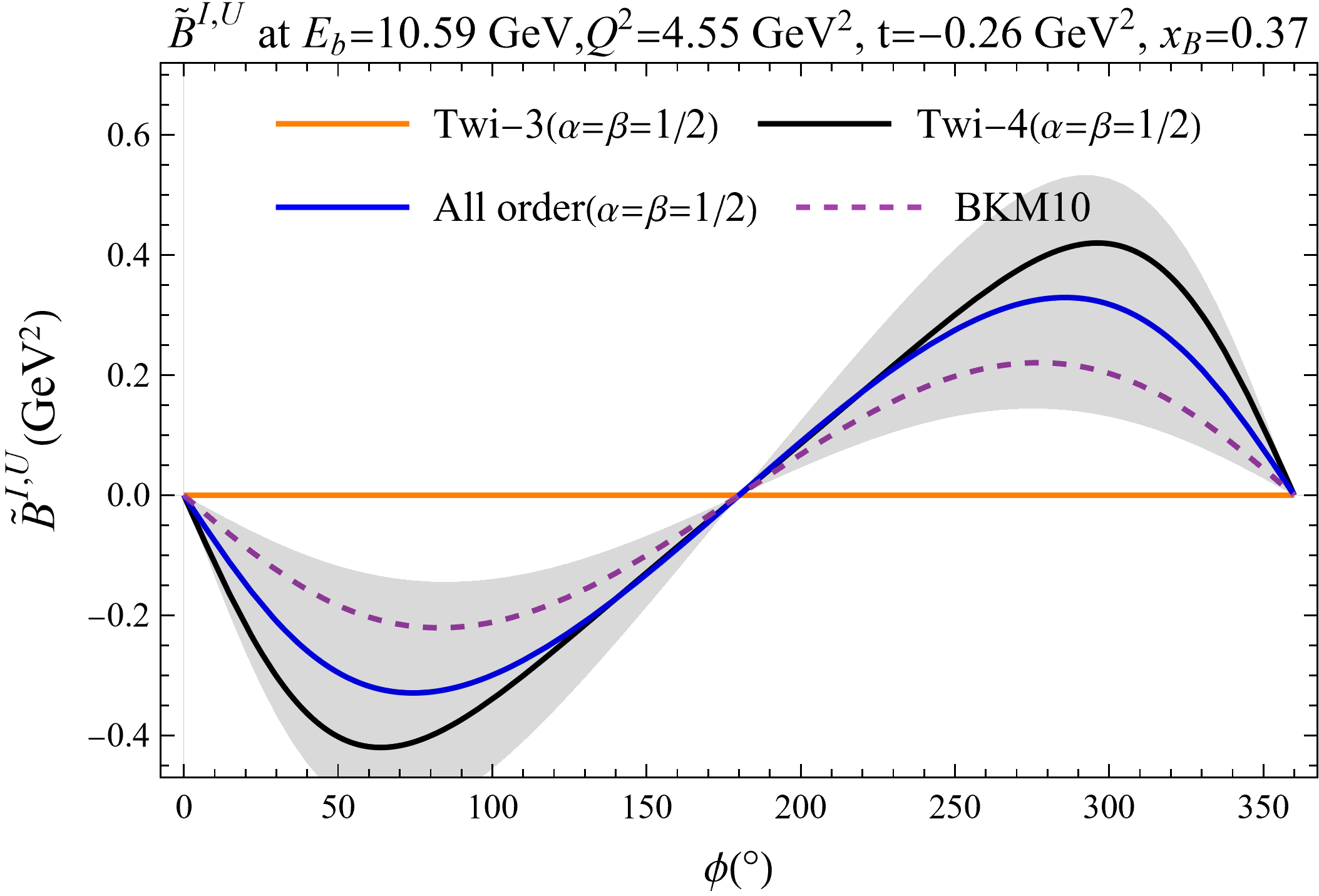}
\end{minipage}
\begin{minipage}[b]{\textwidth}
\includegraphics[width=0.5\textwidth]{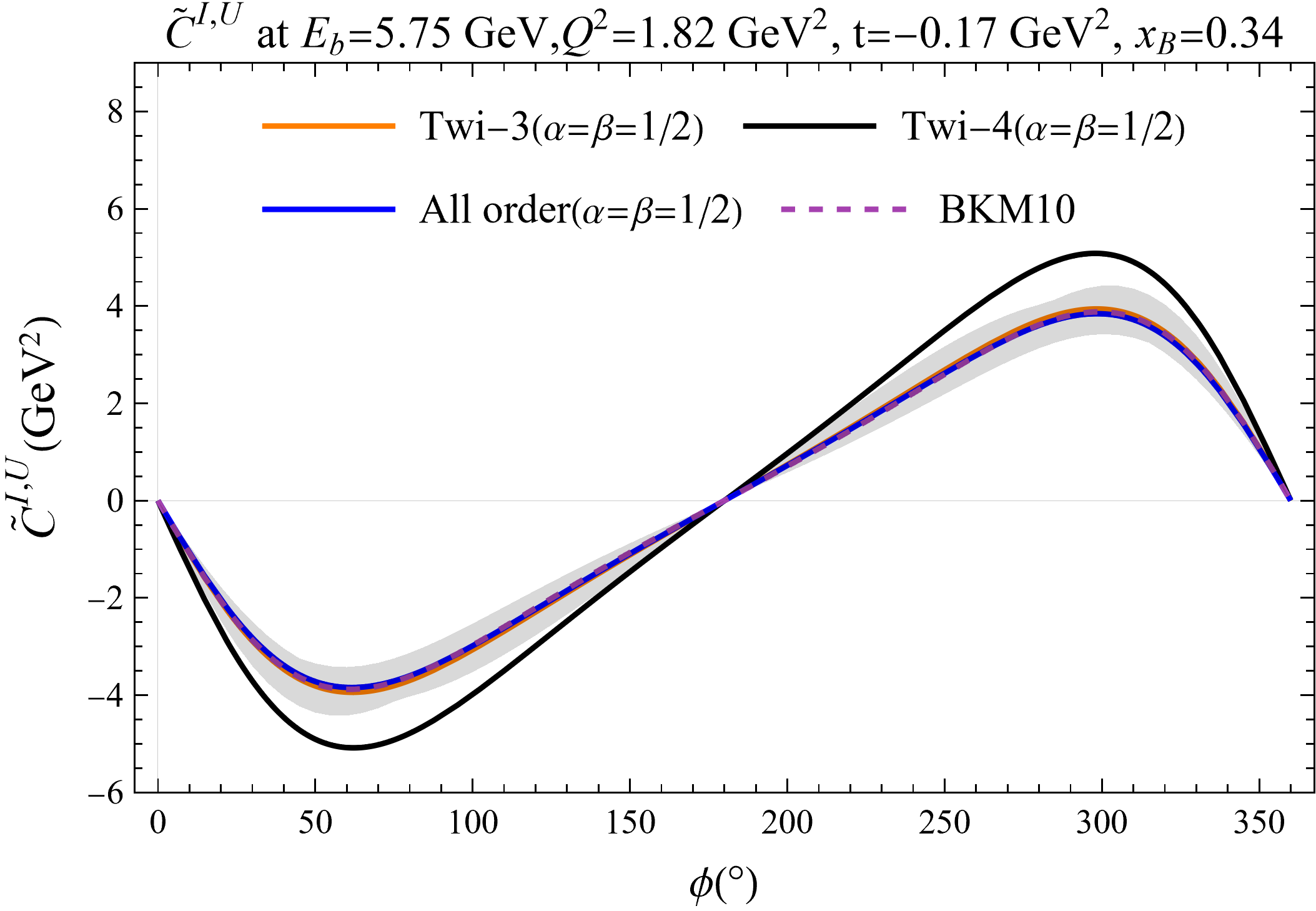}
\includegraphics[width=0.5\textwidth]{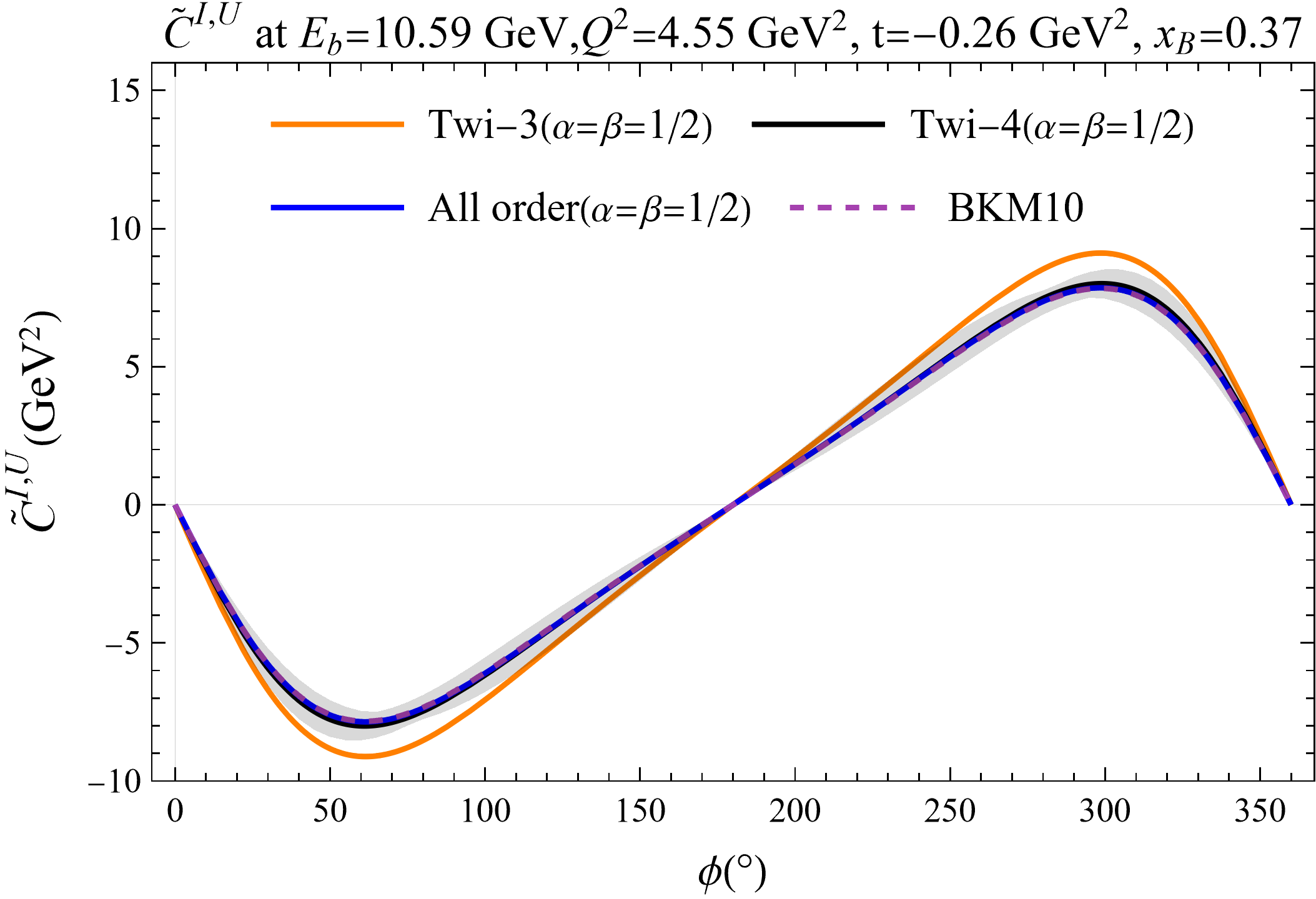}
\end{minipage}
\caption{\label{fig:structurefunc2} Comparison of the coefficients $\tilde{A}^{I,\rm{U}},\tilde{B}^{I,\rm{U}},\tilde{C}^{I,\rm{U}}$ from light cone calculation with the BKM10 result.The orange line stands for the results with only twist-three kinematical accuracy, the black line has twist-four kinematical accuracy, while the blue line contains kinematics of all order. The dashed line is the BKM10 result. The gray band consists of all possible values of those coefficients with $0\le \alpha\le 1$ and $0\le\beta\le 2$.}
\end{figure}

The pure DVCS cross-section does not have leading-twist contribution, as we discussed in subsection \ref{subsec:puredvcs}. Therefore, we focus on comparing the interference cross-section. In Eq. (\ref{eq:FIUL}), we show that the cross-section for unpolarized beam and longitudinally polarized target can be expressed in terms of the other three coefficients $\tilde A^{I,\rm{U}}$, $\tilde B^{I,\rm{U}}$ and $\tilde C^{I,\rm{U}}$. In Ref. \cite{Belitsky:2010jw}, the same hadronic structure shows up, and a comparison of those coefficients would suffice. Again, in order to compare them, we need to convert the BKM result to the form of ours, which are given in Eqs. (\ref{eq:atiubkm}) - (\ref{eq:ctiubkm}). We perform a twist expansion for those three coefficients in Eqs. (\ref{eq:atiutwist}) - (\ref{eq:ctiutwist}) and compare both the expanded formulas and the all-order formulas numerically, as shown in Fig. \ref{fig:structurefunc2}. Note that there is an overall minus sign difference between ours and the BKM10 result which we take away, since the polarization vectors can always be redefined with a minus sign. 

As we can see from the plot, our results agree well with the BKM10 result \cite{Belitsky:2010jw} when taking $\alpha=\beta=1/2$, corresponding to the choice made there. The larger difference in $\tilde{B}^{I,\rm{U}}$ can be explained for two reasons. For one thing, $\tilde{B}^{I,\rm{U}}$ does not have a leading twist contribution, so the differences at a lesser-dominant higher twist will be more significant in $\tilde{B}^{I,\rm{U}}$. Another reason is that the effective light cone vectors $n_{\rm{BKM}}$ used in \cite{Belitsky:2010jw} is not light-like, while $\tilde{B}^{I,\rm{U}}$ is defined as the contraction of $n^\sigma$ with the leptonic part as shown in Eq. (\ref{eq:intstructfunc}), which depends directly on the light cone vector. Apart from that, we see similar behavior as the unpolarized case that the twist expansion converges much faster for the JLab 12 GeV point, due to its larger $Q^2$. Also, our agreements with the BKM10 results improve as $Q^2$ gets large, since their difference is a higher twist effect.

\begin{figure}[ht]
\centering
\begin{minipage}[b]{0.48\textwidth}
\includegraphics[width=\textwidth]{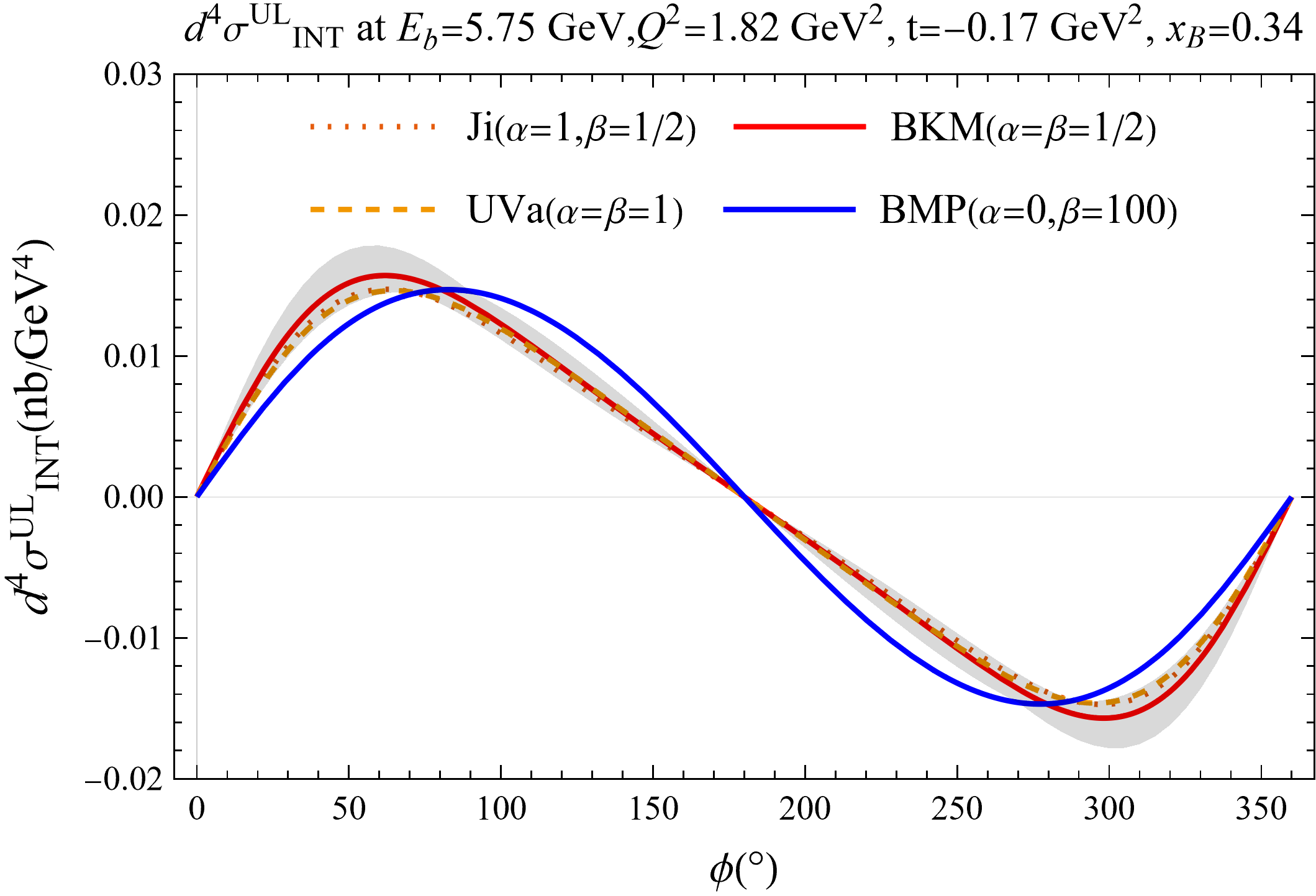}
\end{minipage}
\begin{minipage}[b]{0.49\textwidth}
\includegraphics[width=\textwidth]{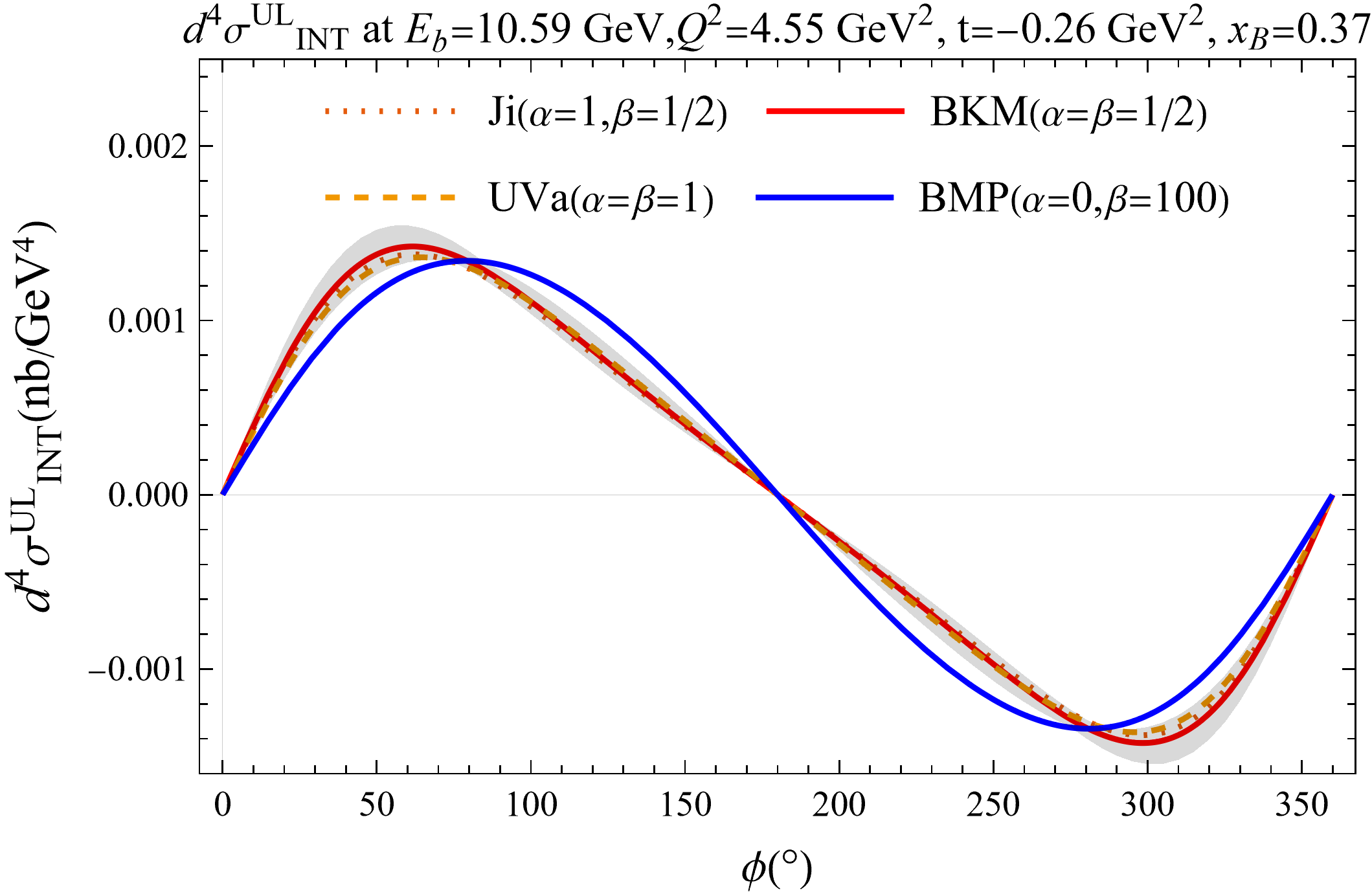}
\end{minipage}
\caption{\label{fig:paraplotlp} Comparison of the four-fold interference cross-section $\text{d}^4 \sigma^{UL}_{\rm{INT}}$ of different choice of $\alpha$ and $\beta$ for longitudinally polarized target and unpolarized beam. The gray band consists of all cross-section results with $0\le\alpha\le 1$ and $0\le\beta\le 2$. The UVa results shown here are obtained by taking $\alpha=\beta=1$ in our set-up and might differ from the actual result in Ref \cite{Kriesten:2019jep}.}
\end{figure}

With those coefficients and nucleon form factors, the cross-sections can be worked out the same way as the unpolarized case. We compare the polarized cross-section with different choice of parameters as shown in Fig. \ref{fig:paraplotlp}. Again, we show that they form a band which estimates the intrinsic theoretical uncertainties of cross-section prediction.

\subsection{Gauge dependence of cross-section}

In the subsection \ref{subsec:gaugeinvariance}, we discussed that the cross-section actually depends on the gauge choice due to the breaking of exact current conservation condition of the Compton tensor: $q'_\mu T^{\mu\nu}\sim T^{\mu\nu} q_\nu\sim \mathcal O\left(\Delta_\perp^2\right)$. Therefore, the cross-section will not be invariant under the gauge transformation $\varepsilon^{\mu}(q',\Lambda') \to \varepsilon^{\mu}(q',\Lambda') + \lambda q'^\mu$. In this subsection, we study how the cross-section will vary under the gauge transformation.

\begin{figure}[ht]
\centering
\begin{minipage}[b]{0.5\textwidth}
\includegraphics[width=\textwidth]{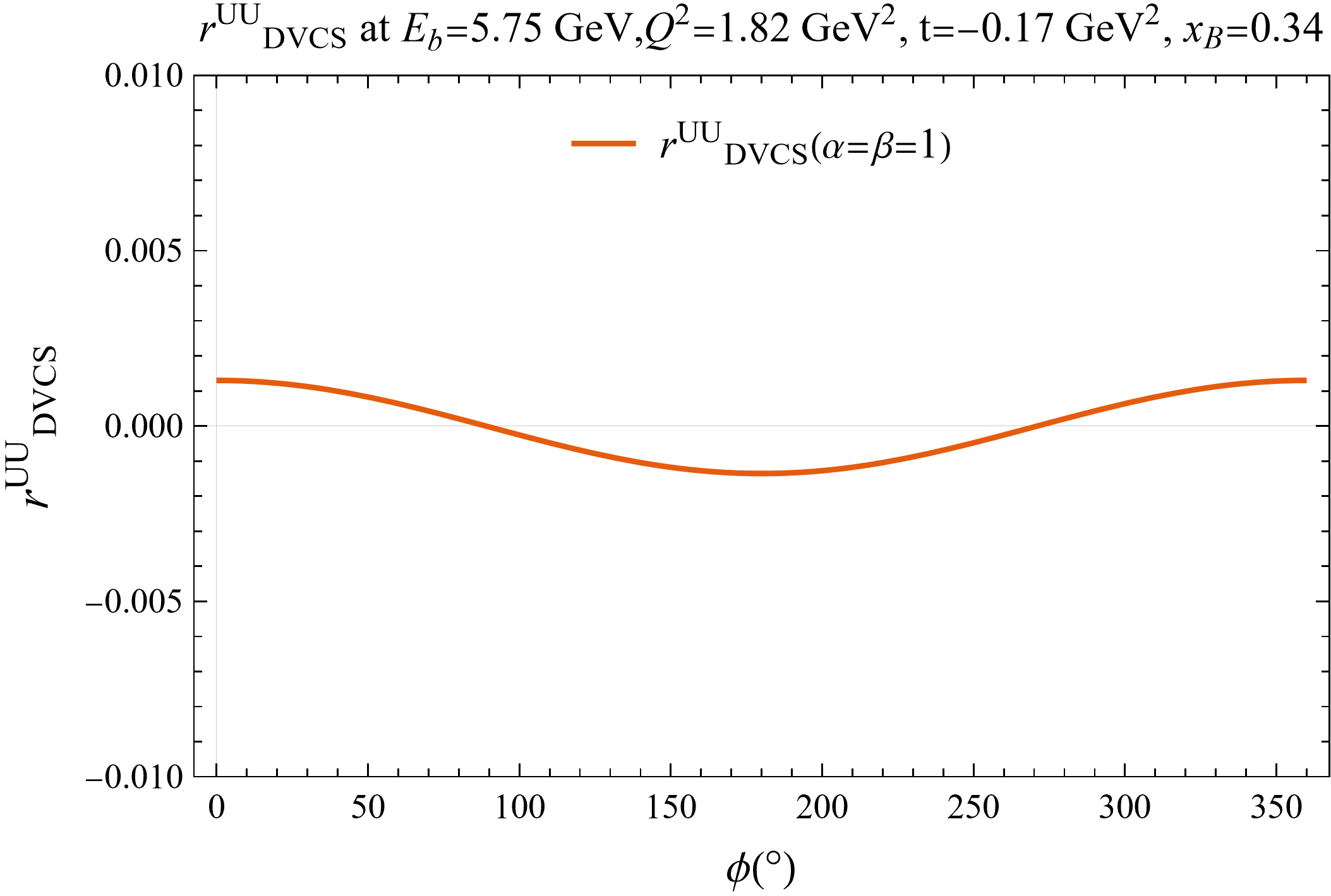}
\end{minipage}
\begin{minipage}[b]{0.49\textwidth}
\includegraphics[width=\textwidth]{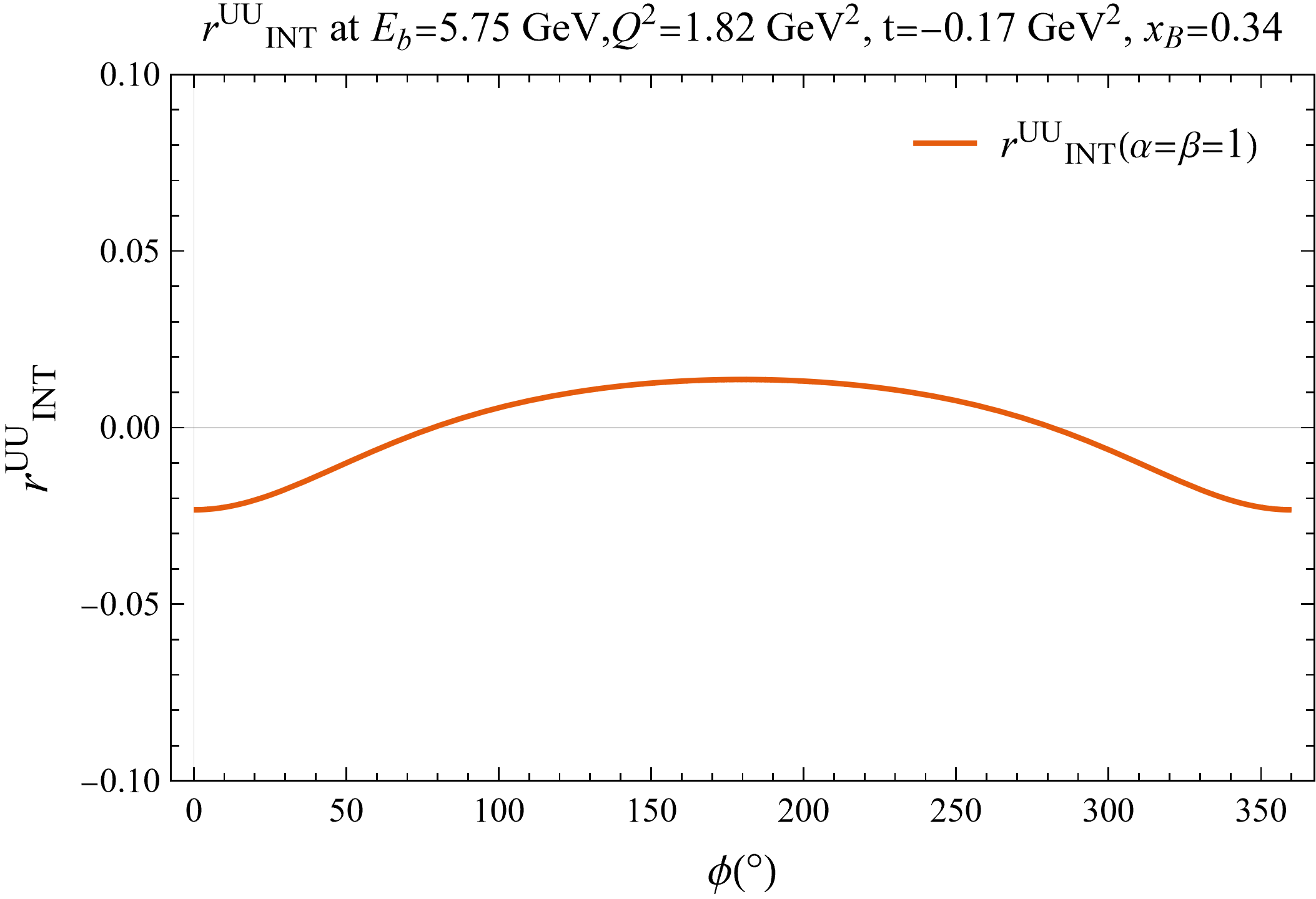}
\end{minipage}
\caption{\label{fig:gaugedep} The ratio $r^{UU}_{\rm{DVCS/INT}}$ defined above in the unpolarized case, calculated at one given kinematic point and radiation gauge in the lab frame with $V=P$. The gauge dependence is in general not significant, though it is more relevant in the case of interference cross-section.}
\end{figure}

Note that the gauge fixing condition $V \cdot \varepsilon(q')=0$ actually requires that $V \cdot q' \not= 0$. Otherwise, the condition $V \cdot \varepsilon(q')=0$ may never be satisfied under the gauge transformation as $V \cdot \left(\varepsilon(q')+\lambda q'\right) = V \cdot \varepsilon(q')$, leaving a pole at $V\cdot q'$ in the polarization sum in Eq. (\ref{eq:completeness2}), which should be avoided when choosing the gauge. For arbitrary gauge choice, the $\mathcal G^{\mu\nu}$ will reduce to $-g^{\mu\nu}$ if the current conservation condition is exactly satisfied. Therefore, we define
\begin{equation}
    \mathcal G^{\mu\nu}_{(0)}=-g^{\mu\nu}\ , \qquad \mathcal G^{\mu\nu}_{(1)}=\frac{V^\mu q'^\nu+V^\nu q'^\mu}{V\cdot q'}-\frac{V^2 q'^\mu q'^\nu}{(V\cdot q')^2}\ ,
\end{equation}
such that $\mathcal G^{\mu\nu}=\mathcal G^{\mu\nu}_{(0)}+\mathcal G^{\mu\nu}_{(1)}$. Then the gauge dependence is all contained $\mathcal G^{\mu\nu}_{(1)}$ and when the Compton tensor is exactly conserved, it can be set to zero. Since both the pure DVCS and interference cross-sections are linear in $\mathcal G^{\mu\nu}$, we can use the ratio 
\begin{equation}
    r_{\rm{DVCS/INT}}=\frac{\text{d} \sigma_{\rm{DVCS/INT}}\left(\mathcal G^{\mu\nu}_{(1)}\right)}{\text{d} \sigma_{\rm{DVCS/INT}}\left(\mathcal G^{\mu\nu}\right)}\ ,
\end{equation}
to estimate the dependence of cross-sections on the gauge choices, and the numerical results with the radiation gauge in the lab frame, corresponding to $V=P$, are shown in Fig. \ref{fig:gaugedep}.

As we can see from the plot, the gauge dependence is indeed a higher order effect. Even for $Q^2=1.82 \text{ GeV}^2$, the ratio is already small enough. Therefore, the gauge dependence of the cross-section will not severely affect the results, though they still add to the intrinsic uncertainties in the theoretical prediction of the cross-section. Some numerical studies on the gauge dependence of the cross-section by choosing different $V$ as $P$, $P'$ and $q$ respectively show that they give almost the same cross-section with difference at the order of $0.1\%$, and thus the gauge dependence is indeed a less relevant effect than the dependence on the choice of light cone vectors.

It is also worth noting that the gauge dependence will be more significant if the leading approximation of Compton tensor coefficients $\mathscr{T}^{\mu\nu}_{(2)}\approx-\frac{1}{2}g_\perp^{\mu\nu},\  \widetilde{\mathscr{T}}^{\mu\nu}_{(2)}\approx\frac{i}{2} \epsilon^{\mu\nu}_\perp$ is used. The reason is simple - if we have the Compton tensor coefficients to all order, it will be exactly conserved, and the contribution of those gauge-dependent terms vanishes. This also indicates that the gauge dependence shown in Fig. \ref{fig:gaugedep} can only be eliminated by calculating the Compton tensor coefficients to a higher order, rather than choosing a specific gauge.

\section{Operator Production Expansion and Covariant Coefficients}
\label{sec:WWrelation}
In previous sections, we show how the cross-section formulas depend on the choice of light cone vectors in the leading twist GPD approximation, where all the higher twist GPDs are set to be zero. However, the light cone dependence persists at twist three even when the Compton tensor coefficients are accurate to twist three, indicating the existence of some other kinematical twist-three correction from higher twist GPDs. These relations between GPDs of different twist are the results of Lorentz symmetry and the equations of motion, which are generally known as the Wandzura-Wilczek (WW) relations, stating that the higher-twist GPDs can be split into the WW parts that are expressible in terms of the lower-twist GPDs and the genuine twist-three parts that are related to higher-twist quark-gluon-quark operators \cite{Belitsky:2000vx,Belitsky:2005qn,Kivel:2000rb,Radyushkin:2000jy,Aslan:2018zzk}. In this section, we show that it is necessary to include the kinematical corrections from twist-three GPDs in order for the dependence on light cone vectors to cancel at twist three.

\subsection{Operator production expansion and covariant Compton tensor coefficients}

The best method to study the twist expansion without light cone vectors is the Operator Production Expansion (OPE), which states that a non-local operator in $z^\mu$ can be expanded in terms of a series of local operators when 
$z^2\ll 1/\Lambda^2_{\rm QCD}$
\begin{align}
    \mathcal {J}_1(z) \mathcal {J}_2(0) =\sum_{i} C_{i}(z^2) \mathcal {O}_i(0)
\end{align}
where $\mathcal {O}_i(0)$ are a set of local operators and the corresponding coefficients $C_{i}(z^2)$ are their Wilson coefficients calculable in perturbative QCD in
strong coupling $\alpha_s$ expansion. As the separation of the non-local operator $z$ approaches light cone as $z^2 \to 0$, local operators in the lowest power of $z$ will dominate, which are defined as the leading twist operators. The twist defined in this case differs from the dynamical twist in light cone power counting, and is known as the geometric twist, which is defined through
\begin{align}
    t\equiv d-s\ ,
\end{align}
with $d$ the mass dimension of the operator and $s$ the Lorentz spin.  For instance, a rank-2 tensor $O^{\mu\nu}$ can in general be decomposed into symmetric traceless, antisymmetric and traceful parts as
\begin{align}
    O^{\mu\nu}= \left(O^{\left(\mu\nu\right)}-\frac{1}{4}g^{\mu\nu}\text{Tr}(O)\right) +O^{[\mu\nu]}+\frac{1}{4}g^{\mu\nu} \text{Tr}(O)
\end{align}
and each of them will be of spin $2$, $1$ and $0$. Consequently, if the operator has a mass dimension 4, the three parts will be twist-two, twist-three and twist-four respectively. Therefore, equations of a general tensor operator will contain operators of different twist, making it possible to express part of the higher twist operators in terms of the lower twist operators through equations of motion. These relations for the twist-three GPDs in DVCS process has been studied and derived in the literatures~\cite{Belitsky:2000vx,Belitsky:2005qn,Kivel:2000rb,Radyushkin:2000jy,Aslan:2018zzk}, which can be written as 
\begin{align}
    W^{\left[\gamma^\mu\right]}(x,\xi,t)&= \frac{\Delta^\mu}{-2\xi} n_\nu W^{\left[\gamma^\nu\right]}(x,\xi,t) +\cdots\ ,\\
    W^{\left[\gamma^\mu\gamma^5\right]}(x,\xi,t)&= \frac{\Delta^\mu}{-2\xi} n_\nu W^{\left[\gamma^\nu\gamma^5\right]}(x,\xi,t) +\cdots\ ,
\end{align}
where a general tensor is projected out with the twist-two part, and the ellipsis stands for twist-two GPDs convoluted with WW kernel and genuine twist-three GPDs. It is worth noting that terms with twist-two GPDs convoluted with WW kernel are in general comparable to the twist-two GPDs themselves, but they are dropped here because they lead to a different set of CFFs which do not mix with the leading twist CFFs we care about in this paper. Therefore, it is sufficient to study the terms above for our purpose, and a more thorough study of the twist-three CFFs is left to future works. While the above relation leads to the trivial identity $n_\mu W^{\left[\gamma^\mu\right]}(x,\xi,t)=n_\nu W^{\left[\gamma^\nu\right]}(x,\xi,t)$ if we project out the leading twist contribution, it also involves a twist-three part in the form of $ W^{\left[\gamma^\mu_\perp \right]}(x,\xi,t)= \Delta^\mu_\perp/(-2\xi) n_\nu W^{\left[\gamma^\nu\right]}(x,\xi,t)+\cdots$. This kinematical twist-three term will modify our Compton tensor coefficients and cancel the light cone dependence at twist three, as we will see.

In order to get the modified Compton tensor coefficients, we need to redo the calculation with the above substitution of the leading twist GPDs.
The result is surprisingly simple and leads to
a modified Compton tensor coefficients as (compared with Eq. \ref{eq:LFcompton}),
\begin{align}
\begin{split}
\mathscr {T}^{\mu\nu}_{(2)} &=-\frac{1}{2}\left[ g^{\mu\nu}-\frac{n^\mu\Delta^\nu +n^\mu \Delta^\nu}{n\cdot \Delta} \right]\ ,\\
\widetilde{\mathscr {T}}^{\mu\nu}_{(2)} &=-\frac{i}{2(n\cdot \Delta)}\epsilon^{\mu\nu n \Delta}\ .
\end{split}
\end{align}
The above expressions reduce to $-\frac{1}{2}g_\perp^{\mu\nu}$ and $\frac{i}{2}\epsilon_{\perp}^{\mu\nu}$ respectively if we choose the light cone vectors such that $\Delta_\perp^\mu=0$, which corresponds to the BMP choice \cite{Braun:2014sta} that $\beta \to \infty$. This indicates the kinematical correction to the leading twist CFFs vanishes at twist-three if we choose the light cone  vectors such that $\Delta_\perp^\mu =0$. On the other hand, the above Compton tensor coefficients can also be promoted with the physical vector $q,q'$ up to twist-four terms as,
\begin{align}
\begin{split}
\label{eq:covCompton}
\mathscr {T}^{\mu\nu}_{(2),\rm{C}} &=-\frac{1}{2}\left[ g^{\mu\nu}-\frac{q^\mu q'^\nu+q^\nu q'^\mu}{q\cdot q'}+\frac{q'^\mu q'^\nu q^2}{(q\cdot q')} \right]\ ,\\
\widetilde{\mathscr {T}}^{\mu\nu}_{(2),\rm{C}} &=\frac{i}{2(q\cdot q')}\epsilon^{\mu\nu q q'}\ ,
\end{split}
\end{align}
which are given in \cite{Braun:2014sta} as well. These promoted Compton tensor coefficients are covariant and apply to a general choice of light cone vectors. Besides, they are also manifestly gauge-invariant, which frees the scattering amplitude from any specific gauge fixing conditions.

\subsection{Comparison of the cross-section formulas with covariant coefficients}

Now we are in a position to go back to the cross-section formulas, and compare different formulas with the kinematical corrections above. Our results above indicates two ways to calculate the cross-section: either to take $\beta\to\infty$ with the light-cone Compton tensor coefficients in Eq. (\ref{eq:LFcompton}) such that all kinematical twist-three corrections vanish, or to use the covariant Compton tensor coefficients defined in Eq. (\ref{eq:covCompton}). While the former approach provides a simple connection to our previous results, it is restrained to one specific choice of light cone vectors and will not be general enough to study the light cone dependence of the cross-section formulas. Therefore, we focus on the latter approach, for which we use the covariant Compton tensor coefficients and study the remnant light cone dependence in the cross-section formulas.

\begin{figure}[ht]
\centering
\begin{minipage}[b]{\textwidth}
\includegraphics[width=0.48\textwidth]{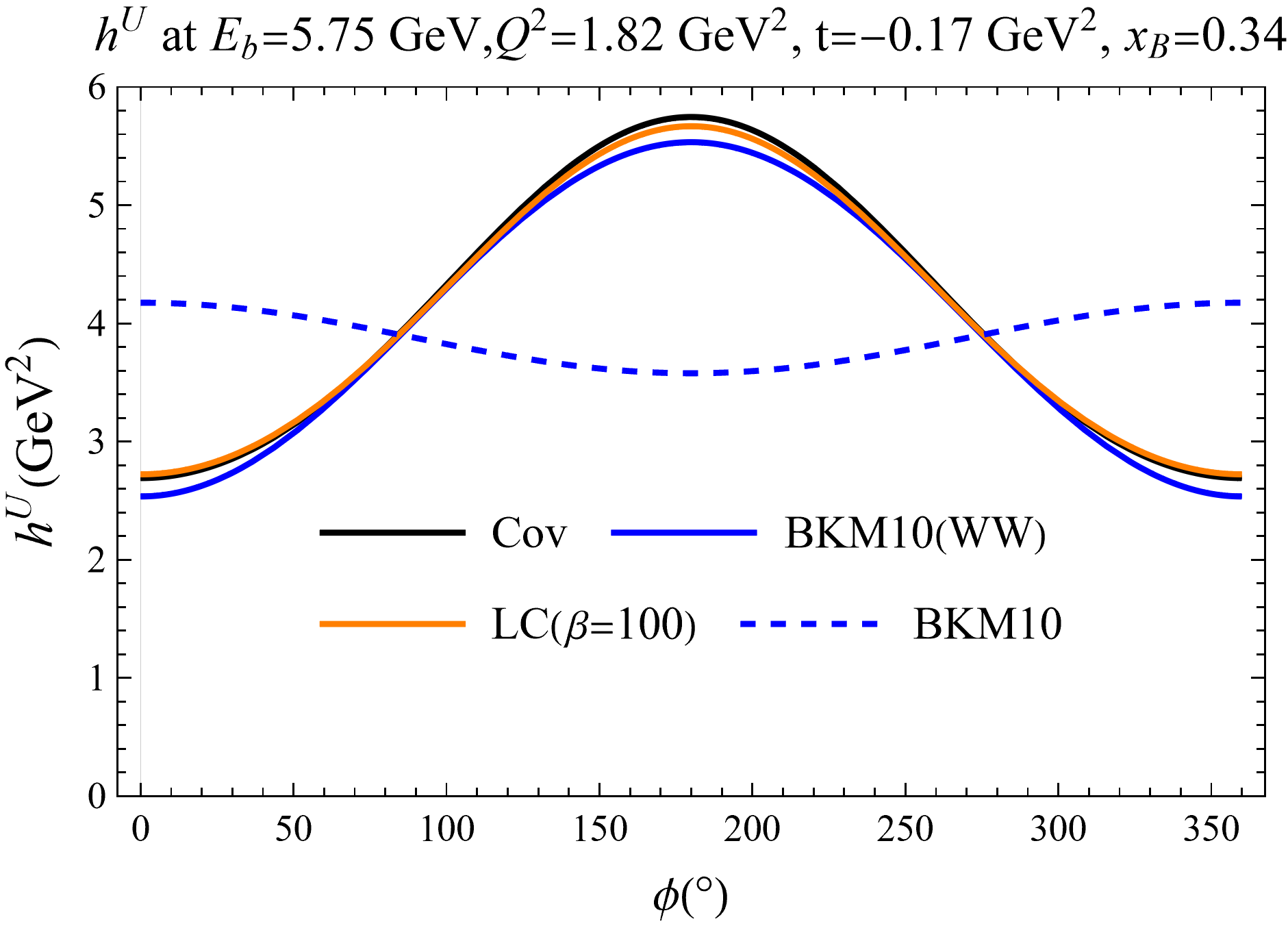}
\includegraphics[width=0.5\textwidth]{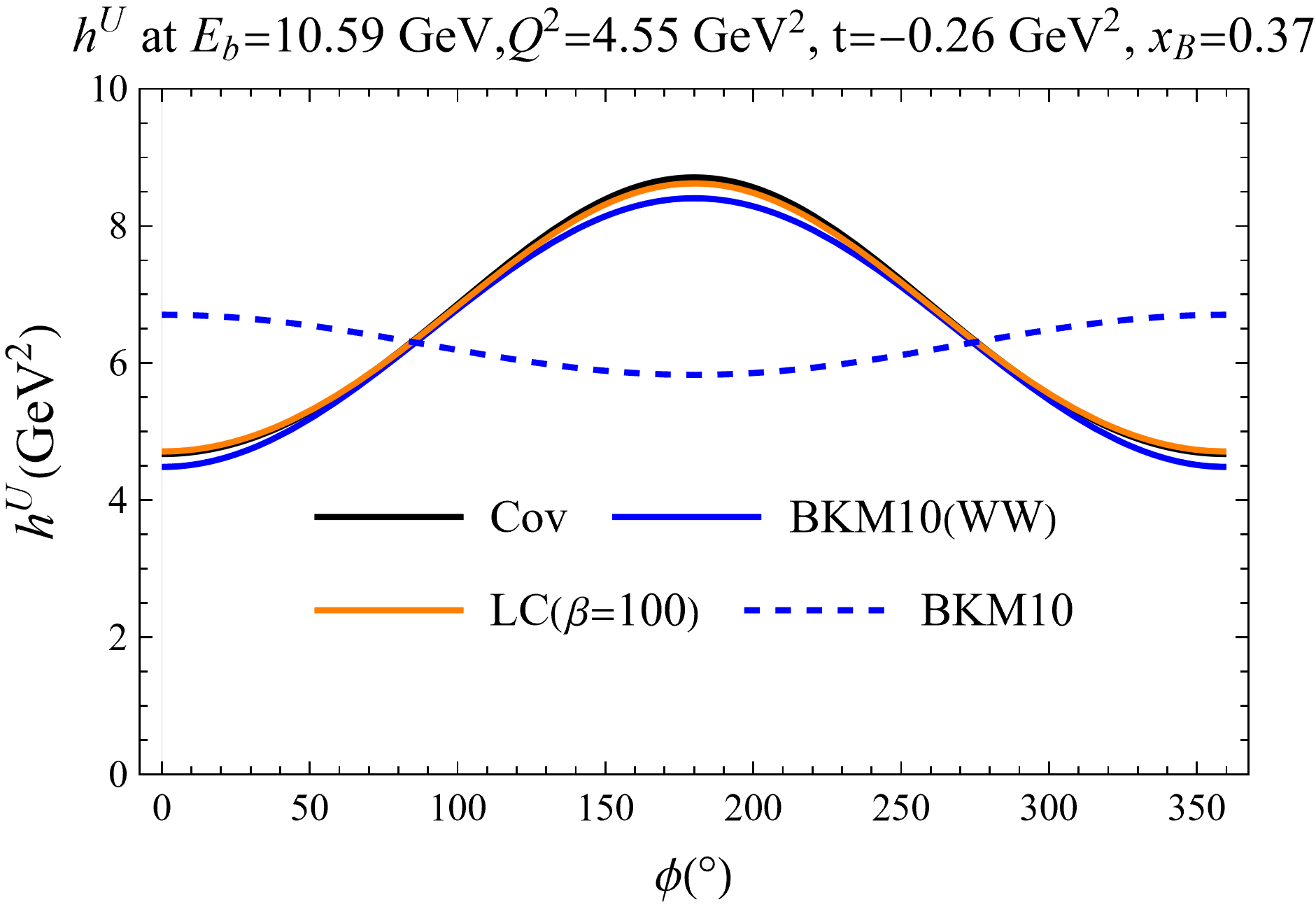}
\end{minipage}
\begin{minipage}[b]{\textwidth}
\includegraphics[width=0.49\textwidth]{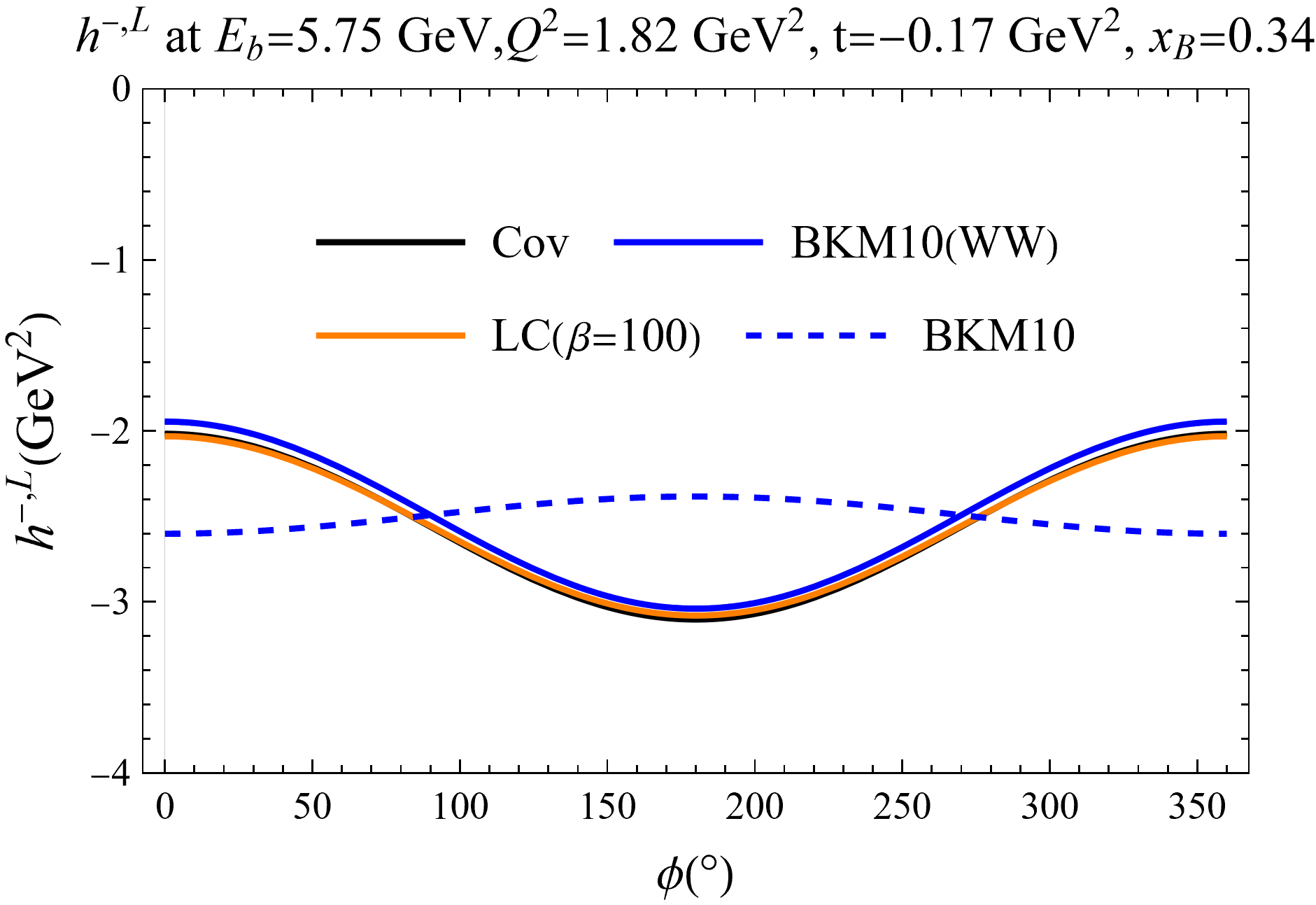}
\includegraphics[width=0.5\textwidth]{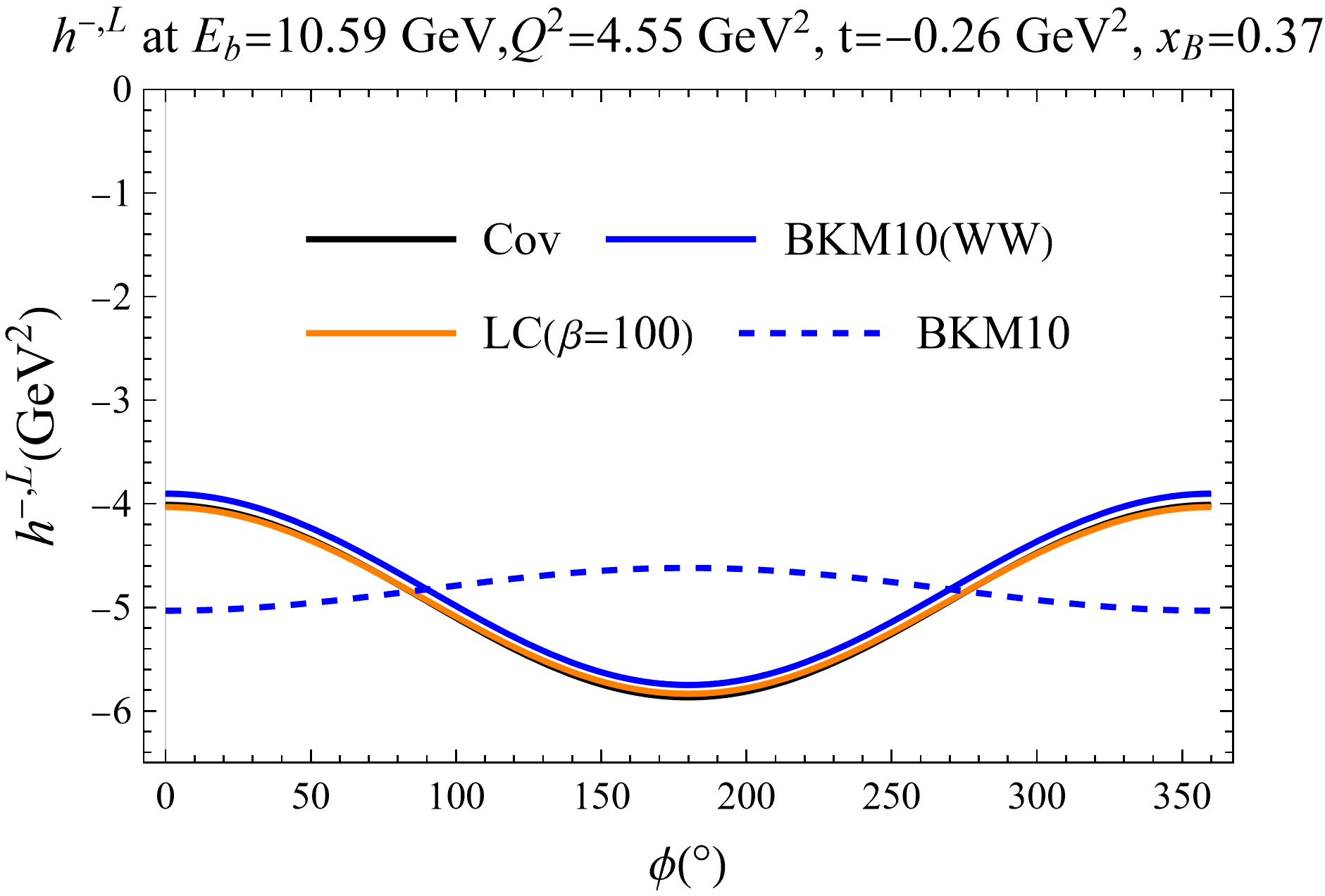}
\end{minipage}
\caption{\label{fig:hcompc} A comparison of the coefficients $h^{\rm{U}}$ and $h^{-,\rm{L}}$ over different formulas. In the plots, we can see that the covariant coefficients above agrees well with the BKM10 results with WW relations and also our light cone calculation with $\beta\to\infty$, while they differ a lot from the result without WW relation. Note that in the $\beta \to \infty$ limit, the value $\alpha$ is actually degenerate and irrelevant.}
\end{figure}

Recall that in previous sections, we defined 4 scalar coefficients $h^{\rm{U}},\tilde{h}^{\rm{U}}, h^{-,\rm{L}}$ and $ h^{+,\rm{U}}$ for DVCS amplitude and 6 complex (or 12 real) scalar coefficients $A^I, B^I,C^I,\tilde A^I,\tilde B^I$ and $\tilde C^I$. A comparison of those coefficients would suffice in order to compare different formulas. For the pure DVCS cross-section, it turns out that with the covariant Compton tensor coefficients, those coefficients have a compact form that is independent of the light cone vectors, for which we have 
\begin{align}
\label{eq:covh}
\begin{split}
h^{\rm{U}}=\tilde{h}^{\rm{U}}&=\frac{2(k\cdot q)(k\cdot q')}{q\cdot q'}-\frac{q^2(k\cdot q')^2}{(q\cdot q')^2}-k\cdot q\ ,\\
h^{-,\rm{L}}&=\frac{-q^2(k\cdot q')}{q\cdot q'}+k\cdot q\  ,\\
h^{+,\rm{U}}&=0\ .
\end{split}
\end{align}
Using those coefficients, we compare different pure DVCS cross-section formulas numerically, as shown in Fig. \ref{fig:hcompc}. 
Apparently, they confirm our observation that the general covariant coefficients agree with the light cone results with $\beta\to \infty$ as well as the BKM10 results with the WW relation, whereas their differences to BKM10 results without WW relation which are twist-three effects are shown to be quite significant.  Note that in the $\beta \to \infty$ limit, the value $\alpha$ is actually degenerate and irrelevant. Therefore, we suppress the choice of $\alpha$ in the limit $\beta\to\infty$ hereafter.

The comparison of the interference cross-section formulas can be done similarly, except that we have more coefficients in the case of interference. The expressions of those scalar coefficients are collected in Appendix \ref{app:structurefunc}, and here we focus on comparing them numerically.

\begin{figure}[ht]
\centering
\begin{minipage}[b]{\textwidth}
\centering
\begin{minipage}[b]{\textwidth}
\includegraphics[width=0.49\textwidth]{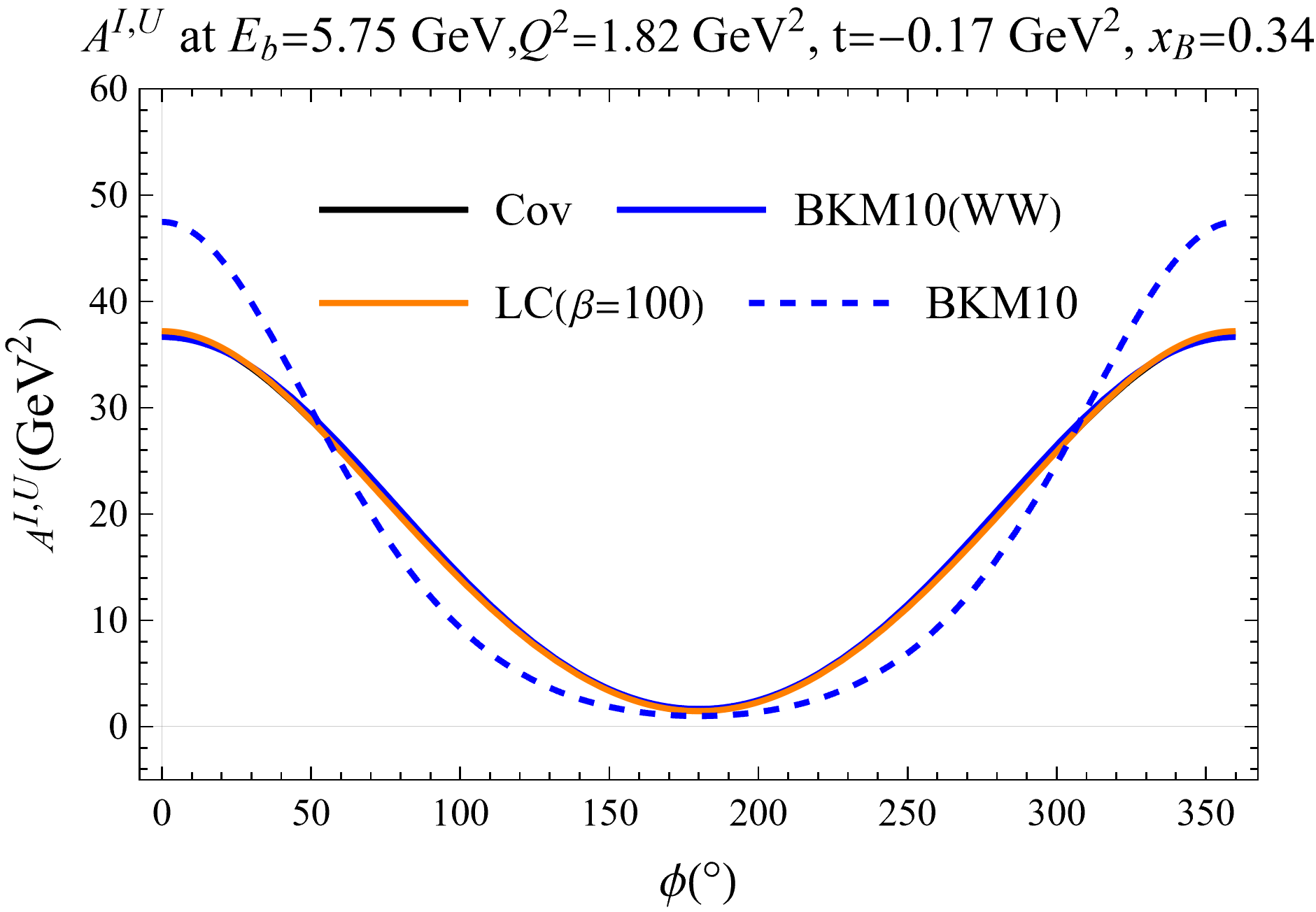}
\includegraphics[width=0.5\textwidth]{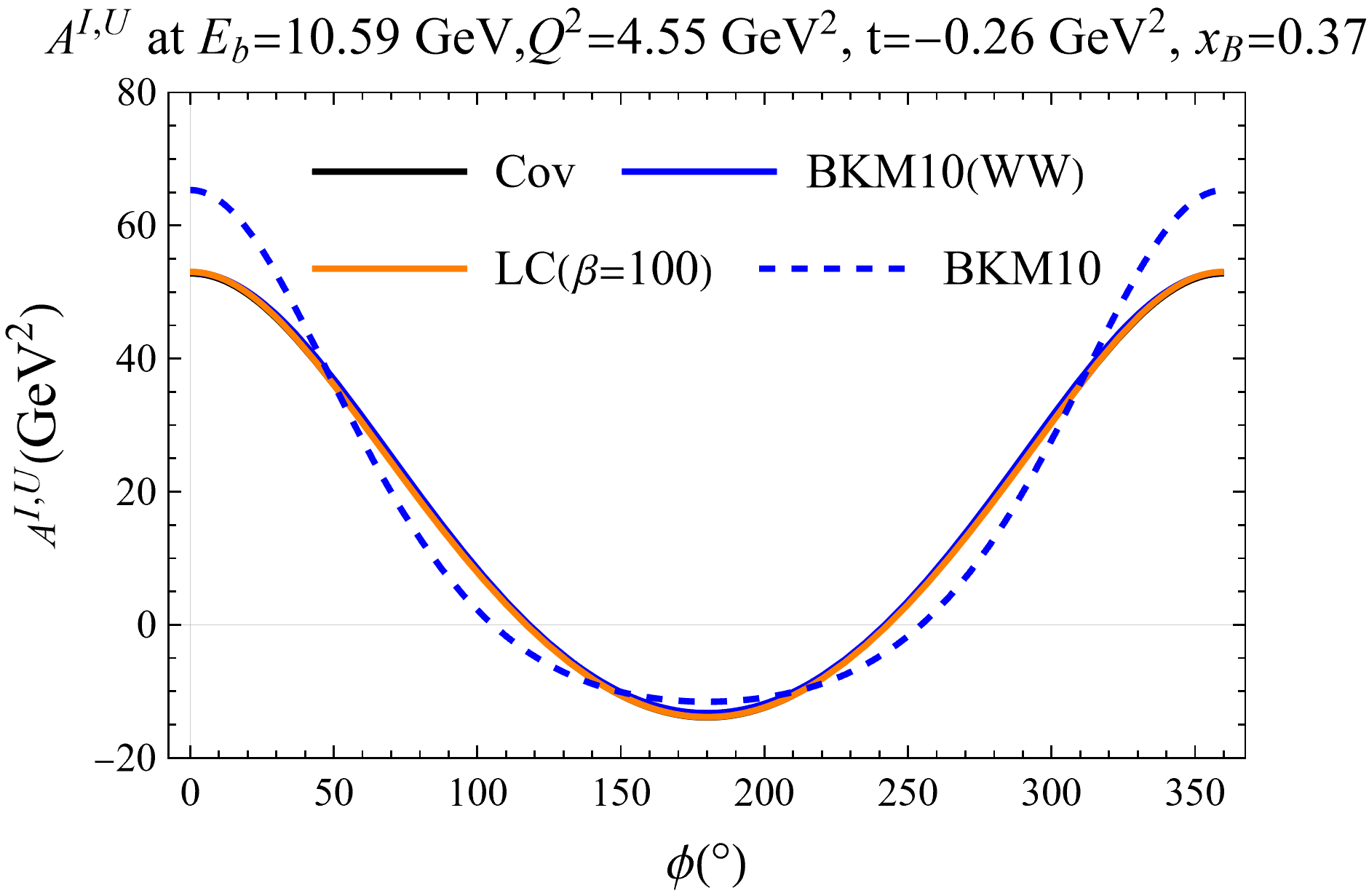}
\end{minipage}
\begin{minipage}[b]{\textwidth}
\includegraphics[width=0.49\textwidth]{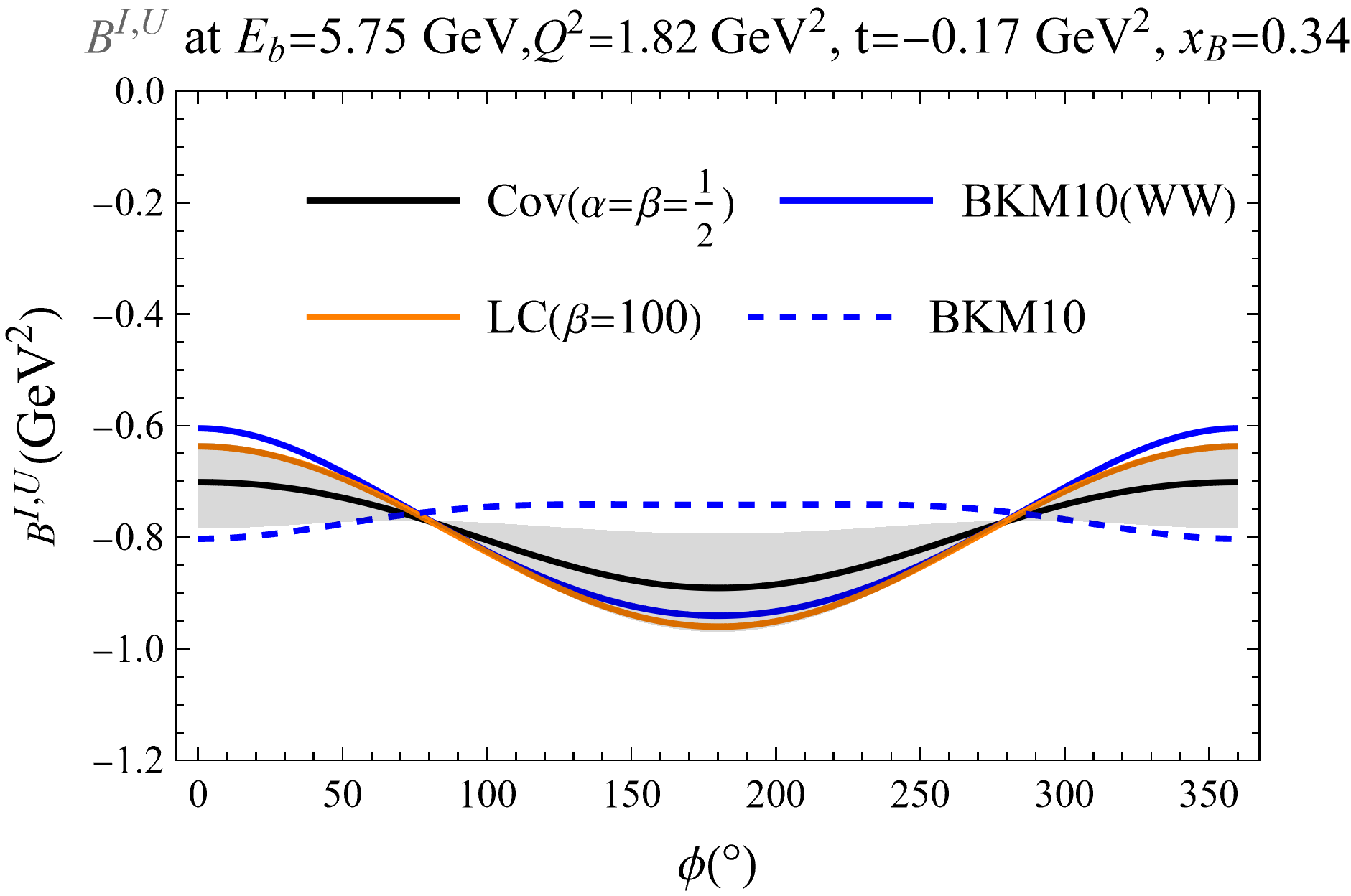}
\includegraphics[width=0.5\textwidth]{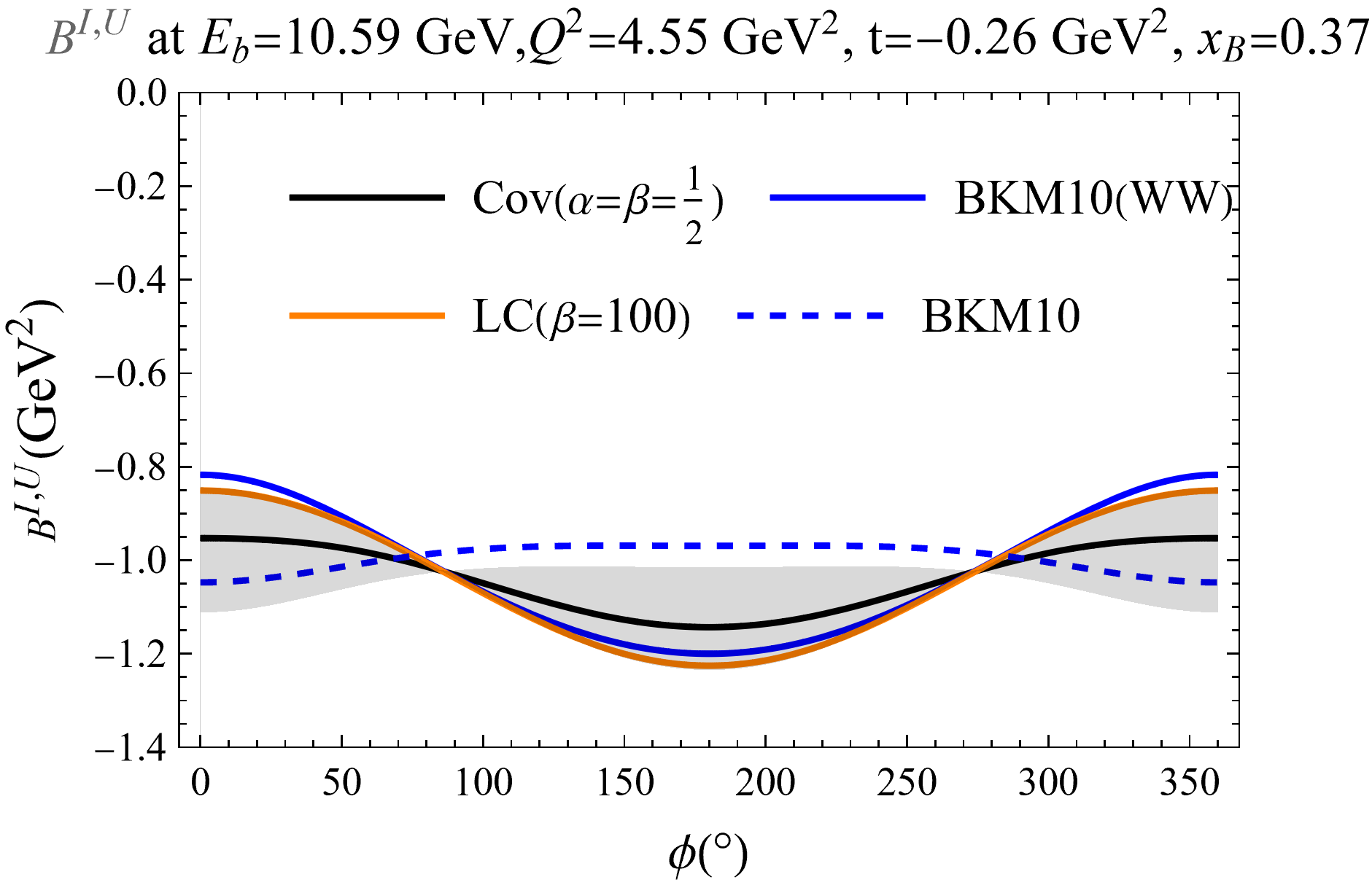}
\end{minipage}
\begin{minipage}[b]{\textwidth}
\includegraphics[width=0.49\textwidth]{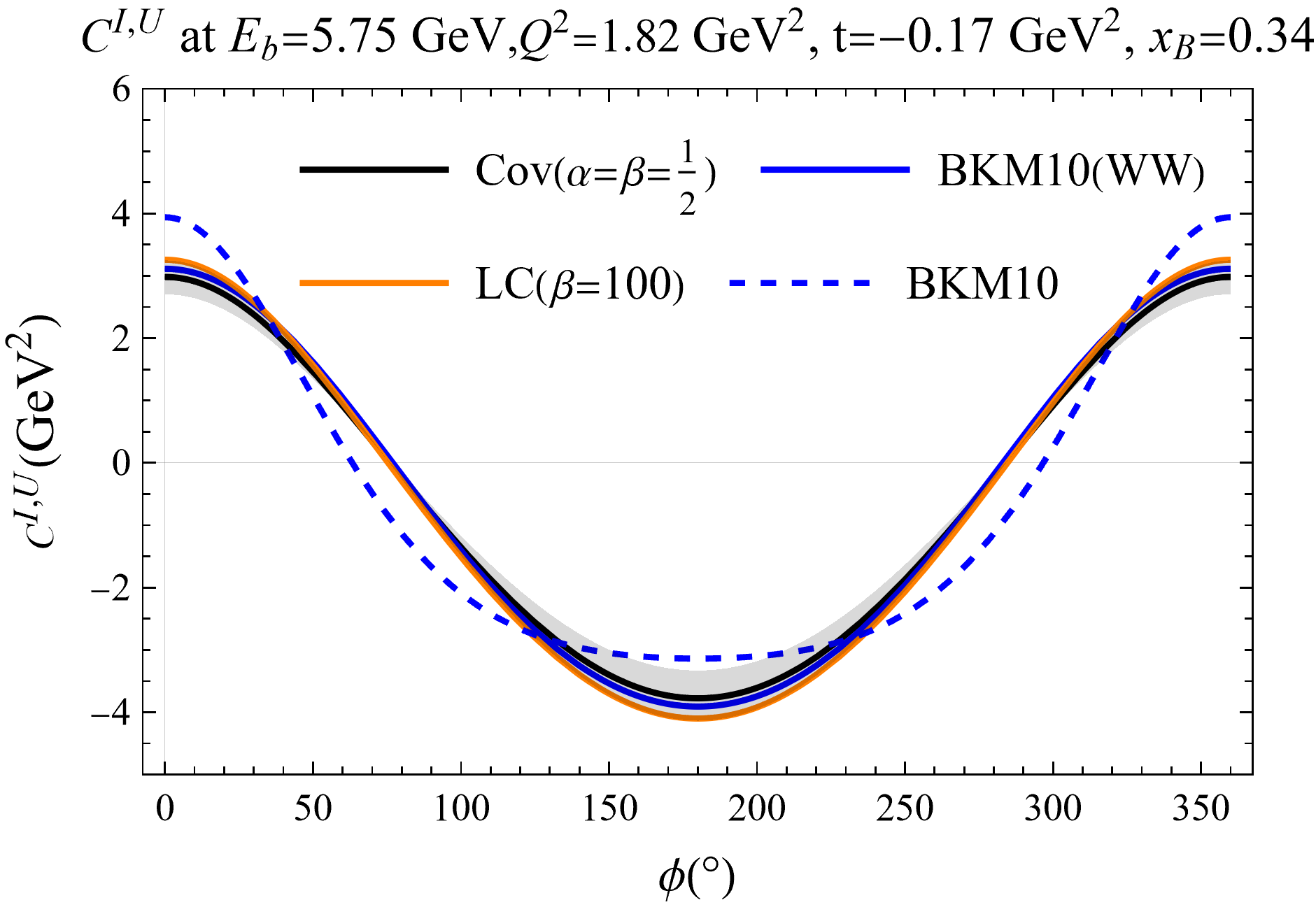}
\includegraphics[width=0.5\textwidth]{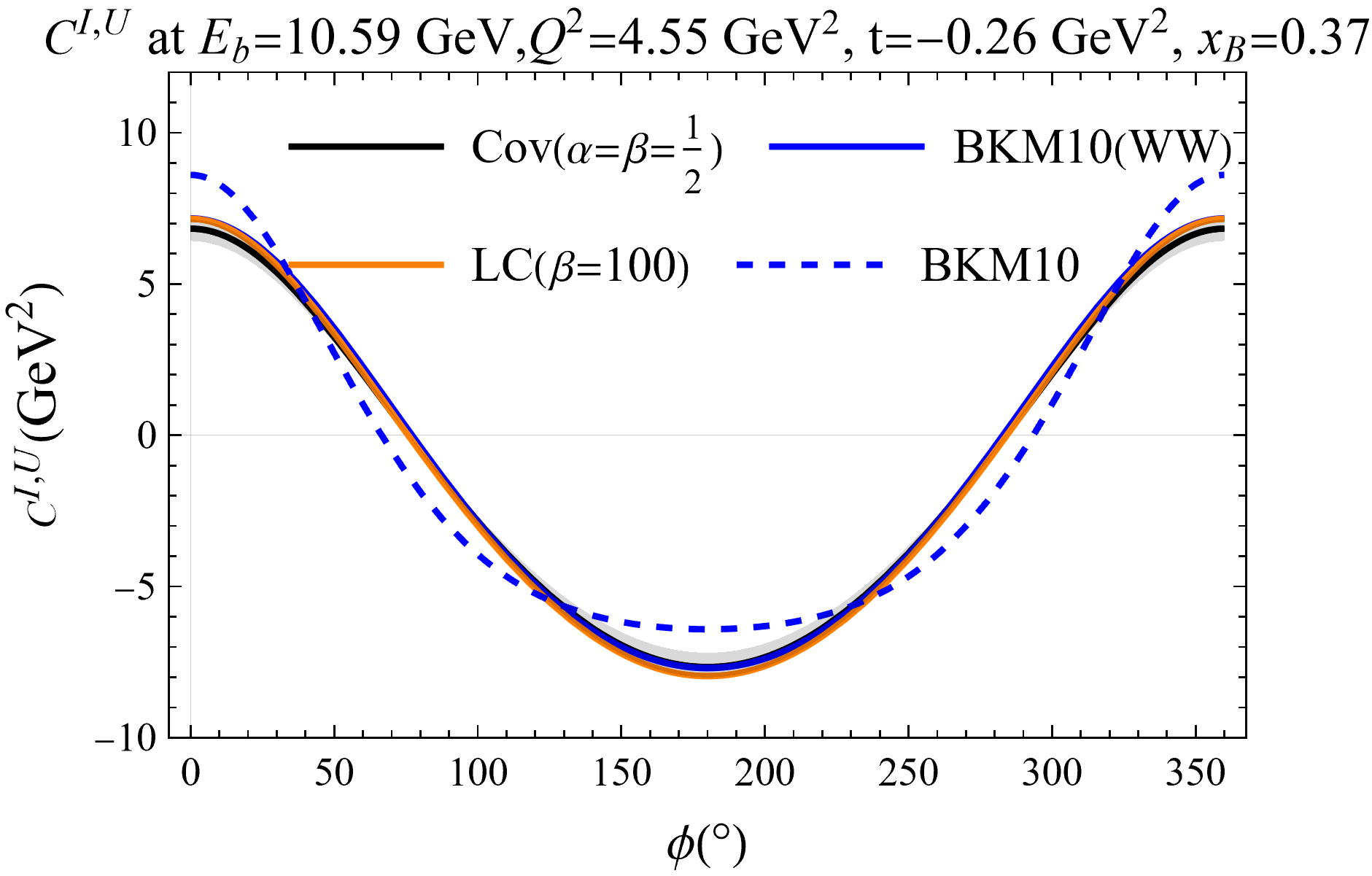}
\end{minipage}
\end{minipage}
\caption{\label{fig:intcompc} A comparison of the coefficients $A^{I,U}$, $B^{I,U}$ and $C^{I,U}$ over different formulas. The coefficients $B^{I,U}$ and $C^{I,U}$ depend on the light cone vectors explicitly, and thus we vary the $\alpha$ and $\beta$ from $0\le\alpha\le 1$ and $0\le\beta\le100$ to get the gray band in the plot. The blue lines in the plot agree well with each other. }
\end{figure}

As shown in Fig. \ref{fig:intcompc}, the covariant coefficients agree well with the BKM10 formula with WW relation and they both agree well with the previous light cone result at $\beta\to \infty$, confirming that the kinematical twist-three corrections vanish when we choose the light cone vectors such that $\Delta_\perp =0$. However, the coefficients $B^{I}$ and $C^{I}$ still depend on the light cone vectors, which is shown in the plot as the gray band for which we vary the $\alpha$ and $\beta$ from $0\le\alpha\le 1$ and $0\le\beta\le100$. We emphasize that after using the covariant Compton tensor coefficients, the remnant light cone dependence is twist-four suppressed, so it will be less significant at large $Q^2$, as indicated by the plots on the right of Fig. \ref{fig:intcompc} which has $Q^2=4.55 \text{GeV}^2$, much larger than the left plots with $Q^2=1.82\text{GeV}^2$.
 
\subsection{Light cone dependence of cross-sections with WW relations}

\begin{figure}[t]
\centering
\begin{minipage}[b]{0.48\textwidth}
\includegraphics[width=\textwidth]{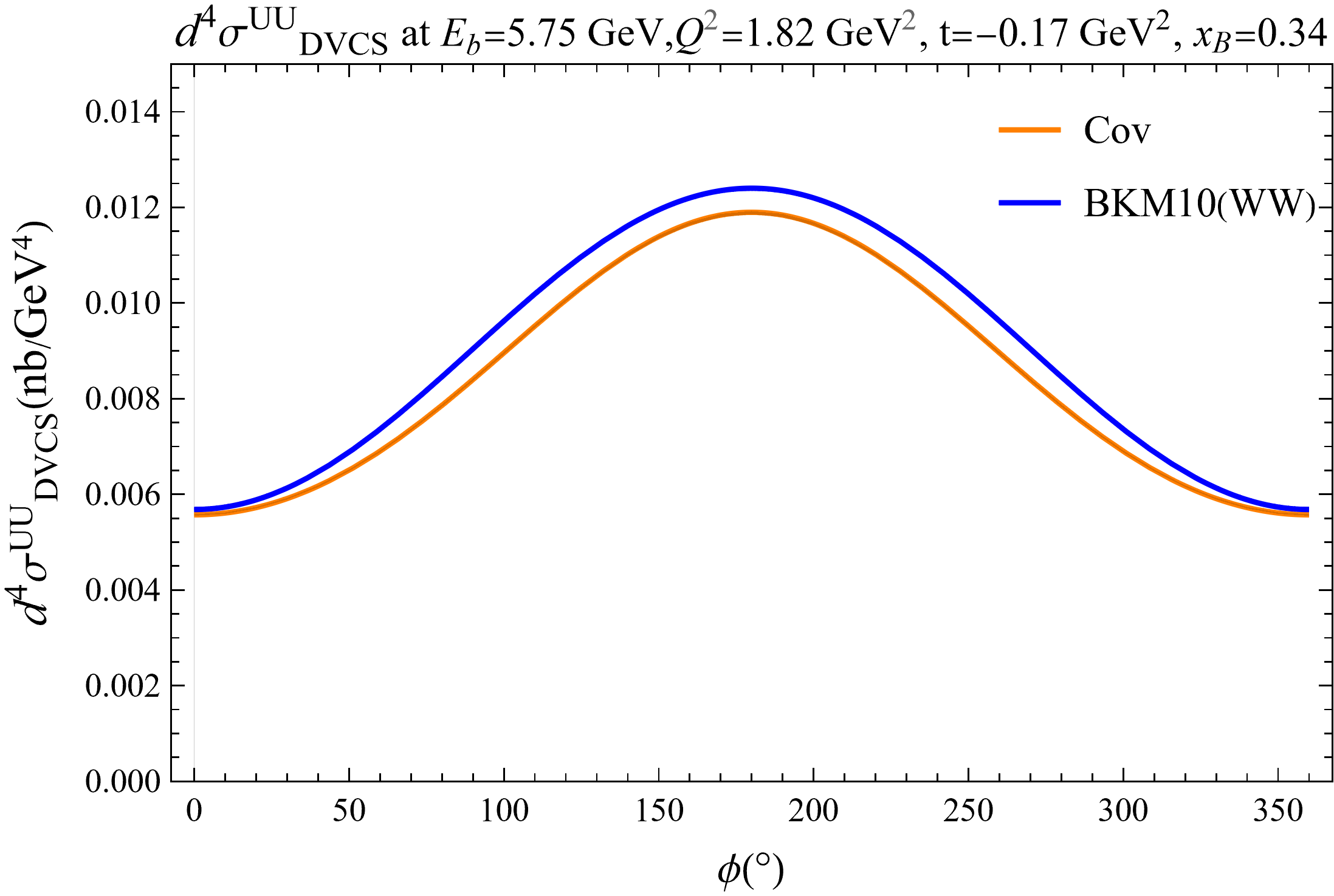}
\end{minipage}
\begin{minipage}[b]{0.5\textwidth}
\includegraphics[width=\textwidth]{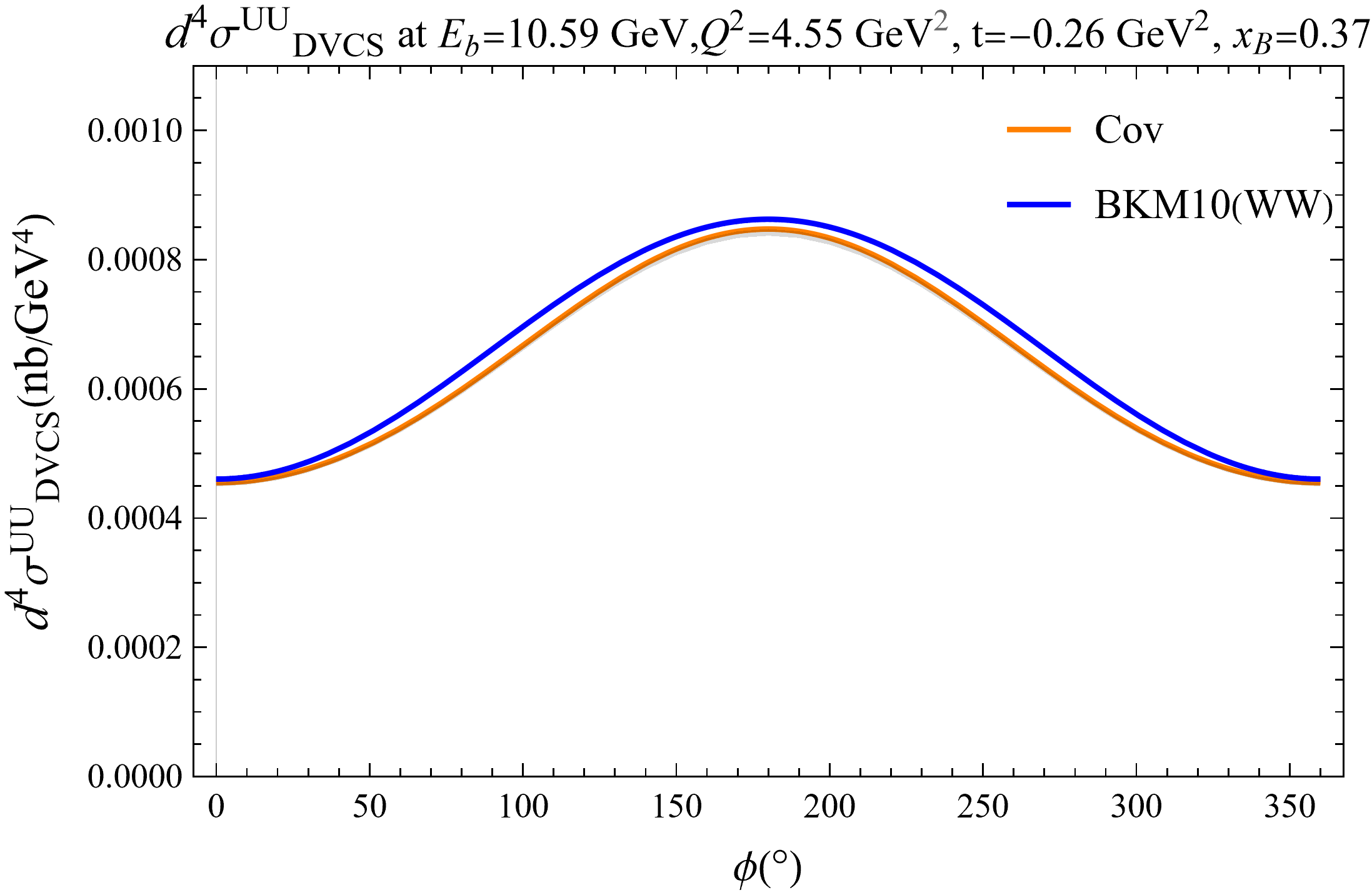}
\end{minipage}
\caption{\label{fig:DVCSxsectionc} Comparison of the unpolarized four-fold pure DVCS cross-section $\text{d}^4 \sigma^{UU}_{\rm{DVCS}}$. Our pure DVCS cross-sections almost have no light cone dependence, while they differ a bit from the BKM10 results with WW relations.}
\end{figure}

With those scalar coefficients compared, we can now move on to compare the cross-section formulas with our covariant coefficients and find out how the cross-section depends on the choice of light cone vectors. In Fig. \ref{fig:DVCSxsectionc}, we compare the pure DVCS cross-section numerically. It turns out that our covariant results for the pure DVCS cross-sections has almost no light cone dependence. We get almost identical results by varying $\alpha$ and $\beta$ in the region $0\le\alpha\le 1,0\le \beta\le 100$. Therefore, we suppress the $\alpha$ and $\beta$ when presenting the pure DVCS cross-sections. On the other hand, our results still differ from the BKM10 results with a twist-four effect. As we discussed in the previous section, the main reason is that BKM10 used the effective light cone vectors that are not on the light cone, leading to the differences at twist-four level.

\begin{figure}[t]
\centering
\begin{minipage}[b]{\textwidth}
\includegraphics[width=0.5\textwidth]{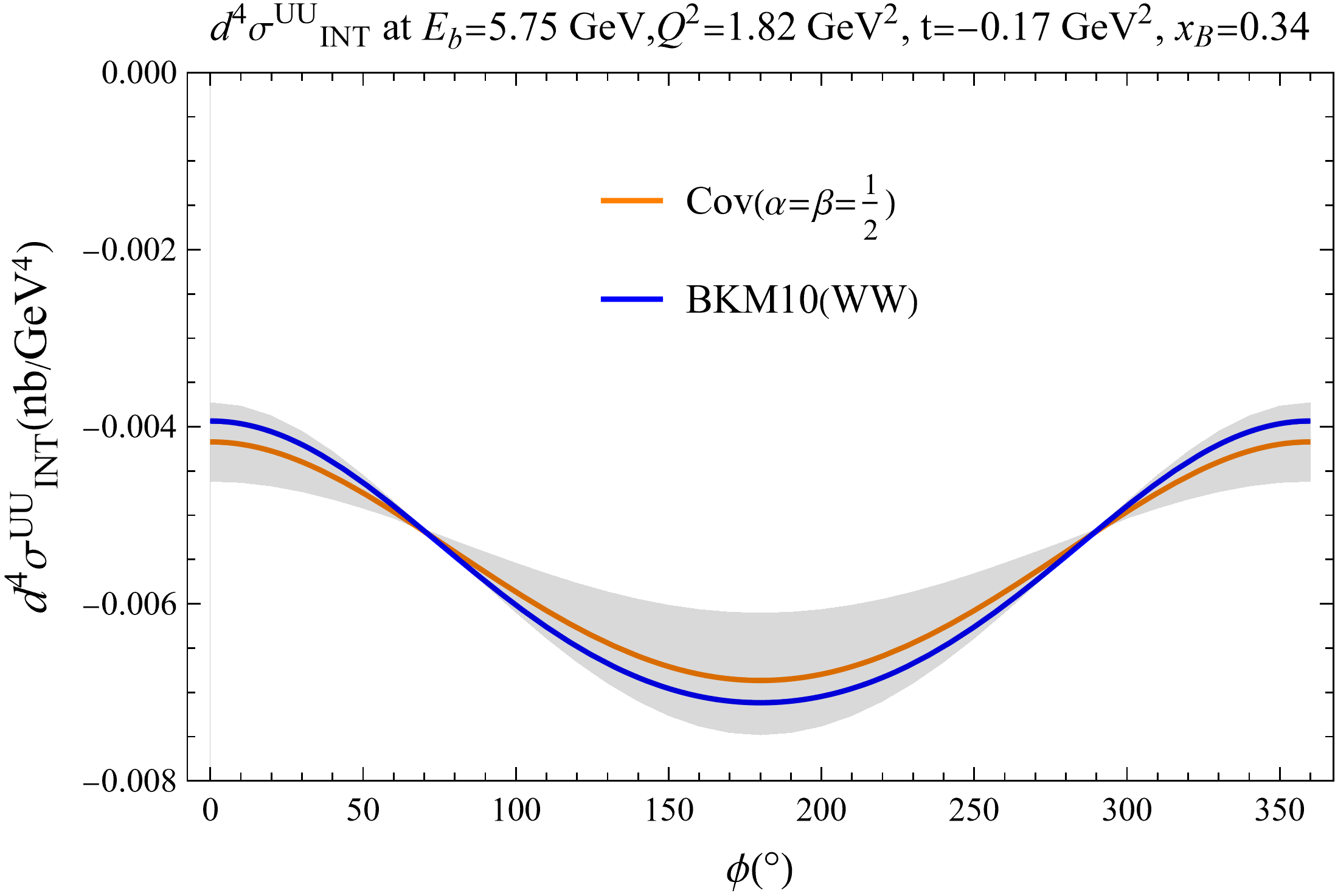}
\includegraphics[width=0.5\textwidth]{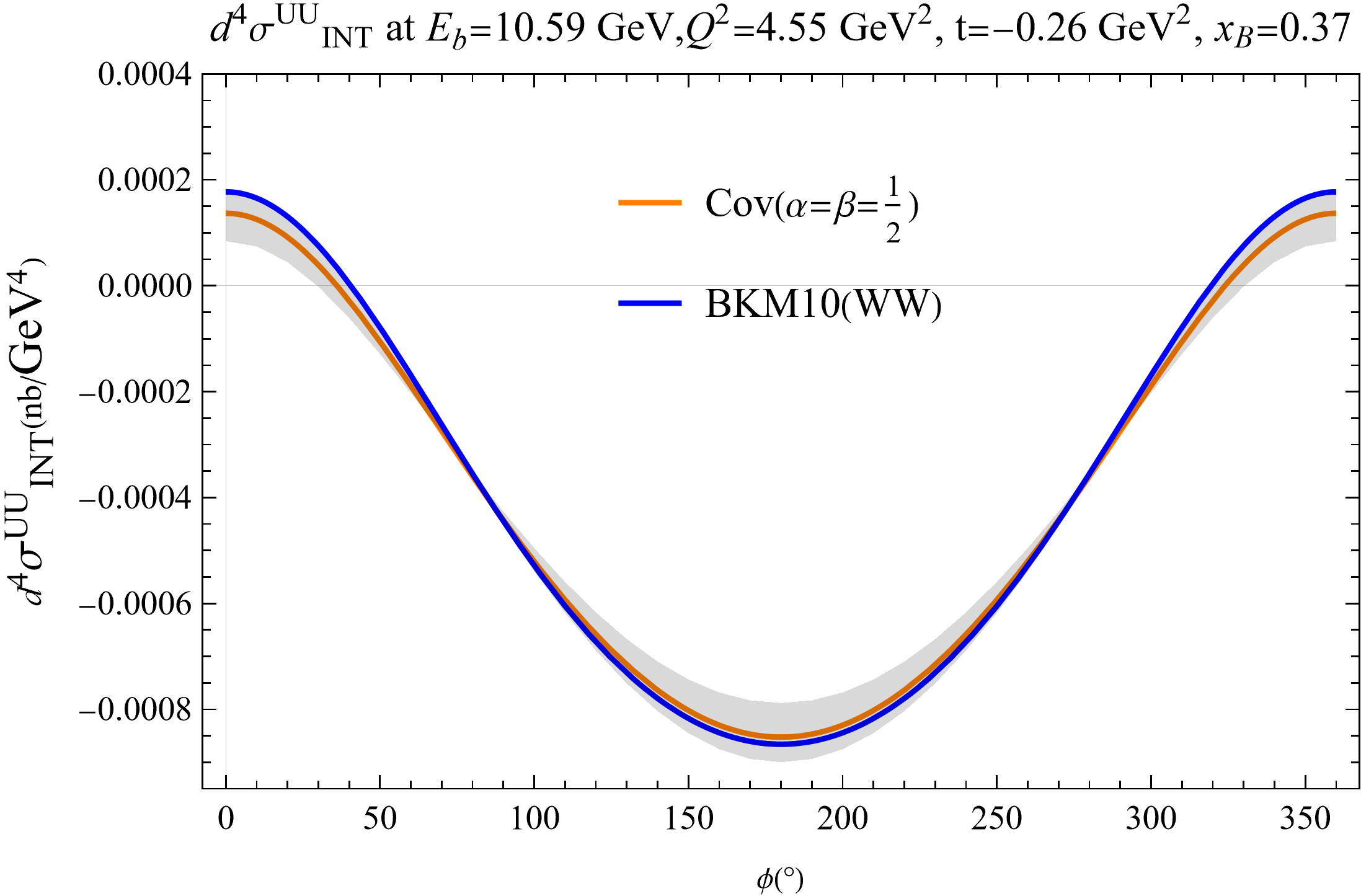}
\end{minipage}
\begin{minipage}[b]{\textwidth}
\includegraphics[width=0.5\textwidth]{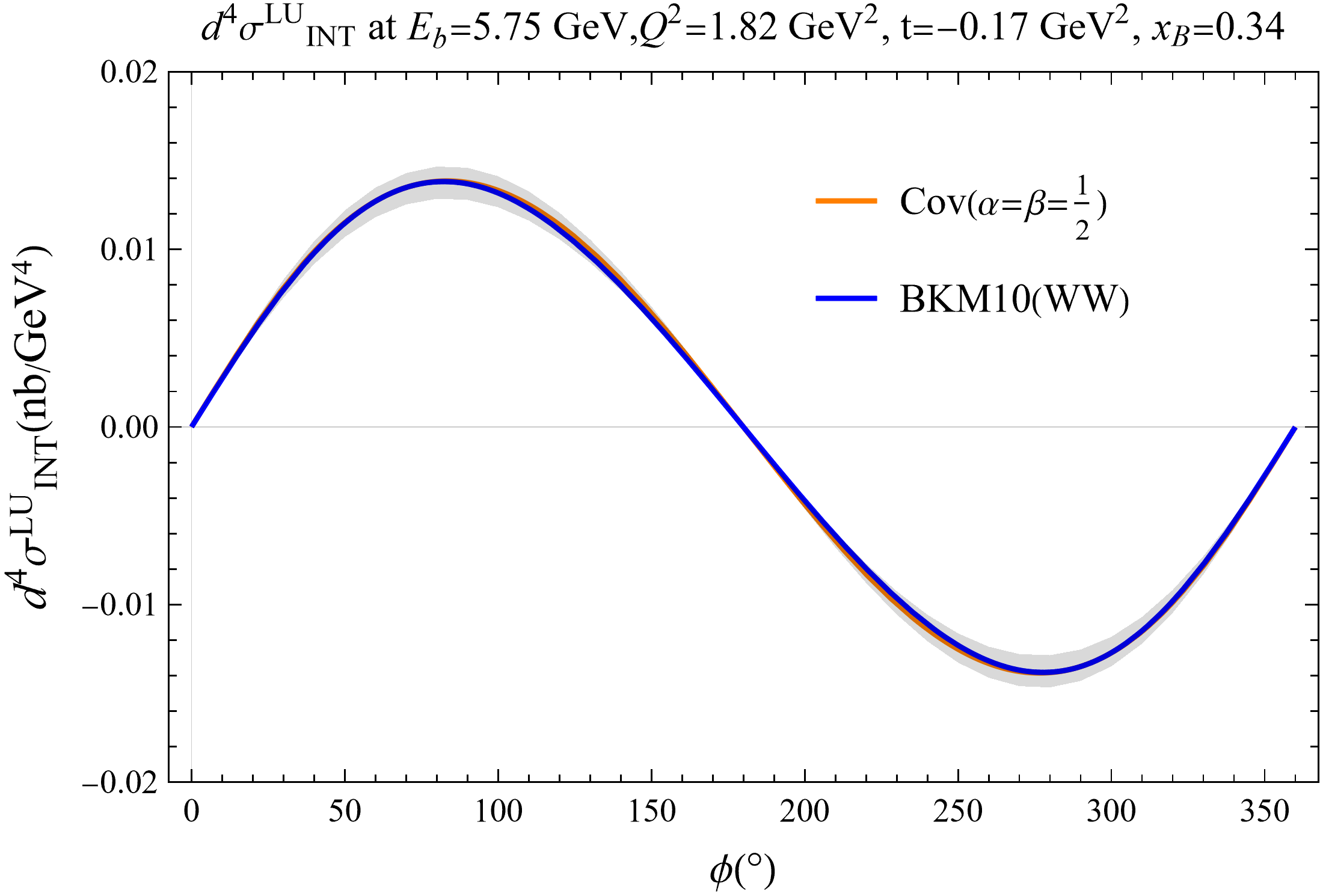}
\includegraphics[width=0.5\textwidth]{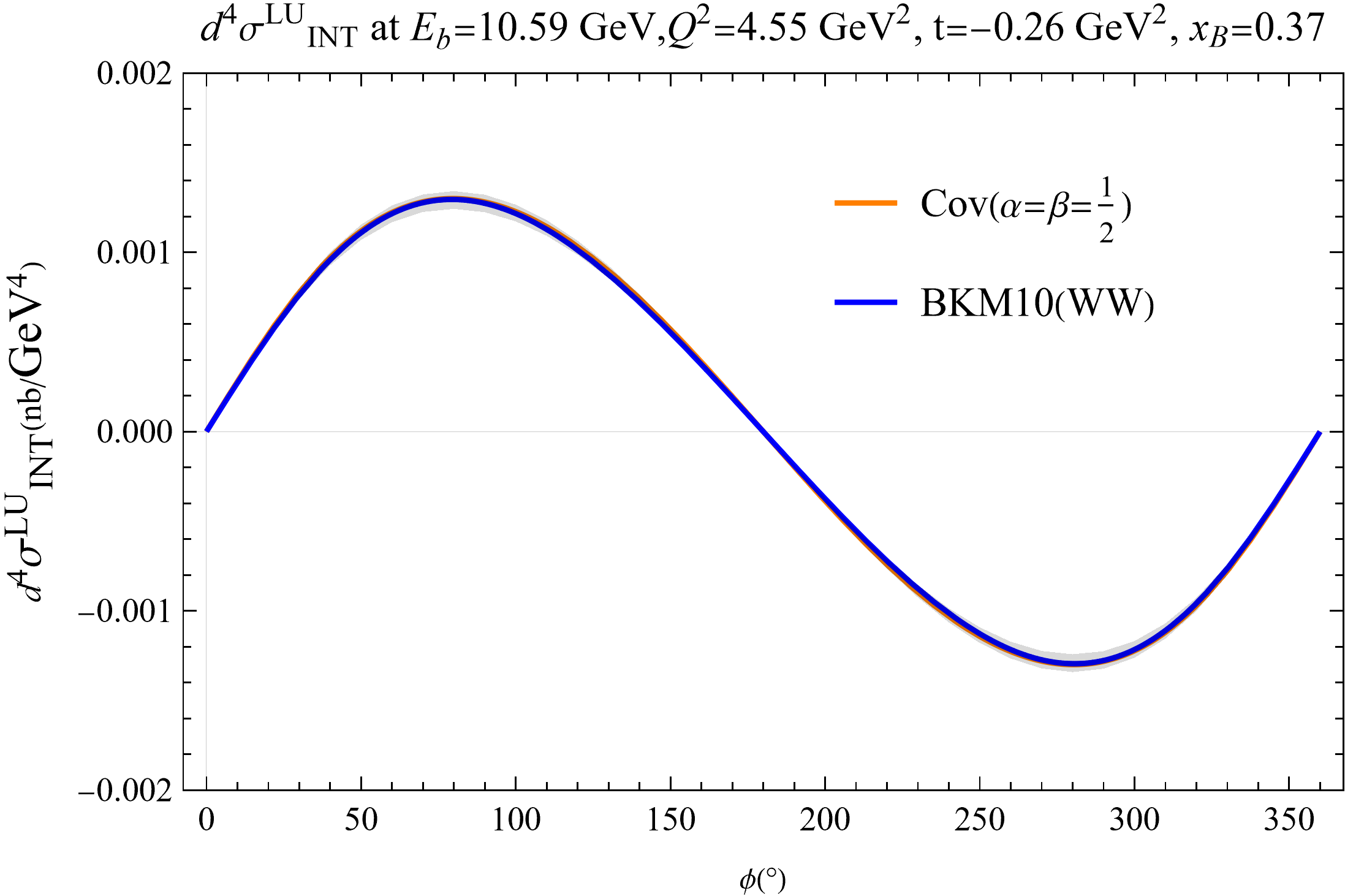}
\end{minipage}
\caption{\label{fig:INTxsectionc} Comparison of the four-fold interference cross-section $\text{d}^4 \sigma^{UU}_{\rm{INT}}$ and $\text{d}^4 \sigma^{LU}_{\rm{INT}}$. Our interference cross-sections agree well with the BKM10 results with WW relations. Different from the case of pure DVCS cross-sections, the interference cross-sections still show dependence on the choice of light cone vector, which is indicated by the gray band for which we vary $\alpha$ and $\beta$ in the region $0\le\alpha\le 1,0\le \beta\le 100$.}
\end{figure}

Besides, we also compare the interference cross-sections for both unpolarized and longitudinally polarized beam and unpolarized target as shown in Fig. \ref{fig:INTxsectionc}. Due to the explicit light cone dependence in the coefficients $B^{I,U}$ and $C^{I,U}$ (also $\tilde{B}^{I,U}$ and $\tilde{C}^{I,U}$ in the case of a longitudinally polarized beam), the interference cross-sections still show some dependence on the choice of light cone vectors. Again, we emphasize that after using the covariant Compton tensor coefficients, the light cone dependence vanishes at twist three, so the light cone dependence indicated by the gray bands are twist-four effects.

\section{Conclusion}
\label{sec:conclusion}
We study the high order kinematical effects in DVCS related to the choice of light cone vectors and gauge fixing conditions. We show that different choices of light cone coordinates can lead to differences at twist three in both the helicity amplitudes and the cross-sections, whereas the reliance on the gauge fixing conditions has less effect. Our general formula reproduces the BKM results \cite{Belitsky:2001ns,Belitsky:2010jw} both numerically and analytically up to twist-four terms, when specific choice of light cone vectors corresponding to the set-up in \cite{Belitsky:2001ns,Belitsky:2010jw} is made.

We considered different conventions that have been used in the literature and compared their differences. Those differences between various conventions can be considered as the theoretical uncertainties in the cross-section formulas, since the choice of light cone vectors are conventional. Consequently, one must take them into account when reconstructing the CFFs from experimental data of DVCS cross-sections.

We also study the polarized cross-sections, and show that our formulas apply to the polarized case as well. We show that there are four different polarization configurations for pure DVCS cross-sections and eight different polarization configurations for interference cross-sections, which agrees with the arguments in \cite{Kriesten:2019jep}. Since each of those configurations is associated with a different combination of nucleon form factors, a combined measurement of those different polarization configurations can help get more information of the CFFs.

In addition, we studied the kinematical corrections of twist-three GPDs with the OPE and showed that the light cone dependence gets canceled at twist three with the kinematical corrections of twist-three GPDs. With those corrections from higher twist GPDs, we also developed a set of covariant scalar coefficients which applies to all different frames and kinematics.

\section*{Acknowledgments}
We thank V. Braun, C. Hyde, B. Kriesten and S. Liuti for discussions related to the subject of this paper. This research is supported by the U.S. Department of Energy, Office of Science, Office of Nuclear Physics, under contract number DE-SC0020682, and the Center for Nuclear Femtography, Southeastern Universities Research Association, Washington D.C. Y. Guo is partially supported by a graduate fellowship from Center for Nuclear Femtography, SURA, Washington DC.

\newpage
\appendix

\section{Lab Frame Kinematics}
\label{app:labframe}
Here we define all the physical four momenta in the lab frame, which can be written in terms of their components $V^\mu=(V^0,V^x,V^y,V^z)$ as
\begin{align}
\begin{split}
      P=&(M,0,0,0) \ ,\\
  q=&\frac{Q}{\gamma}(1,0,0,-\sqrt{1+\gamma^2}) \ ,\\
  k=&k_0\left(1,\sin\theta_l,0,\cos\theta_l\right)\ , \\
   k'=&k'_0\left(1,\sin\theta'_l,0,\cos\theta'_l\right)\ ,\\
  P'=&\left(M-\frac{t}{2M},|\vec P'|\sin\theta'\cos\phi,|\vec P'|\sin\theta'\sin\phi,|\vec P'|\cos\theta'\right)\ ,\\
   q'=&q'_0\left(1,\sin\theta\cos\phi,\sin\theta\sin\phi,\cos\theta\right)\ .
\end{split}
\end{align} 
The relevant constants can be solved with the kinematical invariants such as $Q,t,x_B,~\cdots$ which give \cite{Kriesten:2019jep}
\begin{align}
\begin{split}
     |\vec P'|=\sqrt{-t\left(1-\frac{t}{4M^2}\right)}\ ,&\quad q'_0 =\frac{Q^2+x_B t}{2 M x_B}\ ,\quad k_0 =\frac{Q}{\gamma y}\ ,\quad k'_0 =\frac{Q(1-y)}{\gamma y}\ ,\\
    \cos\theta =-\frac{1+\frac{\gamma^2}{2}\frac{Q^2+t}{Q^2+x_B t}}{\sqrt{1+\gamma^2}}\ ,&\quad \cos\theta' =-\frac{\gamma^2(Q^2-t)-2x_B t}{4 x_B M\sqrt{1+\gamma^2}\sqrt{-t(1-\frac{t}{4M^2})}}\ ,\\
    \cos\theta_l =-\frac{1+y \frac{\gamma^2}{2}}{\sqrt{1+\gamma^2}}\ ,&\quad  \cos\theta'_l =-\frac{1-y-\gamma^2 y/2}{(1-y)\sqrt{1+\gamma^2}}\ .
\end{split}
\end{align}
Note that those angles are not complete fixed by their cosines, and we have the extra conventional requirements that $\sin\theta_l>0$, $\sin\theta>0$ and consequently $\sin\theta'_l<0$, $\sin\theta'<0$ from momentum conservation.

\section{Pure DVCS Structure Functions}
\label{app:dvcsstructurefunc}
Here we express all the pure DVCS structure functions explicitly. With their definition in Eq. (\ref{eq:dvcsstruct}), we can perform the contraction and express them in terms of the scalar amplitude defined in Eq. (\ref{eq:dvcsscalaramp}) as,
\begin{align}
    \begin{split}
        &F_{UU}=4\Bigg[(1-\xi^2)\left(h^{\rm{U}}\mathcal{H}^*\mathcal{H}+\tilde{h}^{\rm{U}}\widetilde{\mathcal{H}}^*\widetilde{\mathcal{H}}\right) -\frac{t}{4M^2}\left( h^{\rm{U}}\mathcal{E}^*\mathcal{E}+\xi^2 \tilde{h}^{\rm{U}}\widetilde{\mathcal{E}}^*\widetilde{\mathcal{E}}\right)\\
   &\qquad\qquad\quad-\xi^2\left(h^{\rm{U}}\mathcal{E}^*\mathcal{E}+h^{\rm{U}} (\mathcal{E}^*\mathcal{H} +\mathcal{H}^*\mathcal{E})+\tilde{h}^{\rm{U}} (\widetilde{\mathcal{E}}^*\widetilde{\mathcal{H}} +\widetilde{\mathcal{H}}^*\widetilde{\mathcal{E}})\right)\Bigg]\ ,
    \end{split}\\
     &F_{UT,\rm{out}}=N 4\text{Im}\Bigg[h^{\rm{U}}\mathcal{H}^*\mathcal{E}- \tilde{h}^{\rm{U}}\xi \widetilde{\mathcal{H}}^*\widetilde{\mathcal{E}}\Bigg]\ , \\
     &F_{UL}=8 h^{+,\rm{U}}\text{Im}\Bigg\{\left(1-\xi^2\right)\widetilde{\mathcal{H}}^*\mathcal{H}-\xi^2 \left(\widetilde{\mathcal{H}}^*\mathcal{E}+\widetilde{\mathcal{E}}^*\mathcal{H}\right)-\left(\frac{\xi^2}{1+\xi}+\frac{t}{4M^2}\right)\xi\widetilde{\mathcal{E}}^*\mathcal{E}\Bigg\}\ ,\\
      &F_{LL}=8 h^{-,\rm{L}}\text{Re}\Bigg\{\left(1-\xi^2\right)\widetilde{\mathcal{H}}^*\mathcal{H}-\xi^2\left(\widetilde{\mathcal{H}}^*\mathcal{E}+\widetilde{\mathcal{E}}^*\mathcal{H}\right)-\left(\frac{\xi^2}{1+\xi}+\frac{t}{4M^2}\right)\xi\widetilde{\mathcal{E}}^*\mathcal{E}\Bigg\}\ ,\\
      &F_{UT,\rm{in}}=4N h^{+,\rm{U}} \text{Im}\Bigg\{\widetilde{\mathcal{H}}^*\mathcal{E}-\xi\widetilde{\mathcal{E}}^*\mathcal{H}-\frac{\xi^2}{1+\xi}\widetilde{\mathcal{E}}^*\mathcal{E} \Bigg\}\ ,\\
      &F_{LT,\rm{in}}=4N h^{-,\rm{L}}\text{Re}\Bigg\{\widetilde{\mathcal{H}}^*\mathcal{E}-\xi\widetilde{\mathcal{E}}^*\mathcal{H}-\frac{\xi^2}{1+\xi}\widetilde{\mathcal{E}}^*\mathcal{E} \Bigg\}\ ,
\end{align}
Therefore, it will be sufficient to understand those four coefficients $h^{\rm{U}}$, $\tilde{h}^{\rm{U}}$, $h^{+,\rm{U}}$ and $h^{-,\rm{L}}$ in order to get the pure DVCS cross-section besides the CFFs. 
Similar structures show up in the BKM10 \cite{Belitsky:2010jw} and the conversion relation can be written as
\begin{equation}
\label{eq:hbkm}
\begin{aligned}
    h^{\rm{U}}_{\rm{BKM}}=\tilde{h}^{\rm{U}}_{\rm{BKM}}=\frac{Q^2}{4y^2}\Bigg[&2\frac{2-2y+y^2+\gamma^2 y^2/2}{1+\gamma^2}+\frac{16 K^2}{(2-x_B^2)(1+\gamma^2)}\left(\frac{-2\xi_{\rm{BKM}}}{1+\xi_{\rm{BKM}}}\right)^2\\
    &+\frac{8 K(2-y)}{(2-x_B)(1+\gamma^2)}\frac{-2\xi_{\rm{BKM}}}{1+\xi_{\rm{BKM}}}\cos\phi\Bigg] \ ,
\end{aligned}
\end{equation}
where the $\xi_{\rm{BKM}}$ differs from our $\xi$ and is defined as
\begin{align}
\label{eq:xibkm}
\xi_{\rm{BKM}}\equiv \frac{x_B\left(1+\frac{t}{2Q^2}\right)}{2-x_B+x_B\frac{t}{Q^2}}\ .
\end{align}
and $K$ are defined from
\begin{align}
\label{eq:bkmk}
K^2=\frac{(t_0-t)}{Q^2}(1-x_B)  \left(1-y-\frac{1}{4} \gamma ^2 y^2\right) \left[\sqrt{\gamma ^2+1}+\left(\frac{\gamma ^2}{4(1-x_B)}+x_B\right)\frac{(t-t_0)}{Q^2}\right]\ ,
\end{align}
with $t_0$ the minimum momentum transfer defined as,
\begin{align}
t_0=-Q^2 \frac{2(1-x_B)(1-\sqrt{1+\gamma^2})+\gamma^2}{4x_B(1-x_B)+\gamma^2}\ .
\end{align}
Note that there are contribution from those so-called effective twist three Compton form factors in BKM which are defined as \cite{Belitsky:2001ns}
\begin{align}
\mathcal F_{\rm{eff}}\equiv -2\xi_{\rm{BKM}}\left(\frac{1}{1+\xi_{\rm{BKM}}}\mathcal F+\mathcal F^{3}_+-\mathcal F^{3}_-\right)\ ,
\end{align}
and they reduce to the $-2\xi_{\rm{BKM}}/(1+\xi_{\rm{BKM}})\mathcal F$ terms in the coefficients when setting the twist-three CFFs $\mathcal F^{3}_+$ and $\mathcal F^{3}_-$ to zero. We also need to mention that when the WW relations are taken into account, one have $\mathcal F_+^{3} = -1/\xi_{\rm{BKM}} \mathcal F +\cdots$, where ellipsis stands for twist-two GPDs with WW kernel and genuine twist-three GPDs. Then we have $F_{\rm{eff}}\to  2 /\left(1+\xi_{\rm{BKM}}\right)\mathcal F$ with WW relations. 

While our covariant scalar coefficients are already presented in Eq. (\ref{eq:covh}), of which the code is available in \cite{Guo:2021git} as well, we also present our scalar coefficients calculated from light cone vectors with twist expansion, for which we define
\begin{equation}
  \mathscr H =\sum_{t=2}^\infty Q^{4-t} \mathscr H_{(t)}\ ,
\end{equation}
with $\mathscr H=\{ h^{\rm{U}}, \tilde{h}^{\rm{U}},h^{-,\rm{L}},h^{+,\rm{U}}\}$. A twist expansion will then yield the following result
\begin{equation}
\label{eq:hutwist}
    \begin{aligned}
          h^{\rm{U}}_{(2)}&= \tilde{h}^{\rm{U}}_{(2)}=\frac{2-2y+y^2}{2 y^2}\ ,\\
          h^{\rm{U}}_{(3)}&= \tilde{h}^{\rm{U}}_{(3)}= \frac{2x_B(y-2)\sqrt{[t(x_B-1)-M^2x_B^2](1-y)}(\beta-1)\cos\phi}{[1+(\beta-1)x_B]y^2}\ ,\\
          h^{\rm{U}}_{(4)}&= \tilde{h}^{\rm{U}}_{(4)}=\frac{1}{2[1+(\beta-1)x_B]^2y^2} \\
          &\qquad\qquad\times\Bigg[8 (\beta -1)^2 x_B^2 (y-1) \cos ^2\phi \left(M^2 x_B^2-t x_B+t\right)\\
          &\qquad\qquad\qquad- x_B^2 (y-2)^2 \left(M^2 \left([2 (\beta -1) x_B+1]^2+1\right)-2(\beta -1)^2 t (x_B-1)\right)\Bigg]\ ,
    \end{aligned}
\end{equation}
and for $h^{-,\rm{L}}$,
\begin{equation}
\label{eq:hminusutwist}
    \begin{aligned}
          h^{-,\rm{L}}_{(2)}&=\frac{y-2}{2y}\ ,\\
         h^{-,\rm{L}}_{(3)}&=\frac{2x_B\sqrt{[t(x_B-1)-M^2x_B^2](1-y)}(\beta-1)\cos\phi}{[1+(\beta-1)x_B]y} \ ,\\
         h^{-,\rm{L}}_{(4)}&=\frac{x_B^2 (y-2) \left(M^2 \left(2 (\beta -1)^2 x_B^2+2 (\beta -1) x_B+1\right)-(\beta -1)^2 t (x_B-1)\right)}{y [(\beta -1) x_B+1]^2}\ ,
    \end{aligned}
\end{equation}
and for $h^{+,\rm{U}}$,
\begin{equation}
          h^{+,\rm{U}}_{(2)}=h^{+,\rm{U}}_{(3)}=h^{+,\rm{U}}_{(4)}=0\ .
\end{equation}
Note that with twist-four accuracy, we have the relation $h^{\rm{U}} \approx \tilde{h}^{\rm{U}}$ and $h^{+,\rm{U}}\approx 0$, which can lead to some simplification to the pure DVCS cross-section formula.

\section{Interference Structure Functions}

\label{app:structurefunc}
In the Subsection \ref{subsec:intxsection}, we write down the interference cross-section for eight different polarization configurations. Here we put down their expressions explicitly. By simply contracting according to the definitions in Eq. (\ref{eq:intxsecdef}), we have the following expression for all the $F^I$s
\begin{align}
\begin{split}
    &F^{I}_{UU}=\text{Re}\Bigg[A^{I,\rm{U}} \left(F_1 \mathcal {H}^*-\frac{t}{4M^2} F_2 \mathcal {E}^*\right)+B^{I,\rm{U}} (F_1+F_2)(\mathcal {H}^*+\mathcal {E}^*) \\
    &\qquad\qquad\qquad+
     C^{I,\rm{U}}(F_1+F_2)\widetilde{\mathcal {H}}^*\Bigg]\ ,
\end{split}
\end{align}
\begin{align}
\begin{split}
    &F^{I}_{LU}=-\text{Im}\Bigg[A^{I,\rm{L}} \left(F_1 \mathcal {H}^*-\frac{t}{4M^2} F_2 \mathcal {E}^*\right)+B^{I,\rm{L}} (F_1+F_2)(\mathcal {H}^*+\mathcal {E}^*) \\
    &\qquad\qquad\qquad+
     C^{I,\rm{L}}(F_1+F_2)\widetilde{\mathcal {H}}^*\Bigg]\ ,
\end{split}
\end{align}
\begin{align}
\label{eq:FIUL}
\begin{split}
    &F^{I}_{UL}=- \text{Im}\Bigg\{\tilde A^{I,\rm{U}}\left[F_1\left(\widetilde{ \mathcal {H}}^*-\frac{\xi^2}{1+\xi}\widetilde{ \mathcal {E}}^* \right)-F_2\frac{t}{4M^2}\xi\widetilde{ \mathcal {E}}^*\right]\\
       &\qquad\qquad+\tilde B^{I,\rm{U}} (F_1+F_2)\left(\widetilde{ \mathcal {H}}^*+\frac{\xi}{1+\xi} \widetilde{\mathcal {E}}^* \right)-\tilde C^{I,\rm{U}}(F_1+F_2)\left( \mathcal {H}^*+\frac{\xi}{1+\xi} \mathcal {E}^* \right)\Bigg\}\ ,
\end{split}
\end{align}
\begin{align}
\begin{split}
    &F^{I}_{LL}= -\text{Re}\Bigg\{\tilde A^{I,\rm{L}}\left[F_1\left(\widetilde{ \mathcal {H}}^*-\frac{\xi^2}{1+\xi}\widetilde{ \mathcal {E}}^* \right)-F_2\frac{t}{4M^2}\xi\widetilde{ \mathcal {E}}^*\right]\\
       &\qquad\qquad+\tilde B^{I,\rm{L}} (F_1+F_2)\left(\widetilde{ \mathcal {H}}^*+\frac{\xi}{1+\xi} \widetilde{\mathcal {E}}^* \right)-\tilde C^{I,\rm{L}}(F_1+F_2)\left( \mathcal {H}^*+\frac{\xi}{1+\xi} \mathcal {E}^* \right)\Bigg\}\ ,
\end{split}
\end{align}
\begin{align}
\begin{split}
    &F^{I}_{UT,\rm{in}}=-\frac{2}{N}  \text{Im}\Bigg\{\tilde A^{I,\rm{U}}\Bigg[\xi F_1\left(\xi \widetilde{ \mathcal {H}}^*+\left(\frac{\xi^2}{1+\xi}+\frac{t}{4M^2}\right)\widetilde{ \mathcal {E}}^* \right)+F_2\frac{t}{4M^2}\left((\xi^2-1)\widetilde{\mathcal {H}}^* +\xi^2\widetilde{ \mathcal {E}}^*\right)\Bigg]\\
       &\qquad\qquad\qquad\qquad+\tilde B^{I,\rm{U}}(F_1+F_2)\left[\widetilde{ \mathcal {H}}^*+\left(\frac{t}{4M^2}-\frac{\xi}{1+\xi}\right)\xi\widetilde{\mathcal {E}}^* \right]\\
       &\qquad\qquad\qquad\qquad+\tilde C^{I,\rm{U}} (F_1+F_2)\left[\xi \mathcal {H}^*+\left(\frac{\xi^2}{1+\xi}+\frac{t}{4M^2}\right) \mathcal {E}^* \right]  \Bigg\}\ ,
\end{split}
\end{align}
\begin{align}
\begin{split}
    &F^{I}_{LT,\rm{in}}= -\frac{2}{N}  \text{Re}\Bigg\{\tilde A^{I,\rm{L}}\Bigg[\xi F_1\left(\xi \widetilde{ \mathcal {H}}^*+\left(\frac{\xi^2}{1+\xi}+\frac{t}{4M^2}\right)\widetilde{ \mathcal {E}}^* \right)+F_2\frac{t}{4M^2}\left((\xi^2-1)\widetilde{\mathcal {H}}^* +\xi^2\widetilde{ \mathcal {E}}^*\right)\Bigg]\\
       &\qquad\qquad\qquad\qquad+\tilde B^{I,\rm{L}}(F_1+F_2)\left[\widetilde{ \mathcal {H}}^*+\left(\frac{t}{4M^2}-\frac{\xi}{1+\xi}\right)\xi\widetilde{\mathcal {E}}^* \right]\\
       &\qquad\qquad\qquad\qquad+\tilde C^{I,\rm{L}} (F_1+F_2)\left[\xi \mathcal {H}^*+\left(\frac{\xi^2}{1+\xi}+\frac{t}{4M^2}\right) \mathcal {E}^* \right]  \Bigg\}\ ,
\end{split}
\end{align}
\begin{align}
\begin{split}
    &F^{I}_{UT,\rm{out}}=\frac{2}{N} \text{Im}\Bigg\{A^{I,\rm{U}} \left[F_1\left(\xi^2\mathcal H^*+\left(\xi^2+\frac{t}{4M^2}\right)\mathcal E^* \right)+\frac{t}{4M^2}F_2\left((\xi^2-1)\mathcal H^*+\xi^2\mathcal E^*\right)\right]\\
      &\qquad\qquad\qquad +B^{I,\rm{U}}(F_1+F_2) \left(\mathcal H^*+\frac{t}{4M^2}\mathcal E^*\right)-C^{I,\rm{U}} \xi(F_1+F_2)\left(\widetilde{\mathcal H}^*+\frac{t}{4M^2}\widetilde{\mathcal E}^*\right)\Bigg\} \ ,
\end{split}
\end{align}
\begin{align}
\begin{split}
    &F^{I}_{LT,\rm{out}}=\frac{2}{N} \text{Re}\Bigg\{A^{I,\rm{L}} \left[F_1\left(\xi^2\mathcal H^*+\left(\xi^2+\frac{t}{4M^2}\right)\mathcal E^* \right)+\frac{t}{4M^2}F_2\left((\xi^2-1)\mathcal H^*+\xi^2\mathcal E^*\right)\right]\\
      &\qquad\qquad \qquad+B^{I,\rm{L}}(F_1+F_2) \left(\mathcal H^*+\frac{t}{4M^2}\mathcal E^*\right)-C^{I,\rm{L}} \xi(F_1+F_2)\left(\widetilde{\mathcal H}^*+\frac{t}{4M^2}\widetilde{\mathcal E}^*\right)\Bigg\} \ .
\end{split}
\end{align}
In \cite{Belitsky:2010jw}, similar structures show up, and the BKM10 results can be re-expressed in terms of those coefficients the conversion relations for $A^{I,\rm{U}},B^{I,\rm{U}}$ and $C^{I,\rm{U}}$ are  
\begin{align}
\label{eq:aiubkm}
A^{I,\rm{U}}_{\rm{BKM}} &= \frac{-Q^2 }{x_{B}y^{3} \mathcal{P}_{1}(\phi) \mathcal{P}_{2}(\phi)} \Bigg\{ \sum_{n = 0}^{3}\left[ C_{++}^{\text{unp}}(n)+\frac{\sqrt{2}\tilde K}{(2-x_B)Q}\frac{-2\xi_{\rm{BKM}}}{1+\xi_{\rm{BKM}}}C_{0+}^{\text{unp}}(n)\right]\cos{(n\phi)} \Bigg\}\ ,
\end{align}
\begin{align}
\label{eq:biubkm}
\begin{split}
B^{I,\rm{U}}_{\rm{BKM}} &= \frac{-Q^2 \xi_{\rm{BKM}}}{x_{B}y^{3} \mathcal{P}_{1}(\phi) \mathcal{P}_{2}(\phi)\left(1+\frac{t}{2Q^2}\right)} \Bigg\{  \sum_{n = 0}^{3}\Big[ C_{++}^{\text{unp},V}(n)\\
&\qquad\qquad\qquad\qquad\qquad\qquad+\frac{\sqrt{2}\widetilde K}{(2-x_B)Q}\frac{-2\xi_{\rm{BKM}}}{1+\xi_{\rm{BKM}}}C_{0+}^{\text{unp},V}(n)\Big]\cos{(n\phi)} \Bigg\}\ , 
\end{split}
\end{align}
\begin{align}
\label{eq:ciubkm}
\begin{split}
   C^{I,\rm{U}}_{\rm{BKM}} &= \frac{-Q^2 \xi_{\rm{BKM}}}{x_{B}y^{3} \mathcal{P}_{1}(\phi) \mathcal{P}_{2}(\phi)\left(1+\frac{t}{2Q^2}\right)} \Bigg\{ \sum_{n = 0}^{3} \Big[ C_{+ +}^{\text{unp,A}}(n) + C_{++}^{\text{unp}}(n)\\
   &\qquad\qquad\qquad\qquad+\frac{\sqrt{2}\widetilde K}{(2-x_B)Q}\frac{-2\xi_{\rm{BKM}}}{1+\xi_{\rm{BKM}}}\left(C_{0 +}^{\text{unp,A}}(n) + C_{0+}^{\text{unp}}(n)\right) \Big]\cos{(n\phi)}\Bigg\}\ ,
  \end{split}
\end{align}
while for $\tilde A^{I,\rm{U}}, \tilde B^{I,\rm{U}}$ and $\tilde C^{I,\rm{U}}$, the conversion relations are
\begin{align}
\label{eq:atiubkm}
\tilde A^{I,\rm{U}}_{\rm{BKM}} &= \frac{-Q^2 }{x_{B}y^{3} \mathcal{P}_{1}(\phi) \mathcal{P}_{2}(\phi)} \Bigg\{ \sum_{n = 1}^{3}\left[ S_{++}^{\text{LP}}(n)+\frac{\sqrt{2}\tilde K}{(2-x_B)Q}\frac{-2\xi_{\rm{BKM}}}{1+\xi_{\rm{BKM}}}S_{0+}^{\text{LP}}(n)\right]\sin{(n\phi)} \Bigg\}\ , 
\end{align}
\begin{align}
\label{eq:btiubkm}
\begin{split}
\tilde B^{I,\rm{U}}_{\rm{BKM}} &= \frac{-Q^2 \xi_{\rm{BKM}}}{x_{B}y^{3} \mathcal{P}_{1}(\phi) \mathcal{P}_{2}(\phi)\left(1+\frac{t}{2Q^2}\right)} \Bigg\{  \sum_{n = 1}^{3}\Big[ S_{++}^{\text{LP},A}(n)\\
&\qquad\qquad\qquad\qquad\qquad\qquad+\frac{\sqrt{2}\widetilde K}{(2-x_B)Q}\frac{-2\xi_{\rm{BKM}}}{1+\xi_{\rm{BKM}}}S_{0+}^{\text{LP},A}(n)\Big]\sin{(n\phi)} \Bigg\}\ , 
\end{split}
\end{align}
\begin{align}
\label{eq:ctiubkm}
\begin{split}
\tilde C^{I,\rm{U}}_{\rm{BKM}} &= -\frac{-Q^2 \xi_{\rm{BKM}}}{x_{B}y^{3} \mathcal{P}_{1}(\phi) \mathcal{P}_{2}(\phi)\left(1+\frac{t}{2Q^2}\right)} \Bigg\{ \sum_{n = 1}^{3} \Big[ S_{+ +}^{\text{LP,V}}(n) + S_{++}^{\text{LP}}(n)\\
   &\qquad\qquad\qquad\qquad+\frac{\sqrt{2}\widetilde K}{(2-x_B)Q}\frac{-2\xi_{\rm{BKM}}}{1+\xi_{\rm{BKM}}}\left(S_{0 +}^{\text{LP,V}}(n) + S_{0+}^{\text{LP}}(n)\right) \Big]\sin{(n\phi)}\Bigg\}\ .
  \end{split}
\end{align}
$\mathcal{P}_{1}(\phi)$ and $\mathcal{P}_{2}(\phi)$ are defined from the scalar products in the denominators of the BH lepton amplitude as,
\begin{align}
\mathcal{P}_{1}(\phi)\equiv\frac{(k-q')^2}{Q^2}\ ,\qquad \mathcal{P}_{2}(\phi)\equiv \frac{(k'+q')^2}{Q^2}\ ,
\end{align}
and $\tilde {K}$ are defined from
\begin{align}
\tilde {K}^2=\frac{Q^2K^2 }{1-y+\gamma^2 y^2/4}\ ,
\end{align}
with $K$ given in Eq. (\ref{eq:bkmk}) and $\xi_{\rm{BKM}}$ given in Eq. (\ref{eq:xibkm}). Again, there are contributions from those so-called effective twist three Compton form factors corresponding to the $-2\xi_{\rm{BKM}}/(1+\xi_{\rm{BKM}})\mathcal F$ terms in the coefficients, or $2/(1+\xi_{\rm{BKM}})\mathcal F$ with WW relations taken into account.

In our case, we can derive those 6 coefficients right from their definitions in Eq. (\ref{eq:intstructfunc}), for which the expression of the leptonic tensor $L_{\rm{INT}}^{\mu\rho\sigma}$ is needed. We have split the leptonic tensor into the polarized and unpolarized part, and it will be helpful to further split them into the symmetric and antisymmetric parts such that
\begin{align}
\begin{split}
   L_{\rm{INT,U}}^{\mu\rho\sigma}=L_{\rm{INT,U}}^{(\mu\rho)\sigma}+L_{\rm{INT,U}}^{[\mu\rho]\sigma}\ , \qquad L_{\rm{INT,L}}^{\mu\rho\sigma}=L_{\rm{INT,L}}^{(\mu\rho)\sigma}+L_{\rm{INT,L}}^{[\mu\rho]\sigma}\ .
\end{split}
\end{align}
Those tensors can be written as~\cite{Ji:1996nm} 
\begin{align}
\begin{split}
   L_{\rm{INT,U}}^{(\mu\rho)\sigma}=&\frac{2}{(k-\Delta)^2}\Big[\left[k'^\mu (2k-\Delta)^\rho+k'^\rho (2k-\Delta)^\mu-g^{\mu\rho}k'\cdot (2k-\Delta)\right]k^\sigma\\
   &-(k'^\mu k^\rho+k'^\rho k^\mu-g^{\mu\rho} k\cdot k')\Delta^\sigma
  +(k'^\mu g^{\rho\sigma}+k'^\rho g^{\mu\sigma}-g^{\mu\rho}k'^\sigma)(k\cdot \Delta)\Big]\\
  +&\frac{2}{(k'+\Delta)^2}\Big[\left[(2k'+\Delta)^\mu k^\rho+(2k'+\Delta)^\rho k^\mu-g^{\mu\rho} k\cdot(2k'+\Delta)\right]k'^\sigma\\
  &+(k'^\mu k^\rho+k'^\rho k^\mu-g^{\mu\rho} k\cdot k')\Delta^\sigma-(k^\mu g^{\rho\sigma}+k^\rho g^{\mu\sigma}-g^{\mu\rho}k^\sigma)(k'\cdot \Delta)\Big]\ ,
\end{split}
\end{align}
the antisymmetric unpolarized tensor
\begin{align}
\begin{split}
   L_{\rm{INT,U}}^{[\mu\rho]\sigma}=&\frac{2\left[-(k\cdot k')\left(\Delta^\mu g^{\rho\sigma}-\Delta^\rho g^{\mu\sigma}\right)+(k'\cdot \Delta) \left(k^\mu g^{\rho\sigma}-k^\rho g^{\mu\sigma}\right)-(k^\mu \Delta^\rho-k^\rho\Delta^\mu)k'^\sigma\right]}{(k-\Delta)^2}\\
  &+\frac{2\left[(k\cdot k')\left(\Delta^\mu g^{\rho\sigma}-\Delta^\rho g^{\mu\sigma}\right)-(k\cdot \Delta) \left(k'^\mu g^{\rho\sigma}-k'^\rho g^{\mu\sigma}\right)+(k'^\mu \Delta^\rho-k'^\rho\Delta^\mu)k^\sigma\right]}{(k'+\Delta)^2}\ ,
\end{split}
\end{align}
the symmetric polarized tensor
\begin{align}
\begin{split}
   L_{\rm{INT,L}}^{(\mu\rho)\sigma}=&\frac{2}{(k-\Delta)^2}\Big[g^{\mu\rho}\epsilon^{\sigma k k'\Delta}-k'^\mu \epsilon^{\rho\sigma k \Delta}-k'^\rho \epsilon^{\mu\sigma k \Delta}\Big]\\
   &+\frac{2}{(k'+\Delta)^2}\Big[-g^{\mu\rho}\epsilon^{\sigma k k'\Delta}-k^\mu \epsilon^{\rho\sigma k' \Delta}-k^\rho \epsilon^{\mu\sigma k' \Delta}\Big]\ ,
\end{split}
\end{align}
and the antisymmetric polarized tensor
\begin{align}
\begin{split}
   L_{\rm{INT,L}}^{[\mu\rho]\sigma}=&\frac{2}{(k-\Delta)^2}\Big[\epsilon^{\mu\rho k k'}(2k-\Delta)^\sigma+\epsilon^{\mu\rho k' \Delta} k^\sigma +\epsilon^{\mu\rho\sigma k'}(k\cdot \Delta)\Big]\\
   +&\frac{2}{(k'+\Delta)^2}\Big[\epsilon^{\mu\rho k k'}(2k'+\Delta)^\sigma+\epsilon^{\mu\rho k \Delta} k'^\sigma +\epsilon^{\mu\rho\sigma k}(k'\cdot \Delta)\Big]\ .
\end{split}
\end{align}
Then all the scalar coefficients can be worked out by contracting all the indices with $L^{\rm{INT,U}}_{\mu\rho\sigma}$ and $L^{\rm{INT,L}}_{\mu\rho\sigma}$ and written as scalar products. Here we write down all 12 real scalar coefficients:
\begin{align}
\label{eq:covAIU}
\begin{split}
A^{I,U}=&\frac{8}{(k\cdot q')(q\cdot q')(k-q)\cdot q'}\Bigg\{2(\bar P\cdot q)(k\cdot q')^3\\&-(k\cdot q')^2\Big[2(k\cdot q)(\bar P\cdot q')
-2(k\cdot \bar P)(q^2-q\cdot q')+2(\bar P\cdot q)(q\cdot q')+q^2\bar P\cdot (q-q')\Big]\\&+(k\cdot q')\Big[2(k\cdot \bar P)(q\cdot q')^2-(\bar P\cdot q')(q\cdot q') q^2\\
&\qquad\qquad+k\cdot q\Big((q\cdot q')\bar P\cdot (q-4k)+\bar P\cdot q'(q^2+2 q\cdot q')\Big)\Big]\\
&+(k\cdot q)(q\cdot q')\Big[(\bar P\cdot k)(q\cdot q')-(\bar P\cdot q')(k\cdot q)\Big]\Bigg\}\ ,
\end{split}
\end{align}
\begin{align}
\label{eq:covBIU}
\begin{split}
B^{I,U}=&\frac{2t}{(k\cdot q')(q\cdot q')(k-q)\cdot q'}\Bigg\{2(n\cdot q)(k\cdot q')^3\\&-(k\cdot q')^2\Big[2(k\cdot q)(n\cdot q')
-2(k\cdot n)(q^2-q\cdot q')+2(n\cdot q)(q\cdot q')+q^2n\cdot (q-q')\Big]\\&+(k\cdot q')\Big[2(k\cdot n)(q\cdot q')^2-(n\cdot q')(q\cdot q') q^2\\
&\qquad\qquad+k\cdot q\Big((q\cdot q')n\cdot (q-4k)+n\cdot q'(q^2+2 q\cdot q')\Big)\Big]\\
&+(k\cdot q)(q\cdot q')\Big[(n\cdot k)(q\cdot q')-(n\cdot q')(k\cdot q)\Big]\Bigg\}\ .
\end{split}
\end{align}
\begin{align}
\label{eq:covCIU}
\begin{split}
C^{I,U}=&\frac{-4\Big[2(k\cdot q')(k-q)\cdot q'+(k\cdot q)(q\cdot q')\Big]}{(k\cdot q')(q\cdot q')(k-q)\cdot q'}\Bigg\{(k\cdot P')\Big[(n\cdot q')(P\cdot q)-(n\cdot q)(P\cdot q')\Big]\\
+&(k\cdot P)\Big[(n\cdot q)(P'\cdot q')-(n\cdot q')(P'\cdot q)\Big]+(k\cdot n)\Big[(P\cdot q')(P'\cdot q)-(P\cdot q)(P'\cdot q')\Big]\Bigg\}\ .
\end{split}
\end{align}
\begin{align}
\label{eq:covAtIU}
\begin{split}
\tilde{A}^{I,U}=&\frac{8\Big[2(k\cdot q')(k-q)\cdot q'+(k\cdot q)(q\cdot q')\Big]}{(k\cdot q')(q\cdot q')(k-q)\cdot q'}\epsilon^{k\bar P q q'}\ ,
\end{split}
\end{align}
\begin{align}
\label{eq:covBtIU}
\begin{split}
\tilde{B}^{I,U}=&\frac{2t\Big[2(k\cdot q')(k-q)\cdot q'+(k\cdot q)(q\cdot q')\Big]}{(k\cdot q')(q\cdot q')(k-q)\cdot q'}\epsilon^{k n q q'}\ ,
\end{split}
\end{align}
\begin{align}
\label{eq:covCtIU}
\begin{split}
\tilde{C}^{I,U}=&\frac{4\Big[2(q^2-q\cdot q')(k\cdot q')^2+(k\cdot q)(q\cdot q')^2+2(k\cdot q')(q\cdot q')\left(q'-2k\right)\cdot q\Big]}{(k\cdot q')(q\cdot q')(k-q)\cdot q'}\epsilon^{k n P P'}\ ,
\end{split}
\end{align}
\begin{align}
\label{eq:covAIL}
\begin{split}
A^{I,L}=&\frac{8\Big[(k\cdot q)(q\cdot q')-q^2(k\cdot q')\Big]}{(k\cdot q')(q\cdot q')(k-q)\cdot q'}\epsilon^{k\bar P q q'}\ ,
\end{split}
\end{align}
\begin{align}
\label{eq:covBIL}
\begin{split}
B^{I,L}=&\frac{2t\Big[(k\cdot q)(q\cdot q')-q^2(k\cdot q')\Big]}{(k\cdot q')(q\cdot q')(k-q)\cdot q'}\epsilon^{k n q q'}\ ,
\end{split}
\end{align}
\begin{align}
\label{eq:covCIL}
\begin{split}
C^{I,L}=&\frac{-4(q\cdot q')\Big[(k\cdot q)(q\cdot q')-q^2(k\cdot q')\Big]}{(k\cdot q')(q\cdot q')(k-q)\cdot q'}\epsilon^{k n P P'}\ ,
\end{split}
\end{align}
\begin{align}
\label{eq:covAtIL}
\begin{split}
\tilde{A}^{I,L}=\frac{8}{(k\cdot q')(q\cdot q')(k-q)\cdot q'}\Bigg\{&q^2(k\cdot q')^2 (\bar P \cdot q')\\
&-(k\cdot q')(q\cdot q')\left[q^2(q'-k)\cdot \bar P+(k\cdot q)(\bar P\cdot q)\right]\\
&-(k\cdot q)(q\cdot q')\left[(\bar P\cdot k)(q\cdot q')-(k\cdot q)(\bar P\cdot q')\right]\Bigg\}\ ,
\end{split}
\end{align}
\begin{align}
\label{eq:covBtIL}
\begin{split}
\tilde{B}^{I,L}=\frac{2t}{(k\cdot q')(q\cdot q')(k-q)\cdot q'}\Bigg\{&q^2(k\cdot q')^2 (n \cdot q')\\
&-(k\cdot q')(q\cdot q')\left[q^2(q'-k)\cdot n+(k\cdot q)(n\cdot q)\right]\\
&-(k\cdot q)(q\cdot q')\left[(n\cdot k)(q\cdot q')-(k\cdot q)(n\cdot q')\right]\Bigg\}\ ,
\end{split}
\end{align}
\begin{align}
\label{eq:covCtIL}
\begin{split}
\tilde{C}^{I,L}=\frac{4\left[q^2(k\cdot q')-(k\cdot q)(q\cdot q')\right]}{(k\cdot q')(q\cdot q')(k-q)\cdot q'}\Bigg\{&(k\cdot P')\left[(n\cdot q')(P\cdot q)-(n\cdot q)(P\cdot q')\right]\\
&+(k\cdot P)\left[(n\cdot q)(P'\cdot q')-(n\cdot q')(P'\cdot q)\right]\\
&+(k\cdot n)\left[(P\cdot q')(P'\cdot q)-(P\cdot q)(P'\cdot q')\right]\Bigg\}\ .
\end{split}
\end{align}
Those coefficients are all we need for the interference cross-section formula for any polarization configurations, of which the code is also available in \cite{Guo:2021git}. Note that all the $B^I$ coefficients are related to the $A^I$ coefficients by $4 \bar P^\mu \to t n^\mu   $. We also mention that for a given kinematics where twist-four effects are negligible, one can use the substitution $n^\mu\to q'^\mu/(\bar P\cdot q')$ for an approximate light-cone-independent expression, corresponding to the limit $\beta\to \infty$.

Besides, we also present our scalar coefficients calculated from light cone vectors with twist expansion, for which we define
\begin{align}
\mathcal A^{I,\rm{U/L}}=\frac{1}{\mathcal P_1(\phi)\mathcal P_2(\phi)}\sum_{t=2}^\infty Q^{3-t}\mathcal A^{I,\rm{U/L}}_{(t)}\ ,
\end{align}
with the twist $t$ from $2$ to $\infty$ for each coefficient. Then we have the following results
\begin{equation}
\label{eq:aiutwist}
    \begin{aligned}
          A^{I,\rm{U}}_{(2)}&= \frac{-8(2-2y+y^2)}{x_B y^3}\sqrt{[t(x_B-1)-M^2x_B^2](1-y)}\cos\phi\ ,\\
          A^{I,\rm{U}}_{(3)}&=\frac{-8}{x_B y^3} \Bigg\{(y-2)\left\{-M^2x_B^2(y-2)^2+t\left[-2+2y-y^2+x_B\left(3-3y+y^2\right)\right]\right\}\\
           &\qquad\qquad+ \frac{2 x_B\left[t(x_B-1)-M^2x_B^2\right](y-2)(1-y)(\beta-1)\cos 2\phi}{1+(\beta-1)x_B}\Bigg\} \ ,
    \end{aligned}
\end{equation}
and for $B^{I,\rm{U}}$,
\begin{equation}
\label{eq:biutwist}
    \begin{aligned}
          B^{I,\rm{U}}_{(2)}&=0\ ,\\
          B^{I,\rm{U}}_{(3)}&=\frac{8t x_B(2-3y+y^2)}{(x_B-2)y^3} \ ,\\
          B^{I,\rm{U}}_{(4)}&=\frac{-8t xB\sqrt{[t(x_B-1)-M^2x_B^2](1-y)}}{(x_B-2) (1+(\beta-1)x_B)y^3}\\
          &\qquad\times\Big[2(y-2)^2-(2-2y+y^2)\alpha+2x_B(6-6y+y^2)(\beta-1)\Big]\cos\phi \ ,
    \end{aligned}
\end{equation}
and for $C^{I,\rm{U}}$,
\begin{equation}
\label{eq:ciutwist}
    \begin{aligned}
          C^{I,\rm{U}}_{(2)}&=\frac{8(2-2y+y^2)}{(x_B-2) y^3} \sqrt{[t(x_B-1)-M^2x_B^2](1-y)}\cos\phi \ ,\\
          C^{I,\rm{U}}_{(3)}&=\frac{8(y-2)[t(x_B-1)-M^2x_B^2]}{(x_B-2) y^3}\left[(y-2)^2+ \frac{2x_B(1-y)(\beta-1) \cos 2\phi}{1+(\beta-1)x_B}\right] \ .
    \end{aligned}
\end{equation}
In the case of longitudinally polarized cross-section, the other three coefficients $\tilde A^{I,\rm{U}}$, $\tilde B^{I,\rm{U}}$ and $\tilde C^{I,\rm{U}}$ are needed and their expansions read
\begin{equation}
\label{eq:atiutwist}
    \begin{aligned}
          \tilde A^{I,\rm{U}}_{(2)}&= \frac{-8(2-2y+y^2)}{x_B y^3}\sqrt{[t(x_B-1)-M^2x_B^2](1-y)}\sin\phi\ ,\\
          \tilde A^{I,\rm{U}}_{(3)}&=\frac{16\left[t(x_B-1)-M^2x_B^2\right](1-y)(2-y)(\beta-1)\sin 2\phi}{y^3\left[1+(\beta-1)x_B\right]}\ ,
    \end{aligned}
\end{equation}
and for $\tilde B^{I,\rm{U}}$,
\begin{equation}
\label{eq:btiutwist}
    \begin{aligned}
          \tilde B^{I,\rm{U}}_{(2)}&=0\ ,\\
          \tilde B^{I,\rm{U}}_{(3)}&=0 \ ,\\
         \tilde B^{I,\rm{U}}_{(4)}&=\frac{8 x_B t\sqrt{[t(x_B-1)-M^2x_B^2](1-y)}\left[4(1-y)+(2-2y+y^2)\alpha\right] \sin \phi}{(x_B-2)y^3\left[1+(\beta-1)x_B\right]} \ ,
    \end{aligned}
\end{equation}
and for $\tilde C^{I,\rm{U}}$,
\begin{equation}
\label{eq:ctiutwist}
    \begin{aligned}
          \tilde C^{I,\rm{U}}_{(2)}&=-\frac{8(2-2y+y^2)}{(x_B-2) y^3} \sqrt{[t(x_B-1)-M^2x_B^2](1-y)}\sin\phi \ ,\\
          \tilde C^{I,\rm{U}}_{(3)}&=\frac{16 x_B \left[t(x_B-1)-M^2x_B^2\right](1-y)(2-y)(\beta-1)\sin 2\phi}{(x_B-2)y^3\left[1+(\beta-1)x_B\right]}\ ,
    \end{aligned}
\end{equation}
where some twist-four terms are not shown as the expressions get complicated. According to the comparison shown in subsection \ref{subsect:interferencecomp}, those expansions converge to the all-order results at large $Q^2$.

\bibliographystyle{jhep}
\bibliography{refs.bib}
\end{document}